\documentclass[12pt]{iopart}
\usepackage{epsfig}
\usepackage[latin1]{inputenc}
\usepackage{graphicx}
\usepackage{iopams}

\begin{document}

\title[Relaxation and physical ageing in network glasses]{Relaxation and physical ageing in network glasses}

\author{Matthieu Micoulaut
}

\address{Paris Sorbonne Universit\'es, LPTMC - UPMC, 4 place Jussieu, 75252 Paris cedex 05, France}

\ead{mmi@lptl.jussieu.fr}

\date{\today}

\begin{abstract}
Recent progresses in the description of glassy relaxation and ageing are reviewed for the wide class of network-forming materials such as $GeO_2$, Ge$_x$Se$_{1-x}$, silicates (SiO$_2$-Na$_2$O) or borates (B$_2$O$_3$-Li$_2$O), all of them having an important usefulness in domestic, geological or optoelectronic applications. A brief introduction of the glass transition phenomenology is given, together with the salient features that are revealed both from theory and experiments. Standard experimental methods used for the characterization of the slowing down of the dynamics are reviewed. We then discuss the important role played by aspect of network topology and rigidity for the understanding of the relaxation of the glass transition, while also permitting analytical predictions of glass properties from simple and insightful models based on the network structure. We also emphasize the great utility of computer simulations which probe the dynamics at the molecular level, and permit to calculate various structure-related functions in connection with glassy relaxation and the physics of ageing which reveals the off-equilibrium nature of glasses. We discuss the notion of spatial variations of structure which leads to the picture of "{\em dynamic heterogeneities}", and recent results of this important topic for network glasses are also reviewed. 
\end{abstract}



\maketitle
\section{Introduction}\label{intro}

From glass windows and light bulbs to lenses and fiberglass insulation, advances in glass science and technology have indisputably played a vital role in enabling modern civilization. The performances of every glass product, especially high-technology glasses such as optical fibers, amorphous phase change DVDs or scratch-resistant flat panel displays including cell-phones, are governed by underlying properties of the glass at the atomic scale. 
\par
What is a glass ? This important question is poorly understood, and remains unanswered today beyond the level of general statements, albeit substantial progress in understanding has been made in the recent years. Solving this problem represents a great challenge for the science, technology, engineering communities at large with obvious technological applications, and has led to an intense research activity that spans over vast fields of inquiry, from theoretical physics of liquids to materials science. When a high temperature liquid can be cooled fast enough, it will usually be able to avoid cristallisation at the melting temperature $T_m$ and will become "{\em supercooled}", which represents a thermodynamic metastable state with respect to the corresponding cristal. At very high temperatures, relaxation times to equilibrium are found to be of the order of the typical atomic vibrational period, i.e. of about $\tau$=0.1-1ps, whereas the viscosity $\eta$ is of the order of 10$^{-4}$-10$^{-2}$~Pa.s. Once the melting temperature has been bypassed, upon further cooling the viscosity and the relaxation time increase dramatically to reach $\tau\simeq$ 100s-1000s and $\eta$=10$^{12}$~Pa.s at a reference temperature that is defined in the literature as the glass transition temperature $T_g$. This empirical definition simply signals that below $T_g$, a liquid will be too viscous to flow on a laboratory timescale (i.e. days or years \cite{zanotto}), and the obtained material will be identified with a glass, i.e. a material that displays all the salient microscopic features of a liquid of but has the macroscopic characteristics of a solid. Once a glass has been obtained, there is, however, still thermal evolution towards equilibrium but its complete experimental study is partially out of reach so that glasses are usually considered as being "{\em out off equilibrium}". As a result, properties will evolve slowly with time, and measurements will depend on the waiting time at which they have begun, a phenomenon known as "{\em ageing}". 
\par
\begin{figure}
\begin{center}
\end{center}
\caption{\label{zach} Typical network-forming glasses: a) A stoichiometric glass former (SiO$_2$, B$_2$S$_3$) whose structure and network connectivity can be altered by the addition (b) of 2-fold coordinated atoms (usually chalcogens, S, Se) that lead to cross-linked chains. The structure can also be depolymerized (c) by the addition of a network modifier (alkali oxides or chalcogenides, Na$_2$O, Li$_2$S, etc.). Glassy dynamics depends strongly on the network topology, i.e. the way bonds and angles arrange to lead to a connected atomic network. Note that only chalcogenides can produce a mixture of these three kinds of basic networks, e.g. (1-$x$)Ge$_y$Se$_{1-y}$-$x$Ag$_2$Se \cite{mitkova}, and for the latter system $x$=0 corresponds to case (b), $y$=33\% corresponds to case (c), and both conditions together ($x$=0,$y$=33~\%) to case (a).
}
\end{figure}
\par
Of very special interest is the field of network glasses (Fig. \ref{zach}), probably the most familiar and, historically, those which have attracted early interest. This is due, in part, to the effect of the X-ray determined structure at the local, intermediate or long range order which has appeared to be central to the understanding of many chemical physical properties including those revealing the slowing down of the dynamics. Appropriate (stoichiometric) mixtures of Group III-Group V elements (e.g. silicon, boron, germanium,...) with Group VI oxydes and chalcogenides (oxygen, sulphur, selenium) lead indeed to a network structure that is imposed at the very local level by building geometrical blocks typical of a short range order \cite{zachar}, e.g. the SiO$_{4/2}$ tetrahedron in silicates. The disordered arrangement of such building blocks on longer scales is then representative of glasses wich form a highly cross-linked network of chemical bonds. Addition of alkali and alkaline earth modifiers alter the network structure, and while such elements are usually present as ions, they compensate by nearby non-bridging Group VI atoms which induce a disruption of the network structure. The presence of such non-bridging atoms lowers the relative number of strong bonds in the glass, and, in the liquid state, will lead to an important modification of the melt viscosity, relaxation time, and various dynamic quantities. In fact, an appropriate alloying of such components permits one to tune dynamic quantities of glass forming liquids in a nearly systematic fashion, allowing for the detection of anomalies which provide an increased insight into the glass transition phenomenon.
\par
In this contribution, we review experimental and theoretical methods and studies that have been reported recently on relaxation and ageing of network glasses. Because of lack of space, and although some reported features may have an intrinsic interest for the purpose, we will most of the time discard the vast body of literature on the relaxation of sphere liquids interacting with very simple potentials that are not "{\em realistic}" for any physical system. While non-equilibrium processes have been rather well characterized and some generic behavior revealed from such crude models, the simplifed form of the interaction (in short, a repulsive core and an attractive interaction at long distances) are unable to reproduce basic features and structural properties of network glasses which are dominated by specific diffraction patterns for which all the details matter. This makes the correct structural reproduction a prerequisite to any theoretical description. For such glasses, aspects of structure control, indeed, a large part of the dynamics and the relaxation phenomena taking place in the vicinity of the glass transition. However, as emphasized above, glassiness is not restricted to the archetypal silica system and/or to inorganic glasses because sugars, food, organic polymers, and more generally complex disordered systems will display this phenomenon as well, while the glass transition can be also achieved through an appropriate densification, and this indicates that glassy behavior can be also observed under jamming \cite{nagel00}.

\section{Property changes across the glass transition}\label{propchange}

Although all quantities remain continuous across the glass transition, rapid changes in physical, thermal, rheological, mechanical etc. properties are observed.  As the temperature increases from low temperature to above the glass transition temperature, many of these properties suddenly change, indeed, and manifest e.g. by important variations in heat capacity, thermal expansion coefficient and/or viscoelastic properties. 
There are two broad classes of measurements based either on rheological properties (viscosity $\eta$ and modulus $G$) or on thermal and thermodynamic properties (enthalpy $H$, volume $V$, heat capacity $C_p$, and expansion coefficient $\alpha$), the latter usually allowing for a neat measurement of the glass transition temperature $T_g$ from calorimetric/dilatometric measurements, as discussed below. 
\subsection{Viscosity plots and fragility}
The evolution of viscosity ($\eta$) is probably one of the most spectacular observed changes as the melt is cooled down to its glass transition. In Fig. \ref{visc}a are represented the evolution of the viscosity for different network-forming liquids (selenium, silica, germania, GeSe$_4$,...) which can be formed in the supercooled state. These are also compared to other prototypal glass forming liquids. It is seen that the increase of viscosity is dramatic for certain substances, and, for organic glass-formers such as o-terphenyl (OTP) or toluene, the temperature decrease can lead to a change in $\eta$ by several orders of magnitudes under only tens of degrees temperature change. 
\begin{figure}[t]
\centering
\caption{\label{visc} a) Behavior of the liquid viscosity $\eta$ of different supercooled liquids as a function of temperature. b) Uhlmann-Angell plot of viscosity rescaling the same data with respect to $T_g/T$ where $T_g$ is defined by $\eta(T_g)$=10$^{12}$~Pa.s. Data taken from \cite{visc_data1,visc_data2,visc_data3,r18}.}
\end{figure}
The behavior of network glass-forming viscosities with T appears to be more moderate, although similar viscosities (10$^{12}$~Pa.s) are obtained at a reference temperature $T_g$ that is usually found to be somewhat higher than the one of organic glass-forming liquids. This temperature usually serves to rescale the viscosity data in an appropriate plot, initially introduced by Laughlin and Uhlmann \cite{uhlmann}, and subsequently popularized by Angell \cite{angell}.
\par
In this plot, the inverse temperature is rescaled with respect to this reference $T_g$ at which the liquid reaches 10$^{12}$~Pa.s, and the same viscosity data of Fig. \ref{visc}a are now shown in Fig. \ref{visc}b. 
An immediate inspection of this figure leads to the conclusion that such supercooled liquids behave very differently close to their glass transition. Some of them show a behavior $\eta(T)$ that follows an Arrhenius law of the form $\eta$=$\eta_\infty\exp[E_A/T]$, and typical examples are silica and germania \cite{visc_data1} or GeSe$_4$ \cite{visc_data2}. However, as one moves downwards in the figure, other liquids (e.g. B$_2$O$_3$, 2SiO$_2$-Na$_2$O, As$_2$Se$_3$ or Se) now exhibit a viscosity behavior that shows an important bending \cite{visc_data2} at intermediate values of $T_g/T$, whereas organic glass-formers (OTP and toluene) display the most pronounced curvature and must involve a super-Arrhenius parametrization of the form $\eta$=$\eta_\infty\exp[E_A(T)/T]$ where the activation energy $E_A$ must now explicitely depends on temperature. 
\par
A simple means to separate liquids for which $E_A$ and the underlying relaxation is independent of temperature from those having an explicit temperature dependence, $E_A(T)$, and exhibiting a rapid increase of $\eta$ close to $T_g/T$=1, is provided by the "{\em strong}" versus "{\em fragile}" classification which permits to distinguish the two types of behaviors \cite{angell,angell2,angell3}. This has led to the introduction of a "fragility index" ${\cal M}$ which is defined by the slope of $\log\eta(T)$ vs $T_g/T$ at $T_g$:
\begin{eqnarray}
\label{frag}
{\cal M}\equiv \biggr[{\frac {d\log_{10}\eta}{dT_g/T}}\biggr]_{T=T_g}
\end{eqnarray}
As detected from Fig. \ref{visc}b, large slopes will, indeed, correspond to fragile glass-formers displaying an important curvature, a variable $E_A(T)$ and a rapid evolution of $\eta$ as one approaches $T_g$, while small slopes (i.e. small ${\cal M}$) will correspond to strong glass-formers having a nearly Arrhenius variation involving a constant $E_A$. Once examined over a wide variety of glass forming liquids \cite{mckenna}, ${\cal M}$ is found to vary between a high value \cite{simon} of ${\cal M}$=214 for a polymer to a low value \cite{r13} of 14.8 for the network-forming liquid Ge$_{22}$Se$_{78}$, a value that is actually found to be lower than the usual reported value of silica (${\cal M}$=20 \cite{mckenna}). 
Note also that the terms introduced "{\em strong}" and "{\em fragile}" are somewhat inappropriate given that they neither connect to underlying mechanical properties, nor to the possible inter-atomic interactions, although most of the strong glass formers have a directional iono-covalent interaction but exceptions do exist \cite{doremus}.
\par
Alternative to the definition (\ref{frag}), several fragility indexes have been introduced to characterise the viscosity behavior of liquids, such as the Bruning-
Sutton \cite{bruning}, Avramov \cite{avramov_frag} and Doremus \cite{doremus} fragility parameters. All attempt to obtain within a single parameter the curvature or slope of the viscosity curves. For instance, the Bruning-Sutton approach \cite {bruning} prefers to relate the viscosity behavior of supercooled liquids with an apparent activation energy for viscous flow which is either constant (for strong liquids) or highly temperature dependent for fragile liquids close to the glass transition. 
\par
At high temperature, most of the liquids seem to converge to a value that is close to $\eta_\infty$=10$^{-4}$~Pa.s. An analysis of viscosity curves \cite{mauro_visc} using a convenient fitting formula for silicate liquids and other liquids including metallic, molecular, and ionic systems, has shown that the high-temperature viscosity limit of such liquids is about 10$^{-2.93}$~Pa.s \cite{mauro_visc}. As there seems to be no systematic
dependence of $\eta_\infty$ on composition, at least for the silicates given the narrow spread around the average value of 10$^{-2.93}$~Pa.s, it is believed that $\eta_\infty$ has some kind of a universal character. A similar analysis has been performed by Russell and co-workers \cite{visc_russell} using alternative fitting formulas for a series of silicate melts, and the prediction of the high temperature viscosity limit has been found to be of about 10$^{-4.3\pm0.74}$~Pa.s to 10$^{-3.2\pm0.66}$~Pa.s. While this issue may be considered as secondary for the present purpose, the degree of universality of $\eta_\infty$ appears to be central to the validity of proposed viscosity fitting formulae (see below) which all assume a universal high temperature limit of viscosity $\eta_\infty$ to derive the low temperature behavior of $\eta(T)$ close to the glass transition. Given the highly non-linear behavior of viscosity with temperature, the departure from an Arrhenius scaling (Fig. \ref{visc}b) reflects the influence of the second derivative of $\eta$ with respect to the inverse temperature that might involve a high temperature parametrized limit embedded in $\eta_\infty$, and not only the effect of the low temperature behavior. As seen from equ. (\ref{frag}), a non-Arrhenius character can, indeed, be solely parametrized with the fragility index, ${\cal M}$, but the latter is a low temperature quantity representing only a first-derivative of the viscosity curve at Tg.

\subsection{Fitting functionals}
The temperature dependence of the viscosity data (or relaxation time given that one has $\eta=G_\infty\tau$ with $G_\infty$ the infinite frequency bulk modulus) is often described approximatively by convenient fitting functionals. The most popular one is given by the Vogel-Tammann-Fulcher (VFT) equation \cite{vft}:
\begin{eqnarray}
\label{equ_vft}
\log_{10}\eta=\log_{10}\eta_\infty+ {\frac {A}{T-T_0}}
\end{eqnarray}
where $A$ has the dimension of an activation energy, and $T_0$ a reference temperature that leads to an Arrhenius behavior for $T_0=0$. An alternative and maybe more insightful form of the VFT equation (\ref{equ_vft}) using explicitely the fragility and the glass transition temperature $T_g$ writes:
\begin{eqnarray}
\label{VFT}
\log_{10}\eta(T)=\log_{10}\eta_\infty+{\frac {(12-\log_{10}\eta_\infty)^2}{{\cal M}(T/T_g-1)+(12-\log_{10}\eta_\infty)}}
\end{eqnarray}
It should be noted that for $T=T_0<T_g$, the viscosity will become infinite, and this might indicate some sort of phase transition on which there has been quite some speculation and debate. For instance, it has been stated that $T_0$ is very close to the Kauzmann temperature $T_K$ \cite{kauz}, a temperature at which the excess entropy of the liquid with respect to the corresponding crystal is supposed to vanish. This connects the kinetic view of the glass transition represented by the evolution of $\eta(T)$ with a thermodynamic one. However, this "{\em entropy crisis}" is rather counterintuituive because one does not expect to have the entropy of a glass becoming lower than the one of the corresponding crystal. In addition, upon further cooling, one neither expects the entropy of a liquid to become negative which would violate the third law of thermodynamics. However, apart from the obvious argument stating that an ordered state of matter (the liquid) should not have an entropy lower than the corresponding ordered state (the crystal), there is no general principle ruling the various contributions of liquid- and solid-like materials so that this apparent paradox may well be consistent with the current experimental observation showing the entropy to vary in very similar fashion to first derivatives of the free energy, without indication that it extrapolates to zero at some finite temperature. 
\par
In a rather systematic study on different glass formers, Richert and Angell \cite{richert} have shown that the ratio $T_K/T_0$ is very close to the one for fragile glass-formers but for common network glasses (SiO$_2$, GeO$_2$) which are strong glass-formers and have $T_0\simeq$~0, the fit of the viscosity behavior using equ. (\ref{equ_vft}) and its connection to the Kauzmann temperature, appears to be non conclusive. However, a fit on Ge-Se liquids using the VFT form has shown that $T_0$ goes through a minimum for 22~\% Ge \cite{angell_book}, i.e. at the same composition at which a fragility minimum has been measured \cite{r13}. Since the discusion about the relationship between $T_0$ and $T_K$ depends on the functional used for the viscosity/relaxation time fitting, conclusions regarding the validity of $T_0\simeq T_K$ can be quite contradictory \cite{tanaka,mauro_sci} 
\par
An interesting and insightful link between the configurational entropy of the liquid and the relaxation (or viscosity) has been suggested by Adam and Gibbs \cite{adam}
\begin{eqnarray}
\label{adam_gibbs}
\eta=\eta_\infty\exp\biggl[{\frac {A}{TS_c}}\biggr]
\end{eqnarray}
Equation (\ref{adam_gibbs}) which is central to many investigations of the glass transition (see below) provides an important connection between a kinetic and a thermodynamic viewpoint of the glass transition. In the latter, the configurational entropy variation with decreasing temperature is believed to result from the reduction of the number of possible minima in the complex energy landscape \cite{heuer1,heuer2,heuer3} characterizing the material. According to this picture, the slowing down of the relaxation, the dramatic increase of $\tau$, result from the reduced ability of the system to explore the landscape in order to locate the energy minimum, driven by the strong reduction of the number of accessible energy minima as the temperature is decreased. Ultimately, structural arrest might occur, and since for an ideal glass at $T=T_K$ one has a single energy minimum only, the configurational entropy vanishes and the relaxation time diverges. 
\par
There is actually not necessarily need to have a functional displaying a diverging behavior at some typical/critical temperature $T_0$. Other popular fitting functionals can reproduce the non-exponentiality of the temperature evolution of viscosity, for example the B\"assler law \cite{bass}:
\begin{eqnarray}
\eta=\eta_\infty\exp\biggl[{\frac {D}{T^2}}\biggr],
\end{eqnarray}
which yields a curvature in the Angell representation of liquids, or the Avramov-Milchev \cite{avramov} form:
\begin{eqnarray}
\label{avramov_equ}
\log_{10}\eta=\log_{10}\eta_\infty+\biggl({\frac {A}{T}}\biggr)^\alpha
\end{eqnarray}
\par
where $\alpha$ is the Avramov fragility parameter \cite{avramov_frag} which is equal to $\alpha$=1 for strong liquids, whereas liquids with higher $\alpha$ values become more fragile. Most of these models lead to a systematic error when they are extrapolated to low temperatures. A more recent and interesting contribution is due to Mauro and co-workers \cite{mauro_pnas} which provide a viscosity model with a clear physical foundation based on the temperature dependence of the configurational entropy. It offers an accurate prediction of low-temperature isokoms without any singularity at finite temperature. Using the Adam-Gibbs model for viscosity (equ. (\ref{adam_gibbs})), the configurational entropy can be expressed as a function of topological degrees of freedom \cite{naumis} (see below) that are temperature dependent and thermally activated \cite{mauro_gupta}, and this leads to the Mauro-Yue-Ellison-Gupta-Allan (MYEGA) equation:
\begin{eqnarray}
\label{myega}
\log_{10}\eta=\log_{10}\eta_\infty+{\frac {K}{T}}\exp\biggl[{\frac {C}{T}}\biggr]
\end{eqnarray}
which avoids a divergence at low temperature found in the VFT equation (Fig. \ref{visc_compar}. Equ. (\ref{myega})), and has been tested and compared to alternative viscosity forms (including VFT) over a hundred of network glass-forming liquids (silicates) and organic supercooled liquids.
\begin{figure}
\begin{center}
\end{center}
\caption{\label{visc_compar} Comparison of different viscosity models \cite{mauro_pnas}. (a) Temperature dependence of viscosity for VFT, Avramov-Michaelev (AM) and MYEGA (current) models, assuming ${\cal M}$ = 60 and $\log_{10}\eta_\infty$ = - 4.(b) Plot of The Adam-Gibbs exponential argument $S_c(T)/BT_g$ for $T>T_g$. The AM model yields a divergent configurational entropy in the limit of $T\rightarrow\infty$. (c) Plot of $S_c(T)/BT_g$ for $T < T_g$. As already mentioned, the VFT model predicts a vanishing of the configurational entropy at a finite $T=T_0$.}
\end{figure}
\par
Given the huge number of possible compositions and thermodynamic conditions, it is nearly impossible to provide a full database of viscosity measurements for network glass-forming liquids. Useful references are the handbooks of Borisova \cite{borisova} and Mazurin \cite{mazurin} for oxides and chalcogenides, and the handbook of Popescu on chalcogenides \cite{popescu}. Instead, we prefer focusing in the forthcoming on reported correlations between the fragility index ${\cal M}$ and some insightful physical or chemical properties. 

\subsection{Fragility relationships}

\subsubsection{Fragility-$T_g$ scaling}

An interesting scaling law relating the fragility index to the glass transition temperature is provided by McKenna and co-workers \cite{mckenna}. There is indeed conventional wisdom suggesting that fragility increases with the glass transition temperature \cite{angell} which implicitly underscores the fact that energy barriers for relaxation increase with increasing T$_g$. 
\par
The derivation of this scaling law combines e.g. the VFT form of equ. (\ref{equ_vft}), and the definition of the fragility (equ. (\ref{frag})), and calculates the fragility and the activation energy $E_A$ as a function of the glass transition temperature. One obtains:
\begin{eqnarray}
\label{cor1}
{\cal M}={\frac {AT_g}{(T_g-T_0)^2\ln{10}}}
\end{eqnarray}
and:
\begin{eqnarray}
\label{cor2}
E_a={\frac {AT_g^2}{(T_g-T_0)^2}}
\end{eqnarray}
Because $T_g$ is of the same order as $T_0$, equs. (\ref{cor1}) and (\ref{cor2}) reveal that ${\cal M}$ and $E_a$ will scale with $T_g$ and $T_g^2$, respectively. Note that this scaling law can be also independently derived from  alternative fitting forms for the viscosity, such as the similar Williams-Landel-Ferry form \cite{wlf}. 
\par
Using such scaling laws, Qin and McKenna \cite{qin} have shown that the correlations (\ref{cor1})-(\ref{cor2}) are fulfilled in a large class of hydrogen bonding organics, polymeric and metallic glass formers. All these systems show a linear increase of ${\cal M}$ with T$_g$, and $E_a$ with $T_g^2$, whereas network glass formers do not seem to follow such scaling laws. From this study, ${\cal M}$ appears to be nearly independent of the glass transition temperature for the reported inorganic glass formers \cite{qin}. But, in a systematic study with composition on chalcogenides, Boolchand and co-workers \cite{r13,asse_punit} have demonstrated that this scaling holds in network glasses \cite{qin} at select compositions. 
\par
Figure \ref{scaling} represents the behavior of the fragility index ${\cal M}$ as a function of measured glass transition temperature $T_g$ for various network-forming glasses. An inspection of As$_x$Se$_{1-x}$ and Ge$_x$Se$_{1-x}$ chalcogenides shows that when the off-stoichiometric melts are followed as a function of tiny changes in composition, the scalings laws (\ref{cor1})-(\ref{cor2}) are fulfilled for only selected compositions corresponding to the stressed rigid and intermediate phase compositions (see below) of these glasses \cite{ip,gese}, i.e. one has a linear increase in ${\cal M}$($T_g$) for $x>$22\% in Ge-Se \cite{r13}, and for $x>$27\% in As-Se \cite{asse_punit}. A least-square fit to such compositions yields to ${\cal M}$=-7.53(5)+0.061(7)$T_g$ and to ${\cal M}$=-17.356+0.060(1)$T_g$ for As-Se and Ge-Se, respectively \cite{asse_punit}. The slope of both curves ($\simeq$ 0.06) is found to be somewhat lower than the one obtained \cite{qin} in polymers (0.28), metallic glass-formers (0.17), and hydrogen bonded liquids (0.25).
\par
\begin{figure}
\begin{center}
\end{center}
\caption{\label{scaling} Fragility as a function of glass transition temperature in As-Se (red, Ref. \cite{asse_punit}), Ge-S \cite{chakra} and Ge-Se liquids (blue, Ref. \cite{r13}), together with data for typical network glass formers \cite{r20b,frag2,frag3} and binary glasses \cite{frag4}.}
\end{figure}
For low glass transition temperatures (i.e. selenium rich in Ge-Se or As-Se), a negative correlation is found which obviously cannot be accounted from the VFT equation given that it would lead to unphysical behaviors such as the divergence of relaxation at a temperature $T_0>T_g$ or an increase of relaxation time $\tau$ with temperature \cite{r13}. It has been furthermore detected \cite{asse_punit} that only the VFT equation (\ref{VFT}) can lead to a positive correlation in the scaling law (\ref{cor1}). For other fitting formula such as the the simple Arrhenius law ${\cal M}$=$A/T_g\ln{10}$ or the MYEGA equation (\ref{myega}), one obtains \cite{asse_punit}:
\begin{eqnarray}
{\cal M}={\frac {K}{T_g}}\biggl(1+{\frac {C}{T_g}}\biggr)\exp\biggl[{C/T_g}\biggr]
\end{eqnarray} 
which decrease as $T_g$ increases.

\subsubsection{Qualitative fragility relationships}

In network glass-forming liquids, the fragility index appears to be also deeply related to structural properties.  Such a basic observation has been recently made and extensively documented by Sidebottom \cite{sidebot}. By considering a two-state model for the glass transition separating the intact bond state
from a thermally excited broken bond state \cite{angell_two_state1,angell_two_state2,angell_two_state3}, a general variation of the configurational entropy with network connectivity can be proposed. In this model, the fragility index ${\cal M}$ is then determined solely by the entropy increase, which is associated with the occurence of new configurations that become accessible when bonds are broken under temperature increase. From bonds between atomic species \cite{angell_two_state1}, the construction can be generalized via a coarse-graining approach to bonds between local structures (such as tetrahedral $Q^n$ species in silicates \cite{amin}) or even bonds between intermediate range order (IRO) structures that are found in borates \cite{borates}. A generic behaviour of the index ${\cal M}$ variation is obtained (Fig. \ref{sidebot}) which demonstrates a universal dependence of the glass forming fragility on the topological connectivity of the network. For the special case where intermediate range order is present, the coarse-graining procedure to a bond lattice indicates that the weakest links (i.e. those which connect IRO) are the most relevant in determining the liquid fragility. 
\par
\begin{figure}
\begin{center}
\end{center}
\caption{\label{sidebot}Fragility scaling \cite{sidebot} of various network forming liquids as a function of connectivity (mean coordination number $\bar r$ or mean local connectivity or mean intermediate range order connectivity.
}
\end{figure}
\par
Building on a similar idea, large scale molecular dynamics simulations of network forming liquids \cite{salmon_prl} show that aspects of topology and IRO control the relaxation of the liquid. Here the network topology is changed by varying the anion polarizability \cite{wilson} of the interaction potential, which governs the intertetrahedral bond angle, and, ultimately it is shown that the fragility is correlated to structural arrangements on different length scales. In particular, ${\cal M}$ is found to increase with the number of edge-sharing (ES) tetrahedral motifs in tetrahedral glass-forming liquids. For the special case of Ge-X (X=S,Se) systems however, the link between edge-sharing tetrahedra and fragility index does not follow such a correlation \cite{ijags,chakra}. A direct measurement of ${\cal M}$ and the ES fraction from Raman spectroscopy indicates, indeed, that the trend in the fragility index is essentially governed by the underlying topology, and, in particular, by aspects of rigidity (see below).
\par
An alternative viewpoint is proposed by Luther-Davies and others \cite{lucas_geasse,lucas_geass} who emphasize the role played by chemical order, and especially by deviation from stoichiometry, rather than topology or rigidity. A joint spectroscopic and fragility experiment is, indeed, analyzed in terms of network dimensionality and stoichiometry change. It is suggested that fragility does not follow predictions from rigidity percolation (in As-Se) but instead correlates with structural dimensionality, whereas for the ternary As-Ge-Se a minimum in fragility is claimed to be associated with a maximum in structural heterogeneity consisting of appropriate ratios of Se-chains and GeSe$_{4/2}$ tetrahedra. This claim is actually contradicted by the early work of Angell and collaborators highlighting the connection between topology/rigidity and fragility index in the same chalcogenide liquid (Ge-As-Se \cite{r18}). The minimum in ${\cal M}$ has been obtained at the network mean coordination number $\bar r$=$<$r$>$=2.4 which is the location of the rigidity percolation threshold \cite{r6,mft1}. The correlation with chemical order is also debated by Boolchand and others who have emphasized the link between fragility minima and isostatic compositions \cite{r13,asse_punit,chakra}, i.e. compositions that are close to the rigidity percolation threshold.  This link between topology and fragility is also evidenced from the investigation of ionic diffusion and fragility on a series of iron-bearing alkali-alkaline earth silicate glasses \cite{mauro_topol}. 
\subsubsection{Fragility - structure relationships}
Glass fragility is also found to display a relationship with atomic ordering on intermediate and extended ranges, a relationship that also connects to the notion of dynamic heterogeneities (see below). Specifically, the structure can be characterized in terms of topological and chemical ordering from neutron diffraction experiments in real (pair correlation function $g(r)$) and reciprocal space (static structure factor $S(k)$) \cite{salmon_nature}. It turns out that the ordering for GeO$_2$, SiO$_2$ and ZnCl$_2$ at distances greater than the nearest neighbor lengthscale can be rationalized in terms of an interplay between the relative importance of two length scales \cite{salmon_prl2}. One of these is associated with an intermediate range that is directly accessed from the structure factor $S(k)$, the other lengthscale is associated with an extended range that is characterized from the decay of Bhatia-Thornton pair correlation functions in real space. With increasing glass fragility, i.e. when moving from GeO$_2$ to ZnCl$_2$, it has been found that the extended range ordering dominates \cite{salmon_prl2}.
\par

\begin{figure}
\begin{center}
\end{center}
\caption{\label{compar}Correlation between relaxation properties, calculated diffusivity D (blue) and activation energy $E_A$ for diffusivity (red axis of the upper panels), and structural properties in the glassy state (black curve, width $\Delta k_{FSDP}$ of the FSDP) in systems with changing pressure (2SiO$_2$-Na$_2$O, \cite{bauchy1,bauchy2}, left panels) or changing composition (As-Se, \cite{bauchy_asse}, right panels). Note that for As-Se, an opposite behavior for $\Delta k_{FSDP}$ with composition is found from Reverse Monte Carlo  \cite{hosoka}.
}
\end{figure}

Having such simple structural correlations in hand, it is not surprising that glassy relaxation has also been investigated by diffraction methods in order to follow low wavevector features with temperature, and, specifically, the first sharp diffraction peak (FSDP) of the structure factor $S(k)$. It has been stated \cite{angell} that fragile liquids usually do not have any structural signature of long range correlations so that the absence of a FSDP is indicative of a fragile glass-forming liquid. This correlation has been verified on a certain number of systems such as the very fragile ZrO$_2$ and $Al_2$O$_3$ \cite{fsdp_m1} which do not exhibit a FSDP, in contrast with the less fragile ZnCl$_2$ which shows a well-defined, but not sharp, FSDP \cite{fsdp_m2}, and with other examples of strong glass forming liquids such as germania and silica displaying a sharp FSDP \cite{wilson_prl,salmongeo2}. 
\par 
By analyzing the typical features of the simulated structure factors with changing thermodynamic conditions (density, composition), Bauchy and Micoulaut \cite{bauchy1,bauchy2,bauchy_asse} have found that anomalies (extrema) in properties associated with glassy dynamics (diffusivity D, activation energy $E_A$ which is proportional to fragility if $T_g$ does not vary too much) are correlated with anomalies in structural features, as revealed by the change in FSDP (peak position $k_{FSDP}$ and width $\Delta k_{FSDP}$, Fig. \ref{compar}). The established correlation suggests that in strong glasses typical lengthscales of distance L=2$\pi$/k$_{FSDP}$ involved in the slower variation of viscosity with $T_g/T$ must lead to a grwith of the FSDP because the latter reflects some repetitive characteristic distance between structural units. Also, the broadening of the FSDP is indicative of a correlation length (Scherrer equation \cite{scherrer}) defined by $\xi$=2$\pi$/$\Delta$k$_{FSDP}$ that tends to maximize for strong glass-forming liquids (Fig. \ref{compar}, bottom). 

\subsubsection{Other fragility relationships}
Network glass-forming fragilities display a certain number of other correlations with physical, chemical or rheological properties that have been reported in the literature.
\par
An inspection of Fig. \ref{visc} also indicates that supercooled liquids with a lower fragility index ${\cal M}$ will lead to higher viscosities at a fixed $T/T_g$. It has been observed \cite{richet} that glass-forming tendency is increased for liquids that are able to increase their melt viscosity down to lower temperatures, i.e. for a given class of materials having a similar $T_g$, glass-forming tendency is increased for melts with lower fragilities, this argument being particularly relevant for binary alloys. Indeed, at eutectics where freezing-point depressions exist, glasses will form more easily because these depressions bring the liquid to lower temperatures and higher viscosities. Such observations are furthermore found to correlate rather well when the composition of the eutectic is compared to compositions at which one has a minimum of the critical cooling rate that is needed in order to avoid crystallization \cite{fang}. 
\par
Correlations have been suggested between fragility and nonexponentiality ($\exp(-t/\tau)^\beta)$) of the structural relaxation characterized by a Kohlrausch exponent $\beta$ \cite{kohlrausch} at low temperature and long times. These have been established \cite{r20b} from a combination of experimental techniques (Fig. \ref{fragility_beta}). When all subgroups of glass formers are represented (organic, polymers, networks) a clear relationship appears and indicates that the fragility index ${\cal M}$ decreases with the exponent $\beta$, i.e. as one moves towards the Debye-type one-step relaxation limit ($\beta$=1), the fragility reaches its minimum value (${\cal M}<$~20). A correlation of fragility to the non-ergodic level of the glass has been also found in the liquid phase as measured directly by dynamic light scattering \cite{sidebot_frag}.
\begin{figure}
\begin{center}
\end{center}
\caption{\label{fragility_beta} Correlations between fragility and some physical properties. Fragility of various glass formers (adapted from \cite{r20b}) as a function of the Kohlrausch stetching exponent $\beta$. Minimum fragility (${\cal M}$=20 is represented by a broken line. The inset shows a correlation between fragility and a parameter ($\alpha$) characterizing the nonergodicity factor $f_c(T)$ (adapted from \cite{scopigno}) {\em via} $f_c(T)=(1+\alpha(T/T_g))^{-1}$.
}
\end{figure}
Other authors have emphasized the central role played by elastic properties such as compressibility \cite{buchenau}. Novikov and Sokolov \cite{novikov} have shown that the fragility of a glass-forming liquids is directly linked to the ratio of the instantaneous bulk and shear moduli, or the Poisson's ratio. Since the latter is related to the very local deformations of the cage structure made by neighbouring atoms, these authors argue that the Poisson ratio should also control the non-ergodicity parameter which controls the fast dynamics of the liquid. However, this result has been challenged \cite{yannopoulos1}, and evidence has been brought that ${\cal M}$ should rather scale with the ratio of the transverse and longitudinal sound velocity. Building on a similar relationship, Ruocco and co-workers \cite{scopigno} have emphasized that the fragility should be linked with the elastic properties of the corresponding glass, quantified from the non-ergodicty parameter $f_c$ accessed from inelastic x-ray scattering (inset of Fig. \ref{fragility_beta}). 
\par
Some other authors have also proposed an empirical relation between the fragility and the strength of fast dynamics which can be quantified from Raman spectroscopy of corresponding glasses \cite{sokolov1,sokolov2}. According to this idea, the ratio of the relaxational to vibrational contributions around the Boson peak has been proposed to relate to the fragility of the liquid. This relationship has been also challenged and no such correlation could be recovered \cite{yannopoulos2} from a careful Raman analysis. It should be stressed that most of these correlations are proposed from a literature survey of a variety of glasses, allowing to cover large ranges in fragility (see y-axis in Fig. \ref{fragility_beta}a). However, when a focus is made only on the network-forming inorganic liquids which have typically ${\cal M}<$70, the correlation becomes less obvious because of the reduced fragility index range. It would certainly be instructive to quantitatively test such correlations for the wide subclass of network glasses.
\par
\begin{figure}
\begin{center}
\hspace{0.5cm}
\end{center}
\caption{\label{frag_homo}Effect of glass homogeneity on fragility measurements \cite{bhag}. Left: Effect of the reaction time $t_R$ on the fragility index ${\cal M}$ in Ge$_{10}$Se$_{90}$. Symbols indicate measurements on different batch parts, whereas the red circles indicate batch average. The inset shows the spread of the measurement (variance $\sigma_{\cal M}$ as a function of $t_R$). Permission from J. Wiley and Sons 2015. Right: Compositional variation of the fragility index ${\cal M}$ of the As-Se system, and the effect of measurement type (mDSC \cite{asse_punit, lucas_asse}; viscosity \cite{musgraves}) and sample homogeneity. Using the same measurement method, glasses with Raman verified homogeneity \cite{ijags} lead to lower fragilities (red symbols, \cite{asse_punit}). Permission from AIP Publishing LLC 2015. 
}
\end{figure}

It must finally be stressed that proposed relationships are often based on melt fragility indexes that can sometimes be flawed by inproper sample preparation, especially for strong liquids having the highest viscosities (e.g. GeSe$_4$). A careful study \cite{bhag} of the effect of melt homogeneity on the measurement of ${\cal M}$ shows that inhomogeneous melts can lead to a spread in measurements, and, eventually, to improper established correlations. The variance of the measurement decreases as glasses homogenize (Fig. \ref{frag_homo}), whereas the mean value increases to saturate at values characteristic of homogeneous glasses \cite{r13,asse_punit}. 

\subsection{Stress relaxation}

Given their disordered atomic structure and their out-off equilibrium nature once $T<T_g$, glasses exhibit residual frozen stresses. This is because atoms are randomly placed in the network, and this situation is energetically unfavorable, at least with respect to a regular crystalline lattice. These stresses can be partially released by moderate temperature annealing, a technique, known by the ancient Phoenicians, that prevents stress-induced cracking, and the related relaxation can therefore represent an alternative and interesting way to probe the dynamics of a glass or a deeply supercooled liquid through its glass transition \cite{trach1,trach2,trach3}. 
\par
When a material is subjected to a constant strain ($\varepsilon_0$), there is a gradual decay in the stress $(\sigma(t)$) that can be analyzed as function of time, and reveals the viscoelastic properties as a function of thermodynamic conditions \cite{stress1}. Note that for glasses having a low glass transition temperature (chalcogenides), aspects of visco-elasticity can be also probed at room temperature. In practice, relaxation is embedded in a relaxation function $\Phi(t)$ that relates the relaxation modulus $G(t)=\sigma(t)/\epsilon_0$ to its initial value $G(0)$, provided that the strain is imposed in an instaneous fashion at t=0. 
\par
A certain number of experiments have shown that such measured relaxation functions $\Phi(t)$ can be conveniently fitted with a stretched exponential that seem to decay to zero at $t\rightarrow\infty$ for most of the inorganic glasses, in contrast with crosslinked polymers \cite{stress1b} or crystals \cite{stress1t} which decay to a finite equilibrium modulus $G_{equ}$=$G(\infty)$. The detail of the analysis \cite{stress1q} also signals that during relaxation the viscoelastic deformation under stress can be decomposed into a sum of an elastic part, an inelastic (or viscous) part and a delayed elastic part, the latter being responsible for the non-linear primary creep stage observed during creep tests. In addition, such a delayed elasticity has been found to be directly correlated to the dispersion of relaxations times of the processes involved during relaxation.
\par 
Measurements using different methods have been made on e.g. Ge-Se \cite{stress2,stress3,stress3b}, Te-As-Ge \cite{stress4}, As-Ge-Se \cite{stress5}, that can be related to structural aspects, while also revealing that a significant part of the stress does not release on experimental timescales (months) in certain compositions for given systems (e.g. GeSe$_4$ in Ge-Se \cite{stress3}). Such stress relaxation measurements have some importance in field of ion exchange glasses (e.g. sodium borosilicates) because of the strengthening of the glass surface that is steadily improved \cite{stress5a,stress5b}, while it has also been considered for soda-lime \cite{stress5ba,stress5bb,stress5bb1} or borosilicate glasses \cite{stress5bc}. For the latter \cite{stress5bc}, a long-time study has permitted the first detectable signatures of glass relaxation far below $T_g$ ($T/T_g\simeq$0.3), and the measure of strain with time, in other words, the relaxation of the glass, follows a stretched exponent with a Kohlrausch exponent $\beta$=0.43 that has been predicted from dimensional arguments \cite{rep_progr_phys}.
\par
For the case of chalcogenides, an interesting perspective is provided by the comparison with the generic behavior of organic polymers \cite{stress1b} since amorphous Se is considered as a glassy polymer made of long chains that are progressively cross-linked by the addition of alloying elements. Stress relaxation is also thought to have some impact on the resistance drift phenomena \cite{stress5c} that is crucial for the functionalities of heavier chalcogenides such as amorphous phase change tellurides (Ge$_2$Sb$_2$Te$_5$).

\begin{figure}
\begin{center}
\end{center}
\caption{\label{stress_fig} Relaxation function $\Phi(t)$ in Ge-Se and Te-As-Ge glasses from stress relaxation measurements \cite{stress3}. Lines represent fits using a stretched exponential $\exp[-(t/\tau)^\beta]$ function.}
\end{figure}

\subsection{Thermal changes}

Signatures for the onset of glassy behavior can be also detected from thermal changes. 

\subsubsection{General behavior}

Figure \ref{thermal}a shows a typical behavior of the enthalpy or the volume, from the high temperature liquid down to the glass.  A rapid cooling from the melt avoids the cristallisation at the melting temperature $T_m$ and brings the liquid into the supercooled r\'egime. As the equilibration cannot proceed further on computer or experimental timescale (see below) given the rapid increase of the relaxation time, the volume or enthalpy curve deviates from the high temperature equilibrium line at the fictive temperature $T_f$ which depends on the cooling rate. A faster cooling rate (Q) will lead to a higher fictive temperature $T_f^Q$, whereas a lower cooling rate (q) will produce a lower fictive temperature $T_f^q$ because equilibration can be achieved down to lower temperatures so that the deviation sets in at $T_f^q<T_f^Q$. As both enthalpy and volume display a different slope below or above the fictive temperature, their derivative with respect to temperature (heat capacity, thermal expansion, Fig. \ref{thermal}b) will lead to an abrupt change with a step-like change across $T_f$ that depends on the cooling rate.

\begin{figure}
\begin{center}
\end{center}
\caption{\label{thermal} Changes in thermal properties at the glass transition. a) When fast enough, a cooling (Q or q) from the high temperature melt (blue curves) avoids cristallisation, leads to the supercooled liquid and, ultimately, to a glass with different fictive temperatures $T_f^q$ or $T_f^Q$. Upon reheating, a hysteresis curve appears (red curve) that is related to the relaxation of the glass. When the glass is aged (at $T_w$), the relaxation to lower energies or volumes will even enhance the effect. b) The first derivative of enthalpy and volume exhibit a step-like increase in the glass transition region, with an inflexion point (i.e. $T_g$) depending on the thermal history.}
\end{figure}

However, even in the obtained glassy state, the material will continue to relax to a lower energy state leading to lower volumes or lower enthalpies. As mentionned above, these relaxation processes happen on timescales that exceed now the laboratory timescale by several orders of magnitude. As a result, the enthalpy/volume curve upon reheating (red curve) will be markedly different, and this effect can be even enhanced (black curve) if the glass is maintained at some temperature $T_w$ for a certain time (days, weeks, years), allowing for an increased relaxation. This experimental situation corresponds to physical ageing, and it can be also detected from the heat capacity or thermal expansion change. Calorimetry permits to track such effects, relaxation and ageing, and when the heat capacity is measured during an upscan (red and black curves in Fig. \ref{thermal}b) a hysteresis loop appears, causing also a heat capacity overshoot at the glass transition. This endotherm peak simply reveals that previously frozen degrees of freedom during the quench (blue curves) are now excited, so that the overshoot is a direct manifestation of the relaxation taking place between the laboratory temperature, or the temperature $T_w$ at which the glass is aged, and $T_g$. Fig. \ref{thermal}b also reveals that the conditions under which the glass transition temperature, the heat capacity $C_p$ or the thermal expansion $\alpha$ are measured, are strongly affected by the thermal history of the melt, and by the ageing of the glass. As a result, the glass transition temperature $T_g$ cannot be uniquely defined ($T_g^q$ or $T_g^w$), from e.g. calorimetry, and its value differs slightly from the reference temperature satisfying $\eta(T_g)$=10$^{12}$~Pa.s.
\par
In addition, extrinsic factors due to the kinetic character of glass transition must be taken into account. For instance, the effect of the heating rate $q^+$ on glass transition temperature dependence is rather well documented in the literature, and obeys the phenomenological Kissinger equation \cite{kissinger}:
\begin{eqnarray}
\label{kiss1}
{\frac {d\ln (q^+/T_g^2)}{d(1/T_g)}}=- {\frac {E_A}{R}}
\end{eqnarray}
or, alternatively, the Moynihan equation \cite{moynihan}:
\begin{eqnarray}
\label{kiss2}
{\frac {d\ln q^+}{d(1/T_g)}}=- {\frac {E_A}{R}}
\end{eqnarray}
which translates, via the assumption of an activated process with energy $E_A$ for the relaxation kinetics of the glass transition, that a higher rate will lead to a higher measured $T_g$. Under the assumption that the activation energy involved in equ. (\ref{kiss1}) and (\ref{kiss2}) is the same as the one involved in the relaxation of the viscous liquids, a measurement of $T_g$ at different scan rates $q^+$ leads to a determination of the fragility for strong glass formers {\em via} ${\cal M}=E_A\ln_{10} 2/RT_g$. Applications of such methods to network glasses can be found for a variety of glasses (e.g. Ge-As-Se \cite{lucas_geasse}, Fig. \ref{lucas_asgese}).
\begin{figure}
\begin{center}
\end{center}
\caption{\label{lucas_asgese}a) DSC curves \cite{lucas_geasse} showing the total heat flow $\dot H_{tot}$ of a ternary Ge$_{6.25}$As$_{32.5}$Se$_{61.25}$ glass obtained for changing scan (heating) rates $q^+$ from 30~K/min to 5~K/min. b) Plot of ln q$^+$ as a function of 1000/$T_g$ for different glasses in this ternary As-Ge-Se. A linear fit leads to values for $E_A$, using equs. (\ref{kiss1}) or (\ref{kiss2}) Reprinted with permission from J. Phys. Chem. B {\bf 118}, 1436 (2015).
}
\end{figure}
\par 
A vast body of data exists in the literature on such measurements of the glass transition region given that $T_g$ is generally determined by calorimetry which measures the change in thermodynamic properties (heat capacity) at the glass transition.

\section{Experimental methods}
\subsection{Scanning calorimetry}

The most frequently used technique for determining the glass transition temperature and studying enthalpy relaxation is differential scanning calorimetry (DSC). The technique measures a difference between an electrical power needed to heat a sample at a uniform scan (heating) rate. As the measured heat flow, once the reference signal has been removed, is proportional to $C_p$ of the system, one has access to the heat capacity across the glass transition in order to investigate effects such as those represented in Fig. \ref{thermal}b.
\par
For DSC, one usally uses the definition for $T_f$ for the enthalpy:
\begin{eqnarray}
H(T)=H_e(T_f)-\int_T^{T_f}C_{pg}(T_1)dT_1
\end{eqnarray}
where $C_{pg}$ is the heat capacity of the glass, and $H_e(T_f)$ is the equilibrium value of $H$ at the fictive temperature. One then has access to the heat capacity by differentiating the equation to obtain:
\begin{eqnarray}
{\frac {dT_f}{dT}}={\frac {\biggl[C_p-C_{pg}\biggr]_T}{\biggl[C_{pe}-C_{pg}\biggr]_{T_f}}}\simeq{\frac {\biggl[C_p-C_{pg}\biggr]_T}{\Delta C_p}}=C_p^N
\end{eqnarray}
where $C_p^N$ is the normalized heat capacity, and it is often assumed that $\Delta C_p$ calculated at the fictive temperature is the same as at temperature $T$ so that $dT_f/dT$ equals $C_p^N$. In practice, these DSC signals are scan rate dependent given that the fictive temperature itself depends on the heating rate (Fig. \ref{lucas_asgese}).

\subsection{AC calorimetry and modulated DSC}
This first introduction of this technique (AC calorimetry) is due to Birge and Nagel who added on the DSC linear signal a small oscillation \cite{nagel1,nagel2,nagel4}. It represents an interesting extension since enthalpy relaxation can be measured in the linear region of small temperature changes, thus avoiding possible non-linear responses of the sample. However, most applications have focused on organic liquids such as glycerol \cite{nagel1}, and we are not aware of any measurements for network glass-forming liquids. 
\par
\begin{figure}
\begin{center}
\end{center}
\caption{\label{prb00_mdsc} mDSC scan of a As$_{45}$Se$_{55}$ glass showing the deconvolution of the total heat flow $\dot H_{tot}$ into a reversing and non-reversing part ($\dot H_{nr}$). The area between the setup baseline and $\dot H_{nr}$ permits one to define a non-reversing enthalpy $\Delta H_{nr}$.
}
\end{figure}
From a statistical mechanics viewpoint, one can consider the imaginary part of the heat capacity, $C_p^*(i\omega)$, as a complex response function (similarly to the dielectric permittivity $\epsilon^*(i\omega)$, see below), and this part is usually associated with the absorption of energy from an applied external field. This frequency-dependent heat capacity is complex, a property that is a direct consequence from the fluctuation-dissipation theorem which applies to a function that is proportional to the mean-square fluctuations in entropy, $k_BC_p=\langle S^2\rangle$, which in turn have a spectral distribution. Birge \cite{nagel4} signals that in ac calorimetry there is no net exchange of energy between the sample and its
surroundings but there is a change in the entropy of the surroundings that is proportional to $C_p$, and the second law of thermodynamics ensures that $C_p>$ 0.  
\par
Kob and co-workers \cite{kobac} have given a statistical mechanics description of AC calorimetry by deriving a relationship between the frequency dependent specific heat and the autocorrelation function of temperature fluctuations. Using molecular dynamics simulations of silica, they have shown that both real and imaginary part of $C_p$ exhibit the usual shape of complex response functions, the out of phase (imaginary) component displaying a maximum corresponding to the typical $\alpha$-relaxation peak at $\omega_{max}\tau$=1. The dependence of $\tau$ with temperature has been found to agree with the one determined from the long-time ($\alpha$-relaxation) behavior of the incoherent scattering function. This indicates that AC calorimetry (and its extension to mDSC) can be used as a spectroscopic probe for structural relaxation in glasses \cite{descamps1,descamps2}. 
\par
\begin{figure}
\begin{center}
\end{center}
\caption{\label{mdsc_freq} (a) An example of \textit{in-phase} and \textit{out-of phase} components of complex C$_p$ from modulated-DSC scans as a function of modulation frequency for a  Ge$_x$Se$_{100-x}$ melt at x = 10\% \cite{r13}. (b) Log of relaxation time ($\tau$) as a function of T$_g$/T yielding fragility, m, and activation energy E$_a$ from the slope of the Arrhenius plots at indicated compositions (x). Permission from AIP Publishing LLC 2015.}
\end{figure}

An improved technique, modulated DSC, has appeared nearly two decades ago, and represents a promising extension of Ref. \cite{nagel1}, with frequency ranges being reduced by several decades, and allowing for investigations of thermal conditions with increased relaxation times, close to the glass transition. This technique is rather versatile since measurements are performed in the course of an usual DSC scan. It is thus likely to offer a new convenient way to probe molecular mobility in connection with relaxation.
In practice, and as in AC calorimetry, one superposes a sinusoidal variation on the usual linear T ramp of the form $T(t)=T_{dsc}(t)+\sin(\omega t)$. In direct space, this technique permits to deconvolute \cite{mdsc1,mdsc2} the total heat flow ($\dot H_{tot}$) into a reversing and a non-reversing component. The reversing component ($\dot H_r$) tracks the temperature modulation at the same frequency $\omega$ while the difference term (denamed as non-reversing), $\dot H_{nr}=\dot H_{tot}-\dot H_r$ does not, and captures most of the kinetic events associated the slowing down of the relaxation close to the glass transition (Fig. \ref{prb00_mdsc}).
The decomposition into several heat flow components can be formally written as:
\begin{eqnarray}
\label{dotH}
\dot H_{tot}=\dot H_{rev}+\dot H_{nr}=C_p(T)\dot T+f(t,T)
\end{eqnarray}
where $\dot H_{rev}$ and $\dot H_{nr}$ represent the reversing
heat flow and the non-reversing heat flow respectively. The function
$f(T,t)$ contains most of the time and temperature dependent processes. When studying the glass transition, this function becomes important when one reaches the transition
temperature, because the system needs more and more time to
equilibrate upon temperature change, and this is, in fact, observed in the example displayed (Fig. \ref{prb00_mdsc}). Frequency corrections can be realized to provide a nearly independent measure of $C_p$ and its inflexion point serves to determine the glass transition temperature. When properly placed with respect to the measurement baseline, the area $\dot H_{nr}$ leads to the definition of a non-reversing heat enthalpy ($\Delta H_{nr}$, Fig. \ref{prb00_mdsc}) that has some importance in the field of rigidity transitions (see below). The complex heat capacity $C_p^*(\omega)$ can be linked to the sinusoidal part of the heat flow response contained in both contributions of $\dot H_{tot}$, either through the base frequency ($\dot H_{rev}$) or through the secondary harmonics ($\dot H_{nr}$).
\par
In order to probe the dynamics ranging from very short time scale of pico-to-nanoseconds typical of high temperature, to the low temperature domain of $\mu$s to seconds, different sets of experiments can be used, and these comprise neutron scattering, dielectric and calorimetric spectroscopy. These methods can be seen as complementary given that they do not probe the same timescale, the former focusing essentially on the high temperature r\'egime when $\tau$ is very small.
\par
In a mDSC measurement, a decomposition of the complex $C_p^*(\omega)$ into 
real and imaginary parts leads to curves which have the characteristic forms of the complex susceptibility of a relaxation process (Fig. \ref{mdsc_freq}a), as also accessed from dielectric measurements. In particular, for a given temperature the imaginary part $C_p''$ peaks at a frequency $\omega_{max}\tau=1$ which permits to access to the relaxation time, and this calorimetric method has shown to lead to similar results regarding $\tau(T)$ when compared to dielectric data \cite{descamps1,descamps2}. When such determined relaxation times $\tau=1/\omega_{max}$ are represented in an Arrhenius plot close to the glass transition, the T dependence of $\tau(T)$ permits determining the fragility (Fig. \ref{mdsc_freq}b).

\subsubsection{Dielectric relaxation}

Similarly to modulated DSC, dielectric relaxation permits, via the response of the system to an external and oscillating electric field to provide an information about the relaxation behavior. The complex permittivity $\varepsilon^*=\varepsilon'-i\varepsilon"(\omega)$ can be studied as a function of frequency, and the imaginary part $\varepsilon"(\omega)$ (the loss spectra) which also peaks at $\omega_{max}\tau$=1 can be conveniently fitted in the high temperature r\'egime (Debye) as well as in the supercooled r\'egime using empirical functions (Havriliak-Negami \cite{negami}, Cole-Cole \cite{colecole}) to access the relaxation time as a function of thermodynamic conditions, and, particularly, temperature. 
\par 
While this technique has been largely used for the study of organic glass-formers \cite{crichert1,crichert2,crichert3} due to their increased dielectric strength, the study of network glasses has been mostly restricted to solid electrolytes containing modifier ions (Na, Li,...). In this case, a measurement of the complex conductivity $\sigma^*(\omega)$ permits to determine, {\em via} the electrical modulus $M^*(\omega)$ the frequency behaviour of the permittivity \cite{side1}:
\begin{eqnarray}
\epsilon^*(\omega)={\frac {1}{M^*(\omega)}}={\frac {\sigma^*(\omega)}{i\omega\epsilon_0}}
\end{eqnarray}
Again, the frequency $\omega_{max}$ at which the out-of-phase component $\epsilon"(\omega)$ is maximum leads to a determination of the relaxation behavior of the ions, and related characteristics of glassy relaxation ($\beta$,$E_A$,...). Typical applications to silicate \cite{ngai4,germ_relax0}, borates \cite{germ_relax00,germ_relax000,germ_relax0000} or thioborates \cite{swmartin1}, germanates \cite{germ_relax}, and phosphates \cite{novita1} can be found in the literature. 
\subsubsection{Scattering functions}
Given the same timescale involved (ns-ps), inelastic neutron scattering experiments can provide directly access to relaxation functions that can be compared with statistical calculations using molecular simulations (see below).
Measured double differential cross section are proportional to so-called  scattering functions $S({\bf k},\omega)$ which, by Fourier transform, can be related to the intermediate scattering function $F({\bf k},t)$. The coherent and incoherent parts of the scattering function allows determining a coherent part of the intermediate scattering function $F_{coh}({\bf k},t)$ providing information about collective particle motion:
\begin{eqnarray}
\label{kargl_eq}
F_{coh}({\bf k},t)={\frac {1}{N}}\sum_{i=1}^N\sum_{j=1}^N\langle e^{i{\bf k}.{\bf r}_i(0)}e^{-i{\bf k}.{\bf r}_j(t)}\rangle,
\end{eqnarray}
and an incoherent (self) part $F_{inc}({\bf k},t)$ that focuses on single particle motion:
\begin{eqnarray}
F_{inc}({\bf k},t)=F_s({\bf k},t)={\frac {1}{N}}\sum_{i=1}^N\langle e^{i{\bf k}.{\bf r}_i(0)}e^{-i{\bf j}.{\bf r}_i(t)}\rangle
\end{eqnarray}
The latter, which follows the Fourier components of density correlations, characterizes the slowing down of the relaxation that can be investigated in liquids for different temperatures down to a temperature close to the glass transition. 
\par
\begin{figure}
\begin{center}
\end{center}
\caption{\label{stretchf} Schematic representation of density correlation function (intermediate scattering or incoherent scattering function) $F_s(k,t)$ in a viscous liquids at different temperatures: high temperature liquid (red) with a simple exponential decay, deep supercooled (glass, green), and intermediate temperatures (black) displaying both the $\beta$-relaxation plateau, and the $\alpha$-relaxtion r\'egime which can be fitted by a stretched exponential of the form $\exp[-(t/\tau)^\beta]$.
}
\end{figure}
Such correlation functions (e.g. $F_{s}({\bf k},t)$) display some salient features for most of the glass formers (Fig. \ref{stretchf}). At high temperature, $F_s({\bf k},t)$ decays in a simple exponential way of Debye type that takes only into account the interactions between particles (microscopic r\'egime), and $F_{s}({\bf k},t)$ goes to zero rather rapidly, typically the ps timescale for e.g. a silicate liquid at 1800~K (see Fig. \ref{kargl_fig}). For smaller times, smaller than the typical microscopic times (ps) , the time dependence is quadratic in time and arises direcly from the equation of motion of the moving atoms. As the temperature is decreased however, the decay of $F_{inc}({\bf k},t)$ cannot be described by a simple exponential function, and a plateau sets in at longer times. The time window associated with this plateau is called the "$\beta$-relaxation" and this window increases dramatically as the temperature continues to be decreased, driven by the cage-like dynamics of the atoms which are trapped by slowly moving neighbors. This leads to a nearly constant value for density correlations in Fourier space, and is associated with a non-ergodicity parameter $f_c$ characterized by the plateau value $F_s(k,t)\simeq f_c$ in the $\beta$-relaxation r\'egime. 
However, for times which are much larger than this $\beta$-relaxation r\'egime, atoms can escape from the traps, can relax, and jump to other atomic traps so that $F_{s}({\bf k},t)$ can eventually decay to zero (Fig. \ref{kargl_fig}), and its behavior is appropriately described by a stretched exponential of the form $F_{s}({\bf k},t)\simeq\exp[-(t/\tau)^\beta]$ where $\tau$ is the (structural) relaxation time associated with the so-called "$\alpha$-relaxation" r\'egime, and $\beta$ is Kohlrausch parameter introduced previously. For a full review on the stretched exponential and the nature of the parameter $\beta$, we refer the reader to Ref. \cite{rep_progr_phys}.
\par
\begin{figure}
\begin{center}
\end{center}
\caption{\label{kargl_fig} Density correlation functions of $F_s(q,t)$ of sodium and lithium disilicate melts (1600~K, \cite{kargl}) in the fast alkali relaxation regime at different wavevectors $k$ (circles, $k$=0.4~\AA$^{-1}$; triangles, $k$=1.0~\AA$^{-1}$). The solid lines represent fits with a stretched exponential function.}
\end{figure}
As one finally approaches the glass transition, because of the dramatic increase of the relaxation time $\tau$, this $\alpha$-r\'egime is barely observable, and the $\beta$-relaxation plateau extends to timescales which are of the order or larger than the typical laboratory timescale (green curve in Fig. \ref{stretchf}).
\par
There have been quite a large body of research on inelastic neutron scattering applied to the determination of the glassy dynamics in network forming liquids. Kargl et al. \cite{kargl} have used inelastic scattering in alkali silicate liquids to determine the viscous dynamics, the relaxation time $\tau(T)$ and the non-ergodicity parameter $f_c$. It is found that in such liquids fast relaxation processes happen on a 10 ps timescale (accessed from a neutron time of flight experiment) and are associated with the decay of the Na-Na structural correlations, whereas slower processes are found on a 10 ns timescale, and involve the decay of network forming species related coherent correlations
(Si-O, O-O and Si-Si). Such an observation is actually quite systematic for binary modified glasses which contain an alkali modifier, and a certain number of examples of such investigations can be found in the literature (e.g. sodium aluminosilicates \cite{kargl1}).
\subsubsection{Nuclear Magnetic Resonance}

The investigation of Nuclear Magnetic Resonance (NMR) spectra as a function of temperature and/or composition also permits accessing properties of relaxation \cite{NMR0}. The typical time $T_1$ of spin-lattice relaxation (SLR) can be used to link the dynamics of certain structural fragments resolved by NMR with timescales related with $T_1$. This time is, indeed, associated with the mechanism that couples the equilibration of magnetization for a given linewidth (i.e. a local structure) with the effect of the (lattice) neighbourhood. 
\par
\begin{figure}
\begin{center}
\end{center}
\caption{\label{sen_nmr} Full width at half maximum of a $^{77}$Se NMR resonance associated with Se-Se-Se chains as a function of temperature in Ge$_x$Se$_{1-x}$ glass-forming liquids \cite{nmr0}. Systems have been separated into subclasses satisfying $\bar r$=2+2$x\geq$2.4 or $<r>$~$\leq$~2.34. Here, $\bar r$=2.4 represents the rigidity pecolation threshold \cite{mft1}.}
\end{figure}

In the liquid state, the evolution with temperature of the site associated linewidth and their caracteristics (e.g. full width at half maximum) provides direct indication of how structural fragments impact the evolution with time and temperature \cite{nmr0,nmr00,nmr000,nmr40,nmr50}. Linewidths are, indeed, expected to narrow upon temperature increase and since such linewidths can be associated with specific structural features or species, one can have access to aspects of relaxation, and how the local structure affects the dynamics. For instance, for the case of silicate species \cite{nmr00}, it has been found that the typical NMR timescale involved in Na cation exchange between Si tetrahedral species was identical to the one determined from viscosity measurements. This indicates that the local Si-O bond breaking represents the main contribution to viscous flow in silicate liquids. A similar conclusion has been drawn in borosilicates \cite{nmr60}, and represents the central result of this topic, i.e. the investigation of glass relaxation from NMR studies.
\par
In the glassy state, applications to chalcogenides (Ge-Se) have shown that such SLR time scales are significantly smaller for Se-Se-Se chain environments (10$^{-9}$~s) as compared to Ge-Se-Ge fragments (10$^{-6}$~s, \cite{nmr1,nmr2}), and consistent with the fact that these chains are mechanically flexible, and lead to an enhanced ease to relaxation that is also driven by composition (Fig. \ref{sen_nmr}). However, an opposite behavior is found for a similar system (As-Se, \cite{nmr3}), such contradicting trends being eventually driven by the magnitude of the applied magnetic field $H_0$, and how the corresponding frequency ($\omega=\gamma H_0$, $\gamma$ being the Larmor frequency) lies with respect to the characteristic timescale for dipolar coupling fluctuations \cite{nmr2}.  

\subsubsection{Photoelectron correlation spectroscopy}
There is also a possibility to use photon correlation spectroscopy (PCS) to probe the dynamics of the glassy relaxation \cite{xray19} in order to extend the measurement of correlation functions to the $\mu$s-s timedomain, i.e. very close to the glass transition. Another more recent powerful experimental technique using X-ray induced photoelectrons has also emerged thanks to instrumental developments \cite{xray18}, and to an increased flux and coherence of X-ray beams. For a full review on the technique, see Ref. \cite{rev_xpcs}. At present, the development and the first applications of the technique have mostly focused on metallic glasses \cite{xray20,xray21,xray22}. It has been found that for such systems the dynamics evolves from a diffusive atomic motion in the supercooled liquid, to a stress-dominated dynamics in the glass, characterized by a complex hierarchy of aging regimes.
\par
For the case of network glasses, only selected but promising studies have been reported on liquid selenium \cite{cazzato1}, silicates both in glasses \cite{rruffle,plomb} and deep supercooled liquids \cite{sidebot1}, and phosphates \cite{sidebot2,sidebot2b}. In the silicates, it has been found that even at 300~K both lithium and sodium silicate glasses are able to relax \cite{rruffle} and rearrange their structure on a length scale of a few Angstroms, thus contradicting the general view of an almost arrested dynamics. The measured relaxation time has been found
to be surprisingly fast, even hundreds of degrees below $T_g$, a result that contrasts with the common idea of an ultraslow dynamics but which is consistent with the measured relaxation behaviour \cite{stress5bc} of a borosilicate glass far from the glass transition temperature ($T/T_g\simeq$0.3). The findings seem also to suggest the existence of a distinct atomic scale related relaxation dynamics in glasses, not taken into account by any previous study.
\par
In the binary phosphates Na$_2$O-P$_2$O$_5$, Sidebottom and co-workers \cite{sidebot2} have analyzed the relaxation of the glass-forming liquids, and have shown that the substantial increase in fragility is accompanied by a progressive depolymerization of the network structure, suggesting that the  viscoelastic relaxation in network-forming liquids is controlled only by the topology of the covalent structure. Similarly to the case of silicates \cite{kargl1}, a decoupling of ionic motions from those of the network species seem to occur as the glass transition is approached.
\section{Simple models for enthalpic relaxation}
\subsection{Tool-Narayanaswamy-Moynihan equation}
Probably, the simplest way to quantify enthalpic relaxation due to physical aging and structural relaxation is provided by Tool's concept of fictive temperature \cite{tool} which permits defining the enthalpy of a glass as a function of $T_f$: 
\begin{eqnarray}
\label{debut}
H(T,T_f)=H(T_0,T_f)+\int_{T_0}^{T_f}C_{pm}(T_1)dT_1+\int_{T_f}^TC_{pg}(T_1)dT_1
\end{eqnarray}
where $C_{pm}$ and $C_{pg}$ are specific isobaric heat capacities of the metastable supercooled melt and the glass, respectively, and $T_0$ is an arbitrary sufficiently high reference temperature at which the system is in a metastable thermodynamic equilibrium. Narayanaswamy generalized Tool's model \cite{naray} by incorporating a distribution of relaxation times, and obtained the following expression for the fictive temperature that can be calculated for any thermal history:
\begin{eqnarray}
T_f(t)=T(t)-\int_0^tdt_1\biggl({\frac {dT}{dt}}\biggr)_{t_1}M_H[\xi(t)-\xi(t_1)]
\end{eqnarray}
where $M_H$ is the Kohlrausch-William-Watts (KWW) relaxation function introduced previously:
\begin{eqnarray}
\label{kww}
M_H(\xi)=\exp[-\xi^\beta]
\end{eqnarray}
and, as for the case of the long-time behavior fitting of the intermediate scattering function, the Kohlrausch exponent $\beta$ (0 $<\beta <$ 1) characterizes non-exponentiality. The argument of $M_H$ is supposed to be inversely proportional to the width of a distribution of relaxation times of independent relaxation processes, $\xi$ being a dimensionless reduced relaxation time:
\begin{eqnarray}
\xi(t)=\int_0^t{\frac {dt_1}{\tau(t_1)}}.
\end{eqnarray}
The contribution to the relaxation time $\tau(T,T_f)$ is controlled by a non-linearity parameter $x$ (0$<x<$1) according to the Tool-Narayanaswamy-Moynihan (TNM) equation :
\begin{eqnarray}
\label{fin}
\tau=\tau_0\exp\biggl[{\frac {x\Delta h^*}{RT}}+{\frac {(1-x)\Delta h^*}{RT_f}}\biggr]
\end{eqnarray}
where $\tau_0$ is a constant, $\Delta h^*$ is an apparent activation energy, and $R$ is the universal gas constant. Having set these equations, the time evolution of the normalized molar heat capacity can be obtained, and directly compared to the standard output of a DSC measurement.
\par
The combination of these equations (\ref{debut})-(\ref{fin}) with Boltzmann superposition (i.e. the  Tool-Narayanaswamy-Moynihan (TNM) phenomenology) is the most  frequently used nonlinear phenomenology for the study of enthalpy relaxation.
\begin{figure}
\begin{center}
\end{center}
\caption{\label{richert} Plot of normalized heat capacity $C_p$ for As$_2$Se$_3$ with temperature for cooling rates of  -0.31, 2.5, and 20 K/min and heating at a common rate of +10 K/min. Left: direct TNM modelling \cite{as2se3_moy}. Right: TNM modelling using a heterogeneous dynamics with a distribution of relaxation times \cite{richert_TNM}. Symbols are the experimental DSC data taken from Ref. \cite{as2se3_moy}. Permission from AIP Publishing LLC 2015.
}
\end{figure}

\subsubsection{Applications}
There are many applications of the TNM phenomenology to network glasses using either DSC signals for enthalpic relaxation, or dilatometric measurements for volume relaxation (see Fig. \ref{richert}).
\par
Enthalpic structural relaxation in As$_x$Se$_{100-x}$ glasses from DSC has been described within this TNM model \cite{malek1}, and connections can be made with structural changes. A combination of mercury dilatometry and DSC \cite{malek2} on certain network glasses (Ge$_2$Se$_{98}$ and As$_2$Se$_{98}$) using, again, the TNM model shows that enthalpic and volumetric relaxation are nearly identical and lead to the same $\Delta h^*$ value, a situation that is also met for elemental selenium. In this series of selenide network glasses, there has been a lot of attention \cite{malek2a,malek2b,malek2c,malek2d} on the relaxation of the latter (pure Se) whose network is made of long Se chains \cite{malek3}. The TNM parameters (pre-exponential factor $\tau_0$ and the apparent activation energy $\Delta h^*$) have been found to be very close to the activation energy of viscous flow. Other typical applications of the TNM model to the analyis of enthalpy/volume  relaxation can be found for B$_2$O$_3$ \cite{debolt}, Ge-Sb-Se-Te \cite{malek4}, Ge$_{15}$Te$_{85}$ \cite{malek5}, As$_2$S$_3$ \cite{malek6} or Ge$_{38}$S$_{62}$ \cite{malek7} or Te-Se \cite{malek8}.

\subsubsection{Limitations}

One obvious drawback is that the TNM parameters ($\beta$,$\Delta h^*$) do not seem to be fully independent as emphasized by Hodge \cite{hodge}. The TNM parameters of 30 organic and inorganic glass-formers have been collected, and a strong correlation betwen the parameters emerges (Fig. \ref{limit})
\begin{figure}
\begin{center}
\end{center}
\caption{\label{limit} Plot of TNM parameters showing the $x$ versus $\beta$ correlation indicating the possible correlation between fitting parameters for various glass-forming systems: polymers, polystyrenes, inorganics, fluoropolymer (ZBLA), PMMA. Adapted from Ref. \cite{hodge}.}
\end{figure}
It is suggested \cite{hodgeb} that these correlations are somehow expected because the fitting parameters are not orthogonal in parameter search space, and because the TNM parameters themselves have large uncertainties that are also correlated. In addition, there is obviously lack of a physical model that could provide an interpretation for the parameter and the parameter correlations, the explicit account of a KWW behavior (equ. (\ref{kww}) being also introduced by hand in the theory. Also, relatively subtle distortions of the experimental data lead to evaluated TNM parameters that are highly inconsistent.
\par 
One way to circumvent these problems is to provide other indirect fitting methods allowing, for instance, to evaluate the apparent activation energy of enthalpic relaxation $\Delta h^*$ from the dependence of $T_f$  on the heating rate $q^+$ using the Kissinger formula (\ref{kiss1}), in combination with a determination of $T_f$ using the equal enthalpic area method across $T_g$ \cite{moy1,moy2}. Additional indirect fitting techniques \cite{moy3} use the shift of the relaxation peak with the temperature during so-called intrinsic cycles of the glass transition during which the cooling-to-heating ratio is maintained constant. For more details on alternative fitting techniques, see Ref. \cite{moy4,moy5,moy6}. Application to certain glassy selenides (e.g. Ge$_2$Sb$_2$Se$_5$) shows that a full account of the enthalpic relaxation cannot be achieved from the TNM equation. While the results exhibit a significant dependence on experimental conditions, part of the TNM parameters needs, indeed, to be confirmed by such alternative methods \cite{moy7}.
\par
Other important limitations concern the case of the poor reproduction of huge overshoot peaks that manifested after extremely long annealing  periods, a failure that may result from the simple exponential behavior for $\tau$ (equ. (\ref{fin})). This problem can be solved by assuming a heterogeneous dynamics of dynamically correlated domains which relax in an exponential fashion and almost independently from each other \cite{richert_TNM}. In this case, the enthalpic overshot for a DSC up-scan is substantially improved (Fig. \ref{richert}, right) with respect to the basic modelling \cite{as2se3_moy} using equs. (\ref{debut})-(\ref{fin}). 
Also, the TNM framework does not account for multiple glass transition temperatures that are found in heterogeneous glasses or in glasses having a reduced glass forming tendency, i.e. with $\Delta T=T_x-T_g$ quite narrow, $T_x$ being here the crystallization  temperature which leads to a strong endotherm peak in DSC signals. More references and examples on the TNM model limitations can be found in Refs. \cite{lim1,lim2,lim3,lim4,lim5}. 
\par
A comparative method introduced by Svoboda and Mal\`ek \cite{lim6} builds on the parameter control of the TNM approach through the cycling of all possible theoretically calculated datasets with different relaxation curve profiles. This opens the possibility to apply the TNM equations even to extremely distorted differential scanning calorimetry data \cite{lim7}.

\subsection{Adam and Gibbs theory}

Such modeling procedures are actually self-consistent with other simple thermodynamic approaches as emphasized in some examples (e.g. selenium \cite{malek_adam_gibbs}). For moderate departure from equilibrium, it has been, indeed, shown that volume and enthalpy relax in the same way when analyzed from the TNM approach or from the Adam-Gibbs model which relates the relaxation time to the configuration entropy of the liquid.
\par
As emphasized above, this Adam-Gibbs (AG) model \cite{adam} has a rather large importance in the field of glass transition because it relates the relaxation time towards equilibrium, a crucial quantity in the context of glassy relaxation, with the thermodynamic properties and the accessible states for the liquid. In the initial approach, it is assumed that relaxation involves the cooperative rearrangement of a certain number $z$ of particles. This involves a transition state activation energy between at least two stable configurations so that the configurational entropy $S_c$ must satisfy 
$S_c\geq k_B\ln 2$. The configurational entropy can be exactly calculated under the additional assumption that i) the size of these cooperative arranging regions is independent and ii) that these represent equivalent subsystems of the liquid, and linked with the relaxation time. One then obtains equation (\ref{adam_gibbs}). Such a deep and interesting connection between transport coefficient and entropy has been verified directly, i.e. by representing dynamic properties, e.g. diffusivity, as a function of $A/TS_c$ in a semi-log plot from computer simulations of water \cite{starr,sciortino_ag1} silica \cite{sciortino_ag2} or OTP \cite{sciortino_ag3}. It can also be obtained from a simultaneous measurement of both the viscosity/diffusivity and the heat capacity in silicates \cite{richet_ag1,richet_ag2,richet_ag3,richet_ag4,richet_ag5} and water \cite{angell_gs}, given that one has:
\begin{eqnarray}
S_c=\int_{T_K}^T {\frac {\Delta C_p(T_1)}{T_1}}dT_1
\end{eqnarray}
where $T_K$ represents the Kauzmann temperature at which the entropy vanishes (Fig. \ref{adam_gibbs_verif}). In simulations, $S_c$ has been mostly determined from a general thermodynamic framework taking into account the vibrational contributions \cite{sciortino_ag4} from quenched inherent structures (see below). In the experimental determination for the validity of the AG relation, $S_c$ is determined from calorimetric measurements of i) the crystal heat capacity from low temperature up to the melting temperature $T_m$, ii) the enthalpy of melting of the crystal at the melting point, and iii) the heat capacity of the supercooled liquid from $T_m$ to low temperature. It is found that equation (\ref{adam_gibbs}) is satisfied in several families of silicate melts (Fig. \ref{adam_gibbs_verif} right). Note that such studies have been also realized in fragile organic glass-formers \cite{roland}.
\begin{figure}
\caption{\label{adam_gibbs_verif} Verification of the Adam-Gibbs relation. Diffusivity or viscosity of liquids plotted as a function of $1/TS_c$ in simulated densified water \cite{sciortino_ag1} (left) or experimentally measured silicates \cite{richet_ag1} (right). Permission from Elsevier 2015.}
\end{figure}
The Adam-Gibbs expression (\ref{adam_gibbs}) linking $\tau$ with the configurational entropy gives a good account of the non-linearity observed in enthalpy relaxation of amorphous polymeric, inorganic, and simple molecular materials near and below Tg \cite{heat4}. Equation (\ref{adam_gibbs}) can be also modified if a hyperbolic form is assumed for the heat capacity \cite{heat1} which seems to be fulfilled in selected glass-forming systems. In this case, $S_c$ behaves as $C/(1-T_2-T)$ \cite{heat2,heat3}, and leads directly to a VFT behavior (equ. (\ref{equ_vft})) that has $T_2=T_0=T_K$ \cite{richert,tanaka,mauro_sci}. This simple Adam-Gibbs picture \cite{adam}, although powerful, contains a certain number of obvious limitations that have been discussed in e.g. \cite{dyre} (see also the above discussion on $T_0\simeq T_K$). For instance, the rearrangement of cooperative regions is not restricted to supercooled liquids given that such phenomena also take place in crystals with diffusion of correlated vacancies or interstitials. Similarly, the emergence of divergent lengthscales as revealed by the growing heterogeneous dynamics setting on when one approaches $T_g$ is in contradiction with the assumption of independent and equivalent regions. 

\subsection{Harmonic models}

An alternative path for the description of glassy relaxation is given by the wide class of kinetic constraint models (KCM) for which the thermodynamics is trivial but not the dynamics. Complicated dynamics emerges, indeed, from local time-dependent rules, and is able to reproduce some of the standard phenomenology of the glass transition. Among these models, the simplest one can be based on the linear elasticity of the glass and the corresponding interaction can be considered as harmonic.  
\subsubsection{Kirkwood-Keating approach}
The justification of applicability to covalent amorphous networks can be made on the basis of the Kirkwood-Keating interaction potential that has been introduced to fit elastic and vibrational properties \cite{keat1,keat2,keat3}. It represents a semiempirical description of bond-stretching and bond-bending forces given by
\begin{eqnarray}
\label{kk}
V={\frac {3\alpha}{16d^2}}\sum_{i,j}\biggl({\bf r}_{ij}.{\bf r}_{ij}-d^2\biggr)^2
+{\frac {3\beta}{8d^2}}\sum_{k\{i,i'\}}\biggl({\bf r}_{ki}.{\bf r}_{ki'}+{\frac {1}{3}}d^2\biggr)^2
\end{eqnarray}
\begin{figure}
\begin{center}
\end{center}
\caption{\label{keat_fig} Radial distribution function of amorphous silicon \cite{keat4} modelled using equ. \ref{kk}, and compared to experiments \cite{keat5}.} 
\end{figure}
where $\alpha$ and $\beta$ are bond-stretching and bond-bending force constants, respectively, and $d$ is the strain-free equilibrium bond length. Such models have been widely used for the realistic modelling of structural (Fig. \ref{keat_fig}) and electronic properties of tetrahedral amorphous networks \cite{keat4,keat6,keat7}, and these simple interaction potentials have been also used to investigate the glass transition phenomenology \cite{ritort1,ritort2,ritort3,JPCM2010}. A certain number of salient features can be recovered within a Metropolis dynamics (see next subsection). The interaction potential can be assimilated with a simple harmonic model written as:
$V=(m/2)\sum_{i}\omega^2x_i^2$ where $\omega$ represents a typical vibrational mode related to bond interactions, and inelastic neutron scattering studies of glasses \cite{kamita,vdos1,vdos2} give an information about the order of magnitude of the typical stretching and bending vibrational frequencies (energies), typically of about 20-40 meV. 
\subsubsection{Metropolis dynamics}
From these simplified cases \cite{ritort1,ritort2}, the non trivial dynamics of such potentials (as in equ. (\ref{kk})) can be obtained. Once equ. (\ref{kk}) is reduced to the simple harmonic form, changes in atomic positions from $x_i$ to $x'_i$=$x_i$+$r_i/\sqrt{N}$ for all $i$ are accepted with probability $1$ if the energy decreases, i.e. if $\delta V=V(\lbrace x'_i\rbrace)-V(\lbrace x_i\rbrace)$ is negative. Otherwise, the change is accepted with a Metropolis rule $\exp(-\beta\delta V)$. Here, $\{r_i\}$ is a random variable having a Gaussian distribution of zero mean and finite variance equal to $\Delta^2$. A Gaussian integration \cite{ritort1} leads to the probability distribution $P(\delta V)$ for an energy change $\delta V$:
\begin{eqnarray}
\label{ritort_eq}
P(\delta V)=(4\pi m\omega^2 V \Delta^2)^{-\frac{1}{2}}\exp\bigl(-\frac{(\delta
V-\frac{m\omega^2\Delta^2}{2})^2}{4m\omega^2V\Delta^2}\bigr )
\end{eqnarray}
Because the probability distribution $P(\delta V)$ only depends on the interaction $V$, the Markovian dynamics can be analyzed from an equation for energy change. According to the Metropolis dynamics, the equation of evolution for the energy is indeed equal to:
\begin{eqnarray}
\label{metro}
\tau_0\frac{\partial V}{\partial t}=\int_{-\infty}^0P(x)dx+\int_0^{\infty} x\,x P(x) \exp(-\beta x)dx
\end{eqnarray}
where $\tau_0$ is a typical time that is inversely proportional to an atomic attempt frequency (10$^{-12}$~s). For the simplest cases, i.e. when the bonds (oscillators) have the same frequency $\omega$ (see \cite{ritort3} for mode-dependent solutions), equation (\ref{metro}) reduces \cite{ritort1} to :
\begin{eqnarray}
\label{metro1}
\tau_0\frac{\partial V}{\partial t}={\frac {1}{2\beta}}\biggl[(1-4V\beta)f(t)\biggr]+{\frac {m\omega^2\Delta^2}{4}}erfc\sqrt{\frac {m\omega^2\Delta^2}{16V}}
\end{eqnarray}
where $\beta=1/T$, and :
\begin{eqnarray}
f(t)={\frac {m\beta\omega^2\Delta^2}{2}}\exp\biggl[-{\frac {m\beta\omega^2\Delta^2}{2}}(1-2V\beta)\biggr]erfc\biggl[\sqrt{\frac {m\omega^2\Delta^2}{16V}}(4V\beta-1)\biggr]
\end{eqnarray}
\begin{figure}
\begin{center}
\end{center}
\caption{\label{tg_keat} Energy V(T) of a harmonic oscillator system \cite{JPCM2010} (solution of equation (\ref{metro1})) under cooling (black, upper curve), and annealing (red lower curve) for a rate q=$\pm$1~K.s$^{-1}$. The inset shows the evolution of the heat capacity. Black square indicate the inflexion point of the $C_p$ curve.} 
\end{figure}
and $erfc$ is the complementary error function. 
Equation (\ref{metro1}) has an obvious solution, equipartition ($V=T/2$), corresponding to the equilibrium state for the liquid. Results of this model (equ. (\ref{metro1}), Fig \ref{tg_keat}) show that the glass transition can be reproduced and, at low temperature, the system falls out of equilibrium that manifests by a departure from the equilibrium state $V=T/2$. A decrease of the cooling rate $q$ brings the system to a lower glass energy $V(q,T\rightarrow 0)=V^*(q)$ \cite{ritort1}. Upon reheating, the hysteresis curve signals onset of relaxation, and this leads to a strong exotherm peak in the first derivative ($C_p$, inset).  
Linear extrapolations (Fig. \ref{tg_keat}) permit to determine a fictive temperature as a function of cooling rate $q$. The cross-over between the low-temperature expansion of equation (\ref{metro1}) and the equilibrium line $V=T/2$ leads, indeed, to: 
\begin{eqnarray}
q={\frac {4T_f}{\pi\tau_0\biggr[erf\sqrt{\frac {m\omega^2\Delta^2}{8T_f}}-erf\sqrt{\frac {m\omega^2\Delta^2}{8V^*(q)}}\biggr]}}
\end{eqnarray}
Similarly, the corresponding heat capacity $C_p$ has the observed behavior from DSC (Fig. \ref{richert}) for both the cooling and the heating curves, and the inflection point of the heating curve serves to define a 'calorimetric' $T_g$ (filled box in the inset of Fig. \ref{tg_keat}) as in experiments. 
\par
In such class of models, departure from the equilibrium value results from a low acceptation rate for moves $x_i\rightarrow x'_i$ according to the Metropolis algorithm. In fact, at low temperature most of the changes leading to an increase of the energy will be rejected, and the system has an acceptation rate for moves that decays to zero. Interestingly, the relaxational dynamics associated with this low acceptation rate can be exactly calculated by linearizing equ. (\ref{metro1}) around the equlibrium solution, and this leads to an Arrhenius-like behavior at low temperature for the relaxation time of the form:
\begin{eqnarray}
\tau=\tau_0\sqrt{\frac {2\pi T^3}{m^3\omega^6\Delta^6}}\exp\biggl[{\frac {m\omega^2\Delta^2}{8T}}\biggr]
\end{eqnarray} 
A central result of this approach is that the activation energy for relaxation is directly linked \cite{ritort1,JPCM2010} with the typical vibrational frequency of the bonds, which is a local property of the glass, a result that has been recently extended to elastically interacting spring networks \cite{wyart1}. 
\subsection{Survey of other approaches}

There are many other approaches which attempt to relate the glassy relaxation to some other physical quantities or parameters. Kovacs and co-workers introduce a retardation time for exponential decay able to treat appropriately the stretched exponential decay of the $\alpha$-relaxation r\'egime \cite{kovacs} by a finite series of exponentials, and, under certain assumptions, the approach can be connected to the TNM phenomenology. However, while the formalism is able to reproduce thermal histories of the glass transition, i.e. cooling and heating scans of enthalpy, its application has been essentially limited to polymers. 
\par
In a similar spirit, Ngai et al. \cite{ngai1} have developed a coupling model that identifies the relaxation rate as the relevant variable, and connects the relaxation time of the stretched exponential function with the Kohlrausch parameter $\beta$. This leads to a time-dependent decay function that exhibits non-linearity and a slow down of the dynamics as the temperature is decreased \cite{ngai2}. A certain number of inorganic ionic-conducting glasses have been analyzed from this approach \cite{ngai4,ngai3}. However the rate equation of decay function that leads to glassy dynamics has been found to be inconsistent with Boltzmann superposition principle \cite{ngai5}. 
\par
Similarly to the coupling model, a certain number of approaches use, indeed, the stretched exponential to incorporate some non-linear effects able to reproduce the glass transition phenomenology. For instance, in Ref. \cite{ngai6}, non-linearity is introduced by defining a dependence of the relaxation time on the fictive temperature, and such effects act on the endothermic peak obtained in enthalpy at the glass transition and its subsequent evolution under ageing. For the case of vitreous selenium, a multiordering parameter model \cite{ngai7} uses a continuous distribution of relaxation times defined by a single Kohlrausch parameter $\beta$, able to reproduce experimental DSC data, and to predict the fictive temperature evolution under arbitrary temperature-time histories. The reproduction of DSC data appears to be central to the validation of such simple models, and Yue and co-workers \cite{yue_other} have recently proposed a unified routine to characterize the glass relaxation behavior and determine enthalpic fictive temperature of a glass with arbitrary thermal history. As a result, the enthalpic fictive temperature of a glass can be determined at any calorimetric scan rate in excellent agreement with modeled values.

\section{Role of network topology and rigidity}
In network glasses, the effect of structure and network topology or rigidity appears to be central to the understanding of the effect of composition on $T_g$ and relaxation. 
\subsection{Network connectivity and glass transition temperature}
There are various empirical or theory-based relationships showing that the glass transition temperature strongly depends on the glass structure, and that there is much to learn from the evolution with connectivity of $T_g$. 
\par
Besides thermodynamic or vibrational factors such as the well-known "{\em two-third rule}" stating that $T_g$ scales as 2/3$T_m$ \cite{kauzmann} or the Debye temperature of the phonon spectrum, there are, indeed, structural factors, and, in particular aspects of network connectivity. Tanaka \cite{tanaka1} has given an empirical relationship between $T_g$ and the average valence $Z$ of the involved atoms: $\ln T_g\simeq 1.6Z + 2.3$. Varshneya and co-workers \cite{varshneya1,varshneya2} have also shown that a modified Gibbs-Di Marzio equation \cite{gibbs_dimarzio} intially proposed for cross-linked polymers could predict of $T_g$ in multicomponent chalcogenide glass systems as a function of the average network coordination number $\bar r$, based only on the degree of atomic cross-linking in a polymeric selenium-based glass (e.g. Ge-Se). The parameter $\beta$ used has been shown to dependent on the coordination number $r_B$ of the cross-links (Ge,Si) \cite{epl99}:
\begin{eqnarray}
\label{vgdm}
T_g(\bar r)={\frac {T_g(\bar r=2)}{1-\beta(\bar r-2)}}
\end{eqnarray}
with:
\begin{eqnarray}
\label{beta_eq}
{\frac {1}{\beta}}\ =\ (r_B-2)ln\biggl[{\frac {r_B}{2}}\biggr]
\end{eqnarray}
Using stochastic agglomeration of basic local structures representative of the glass \cite{jncs97,epjb98}, an analytical $T_g$ prediction for binary and ternary glasses has been established that seems to be satisfied for a variety of binary and ternary network glasses (Fig. \ref{group4} left). For the former, the glass transition variation of a weakly modified glass $A_xB_{1-x}$ behaves as:
\begin{eqnarray}
\label{sat1}
\biggl[{\frac {dT_g}{dx}}\biggr]_{x=0}={\frac {T_g(x=0)}{\ln\biggr[{\frac {r_B}{r_A}}\biggr]}}={\frac {T_0}{\ln\biggr[{\frac {r_B}{r_A}}\biggr]}}
\end{eqnarray}
where $r_B$ and $r_A$ are the coordination numbers of the atoms or species $B$ and $A$, respectively, acting as local building blocks of the glass structure, e.g. one has $r_B$=3 and $r_A$=4 in a silicate glass made of $Q^3$ and $Q^4$ tetrahedral units \cite{epjb98}, or $r_B$=4 and $r_A$=2 in binary Ge-Se (Fig. \ref{group4}). For a ternary system, a parameter-free relationship between $T_g$ and the network mean coordination number can be also derived on the same basis \cite{pandalai}:
\begin{eqnarray}
\label{aver}
\bar r=r_Ar_Br_C{\frac {r_Ar_C\alpha^2(1-\gamma)+r_Ar_B\gamma^2(1-\alpha)
+r_Br_C\alpha\gamma(\alpha\gamma-\alpha-\gamma)}{r_Ar_C\alpha+r_Ar_B\gamma
-r_Br_C\alpha\gamma)^2-2r_A^2r_Br_C\alpha\gamma}}
\end{eqnarray}
where 
\begin{eqnarray}
\alpha=\biggl({\frac {r_A}{r_C}}\biggr)^{T_0/T_g},\ \ \gamma=\biggl({\frac {r_A}{r_B}}\biggr)^{T_0/T_g}, \ \ \delta=\biggl({\frac {r_A^2}{r_Br_C}}\biggr)^{T_0/T_g}
\end{eqnarray}
and an excellent agreement has been found with experimental data \cite{wang00,jpcm2005}.
\begin{figure}
\begin{center}
\end{center}
\caption{\label{group4}Left: Glass transition temperature in binary  chalcogenide glasses. Data from Ge-Se \cite{tgGeSe}, Ge-S \cite{tgGeS}, Si-Se \cite{tgSiSe}, Ge-Te \cite{tgGeTe} and Si-Te \cite{tgSiTe}. The solid line corresponds to equation (\ref{sat1}) with [$r_A=2$, $r_B=4$]. The dashed curve correponds to the fitted Gibbs-Di Marzio equation equation (equs. (\ref{vgdm}) and (\ref{beta_eq})), with $\beta^{-1}$=2$\ln 2$=0.73. Right: Prediction of the glass transition temperature from the Naumis model \cite{naumistg} for different binary and ternary chalcogenide glasses. Experimental data are taken from Refs. \cite{r18,tgGeS,tgGeTe}.
}
\end{figure}
Using a slightly different approach based on the general link between the mean square displacement, $\langle r^2(t)\rangle$, and the vibrational density of states
\begin{eqnarray}
\langle r^2(t)\rangle={\frac {k_BT}{m}}\int_0^\infty {\frac {g(\omega)}{\omega^2}}d\omega,
\end{eqnarray}
Naumis \cite{naumistg} has derived from the Lindemann criterion of solidification \cite{lindemann}, using $\langle r^2(t)\rangle$, a relationship predicting the variation of the glass transition temperature (Fig. \ref{group4}, right). These analytical models are helpful in understanding the effect of composition on $T_g$, and emphasize the central role played by network connectivity, and such ideas and relationships actually help in decoding further anomalies ($T_g$ extrema) which do appear in particular systems such as Ge-Se \cite{tgGeSe}, borates \cite{borates1} or germanates \cite{germanates}. Given that the glass-transition temperature is, indeed, an intrinsic measure of network connectivity, $T_g$ maxima in Ge-Se and As-Se glasses have been interpreted \cite{cras1} as the manifestation of nanoscale phase separation that is driven by broken chemical order \cite{cras2} in stoichiometric GeSe$_2$ and As$_2$Se$_3$, and this leads to a reduction of the network connectivity for the Se-rich majority phase at compositions where a $T_g$ maximum is measured.

\subsection{Rigidity theory of network glasses}

In addition to effects of network structure on $T_g$, there is an attractive way to analyze and predict relaxation and glass transition related properties using Rigidity Theory. This theory provides an atomic scale approach to understanding the physico-chemical behavior of network glasses using the network topology and connectivity as basic ingredients, and builds on concepts and ideas of mechanical constraints that have been introduced in the pioneering contributions of Lagrange and Maxwell \cite{r4,r5}. Phillips \cite{r6,r7,r8} has extended the approach to disorded atomic networks, and has recognized that glass forming tendency of covalent alloys is optimized for particular compositions. Specifically, it has been emphasized that stable glasses have an optimal connectivity, or mean coordination number $\bar r$ = $\bar r_c$, which satisfies exactly the Maxwell stability criterion of mechanically isostatic structures, or the condition, n$_c$ = n$_d$, where n$_c$ is the count of atomic constraints per atom and n$_d$ the network dimensionality, usually 3. 
\par
In covalent glasses the dominant interactions are usually near-neighbor bond-stretching (BS) and next-near-neighbor bond-bending (BB) forces (see equ. (\ref{kk})), the number of constraints per atom can be exactly computed in a mean-field way, and is given by:
\begin{eqnarray}
\label{nc0}
n_c={\frac {\sum\limits_{r\geq 2}n_r[{\frac {r}{2}}+2r-3]}{\sum\limits_{r\geq 2} n_r}}
\label{eq1}
\end{eqnarray}
where $n_r$ is the concentration of species being $r$-fold coordinated. The contribution of the two terms in the numerator is obvious because each bond is shared by two neighbors, and one has $r$/2 bond-stretching (BS) constraints for a $r$-fold atom. For BB (angular) constraints, one notices that a 2-fold atom involves only one angle, and each additional bond needs the definition of two more angles, leading to the estimate of (2$r$-3). For one-fold terminal atoms, a special count \cite{r10} is achieved as no BB constraints are involved, and in certain situations some constraints may be ineffective \cite{r10b}. By defining the network mean coordination number $\bar r$ of the network by:
\begin{eqnarray}
\bar r={\frac {\sum\limits_{r\geq 2}rn_r}{\sum\limits_{r\geq 2} n_r}}
\end{eqnarray}
one can reduce (\ref{eq1}) to the simple equation:
\begin{eqnarray}
\label{nc_final}
n_c={\frac {\bar r}{2}}+2\bar r-3
\end{eqnarray}
Applying the Maxwell stability criterion, isostatic glasses ($n_c$=3) are expected to be found at the magic average coordination number \cite{r6} of $\bar r$=2.40 in 3D, corresponding usually to a non-stoichiometric composition where glass-forming tendency has been found to be optimized experimentally \cite{rr11,rr12}. 
\par
The physical origin of this stability criterion has been revealed from the vibrational analysis of bond-depleted random networks \cite{mft1} constrained by bond-bending and bond-stretching interactions (see equ. (\ref{kk})). It has been demonstrated, indeed, that the number of zero frequency (floppy) modes $f$ (i.e. the eigenmodes of the dynamical matrix) is vanishing for $\bar r$=2.38  when rigidity percolates in the network. The Maxwell condition $n_c$ = n$_d$ therefore defines a mechanical stiffness
transition, an elastic phase transition, above which redundant constraints produce internally stressed networks, identified with a stressed-rigid 
phase \cite{mft2,mft3}. For n$_c$ $<$ n$_d$ however, floppy modes can proliferate, and these lead to a flexible phase where local deformations with a low cost in energy (typically 5~meV \cite{kamita}) are possible, their density being given by: $f=3-n_c$. There have been various experimental probes of this peculiar transition from Raman scattering \cite{raman}, stress relaxation \cite{stress} and viscosity measurements (Fig. \ref{tatsumisago}, \cite{r18}), vibrational density of states \cite{kamita}, Brillouin scattering \cite{brillouin1,brillouin2}, Lamb-Mossbauer factors \cite{r20}, resistivity \cite{resistivity}, and Kohlrausch exponents \cite{r18,novita1,stress}. For a full account of experimental probes and early verification of Rigidity Theory, readers should refer to books devoted to the subject \cite{boolbook,traverse,cambridge}.
\par
\begin{figure}
\begin{center}
\end{center}
\caption{\label{tatsumisago} Early verification of the role of rigidity \cite{r18} on the relaxation properties in a network forming liquid (here As-Ge-Se). Left: Comparison of pseudobinary As-Ge-Se liquid viscosities near $T_g$, compared to the strong (GeO$_2$) and fragile (K-Ca-NO$_3$) extremes. The inset zooms into the glass transition region, and shows a strong behavior for a network mean coordination number of $\bar r$=2.4 close to $T_g/T$=1. Right: Behavior with mean coordination number $\bar r$: activation energy determined either from viscosity or enthalpy data (a), heat capacity jump $\Delta C_p$ at the glass transition (b), and excess expansion coefficient $\Delta\alpha$ (c). Permission from American Physical Society. 
}
\end{figure}

\subsection{Rigidity Hamiltonians}

With the prediction of such thresholds and their observation in various properties associated with relaxation in chalcogenides oxides and other disordered glassy networks, the connectivity related flexible to stressed-rigid elastic phase transition has become an interesting means to understand and analyze in depth compositional trends of glassy dynamics and relaxation. However, although it provides a framework to understand many features of a system, thermodynamics is absent in the initial approach. One of the main drawbacks of Rigidity Theory is, indeed, that the enumeration of bonding constraints in equ. (\ref{nc0}) is performed on a fully connected network, in principle at T=0 K when neither bonds nor constraints are broken by thermal activation (see however \cite{mousseau_pebble}), and structural relaxation is obviously absent. The use of the initial theory \cite{mft1,r4,r5,mft2} may be valid as long as one is considering strong covalent bonds or when the viscosity $\eta$ is very large at T$<$T$_g$, given that $\eta$ is proportional to the bonding fraction, but equation (\ref{nc0}) is obviously not valid in a high temperature liquid, and one may wonder to what extent it remains useful for the glassy relaxation at $T\simeq T_g$.
However, NMR-related relaxational phenomena in Ge-Se indicate that the low temperature rigidity concept can be extended from the glass to the liquid in binary chalcogenide melts with confidence \cite{nmr0}. Furthermore, in equ. (\ref{nc0}) a mean-field treatment is implicitely assumed given that an average constraint count is performed over all the atoms of the network. This supposes homogeneity of the system even at the microscopic scale, and neglects the possibility of atomic-scale phase separation or large fluctuations in constraints or coordination numbers as the phase transition is approached. 
\par
An important step forward has been made by Naumis and co-workers \cite{naumis1,naumis2,naumis3}. Prior to the production of a rigidity-related Hamiltonian that could serve as starting point for the statistical mechanics derivation of various thermodynamic quantities \cite{naumis1,naumis2}, one has to realize that the fraction of cyclic variables in phase space are identified with the fraction of floppy modes $f=3-n_c$ because when one of these variables
is changed, the system will display a change in energy that is negligible. This means that in the simplest model \cite{naumis3} for network atomic vibrations in the harmonic approximation, the Hamiltonian can be given by:
\begin{eqnarray}
\label{naumisH}
{\cal H}=\sum\limits_{j=1}^{3N} {\frac {P_j^2}{2m}}+
\sum\limits_{j=1}^{3N(1-f)}{\frac {1}{2}}m\omega_j^2Q_j^2
\end{eqnarray}
where $Q_j$ (position) and $P_j$ (momentum) are the $j$-th normal mode coordinates in
phase space, and $\omega_j$ is the corresponding eigenfrequency of each normal mode. Since it is assumed that floppy modes have a zero frequency, they will not contribute to the energy so that the sum over coordinates runs only up to $3N(1-f)$. \par
From this simple Hamiltonian, a certain number of basic features of thermodynamics in connection with rigidity can be derived. First, from the partition function derived from equ. (\ref{naumisH}), both the free energy ${\cal F}$ of the system and the specific heat are found \cite{naumis3} to depend on the fraction of floppy modes:
\begin{eqnarray}
{\cal F}=-{\frac {3Nk_BT}{2}}\ln\biggl({\frac {2\pi mk_BT}{h^2}}\biggr)-{\frac {kT}{2}}\sum\limits_{j=1}^{3N(1-f)}\ln\biggl({\frac {2\pi k_BT}{m\omega_j^2}}\biggr)
\end{eqnarray}
\begin{eqnarray}
C_v=3Nk_B-{\frac {3Nk_B}{2}}f
\end{eqnarray}
the latter expression indicating that the specific heat in such model systems corresponds to the Dulong-Petit value that is decreased by a floppy mode contribution, and the finite value of the floppy mode frequency \cite{kamita} can be also taken into account \cite{naumis3}. Building on these ideas, an energy landscape treatment of rigidity leads to the conclusion that floppy modes can provide a channel in the energy landscape. Indeed, given that variables associated with $f$ are cyclic variables of the Hamiltonian, the energy of the system does not depend upon a change in a floppy mode coordinate, and for a given inherent structure (i.e. a local minimum characterized by $\omega_j$), the number of channels is given by $f$ which increases the available phase space allowed to be visited. Consequently, the number of accessible states $\Omega(E,V,N)$ can be calculated in the microcanonical ensemble, and using the Boltzmann relation $S=k_B\ln\Omega(E,V,N)$, one finds that the configurational entropy provided by the channels in the landscape is simply given by :
\begin{eqnarray}
\label{entropy_naumis}
S_c = fNk_B\ln V,
\end{eqnarray}
i.e. the floppy mode density is contributing to the configurational entropy and the dynamics of the glass-forming system. From a short range square potential, the basin free energy of a potential energy landscape has been investigated from MD simulations \cite{sciortino_pre}, and it can be separated into a vibrational and a floppy mode component, allowing for an estimate of the contribution of flexibility to the dynamics, and for this particular class of potentials it has been found that $S_c$ scales as $f^3$. 

\subsection{Temperature dependent constraints}

Building on this connection between floppy modes and the configurational entropy $S_c$ (equ. (\ref{entropy_naumis})), Gupta and Mauro have extended topological constraint counting to account explicitely for thermal effects \cite{mauro_gupta} in an analytical model {\em via} a two state thermodynamic function $q(T)$. This function  quantifies the number of rigid constraints as a function of temperature and subsequently modifies equation (\ref{eq1}) to become:
\begin{eqnarray}
n_c(T)={\frac {\sum\limits_{r\geq 2}n_r[q_\alpha^r({T}){\frac {r}{2}}+q_\beta^r({T})(2r-3)]}{\sum\limits_{r\geq 2} n_r}},
\label{eq2}
\end{eqnarray}
where $q^\alpha_r$(T) and $q^\beta_r$(T) are step functions associated with BS and BB interactions of a $r$-coordinated atom (Figure \ref{mauro_gupta_fig}, left) so that $n_c$ now explicitely depends on temperature. 
This function has two obvious limits because all relevant constraints can be either intact at low temperature ($q(T)$=1) like in the initial theory \cite{r6,mft1} or entirely broken ($q$($\infty$)=0) at high temperature. At a finite temperature however, only a fraction of these constraints can become rigid once their associated energy is less than $k_B$T. Different forms can be proposed for $q(T)$ based either on an energy landscape approach \cite{landscape} :
\begin{eqnarray}
\label{q0}
q(T)=(1-exp(-\Delta/T))^{\nu t_{obs}},
\end{eqnarray}
$\nu$ being the attempt frequency and $t_{obs}$ the observation time, or
involving a simple activation energy $\Delta$ for broken constraints \cite{broken_constraints}:
\begin{eqnarray}
\label{q1}
q(T)={\frac {1-e^{\Delta/T}}{1+e^{\Delta/T}}},
\end{eqnarray}
and the general behavior of $q(T)$ can be computed for any thermodynamic condition from MD simulations \cite{dens_prl} (see below). A certain number of thermal and relaxation properties of network glass-forming liquids can now be determined, and a simple step-like function (thick black line in Fig. \ref{mauro_gupta_fig}) with an onset temperature $T_\alpha$ for various constraints allows obtaining analytical expressions for fragility and glass transition temperature \cite{mauro_gupta,mauro_gupta1,yue1,yue2,yue3}, heat capacity \cite{yue4}, and glass hardness \cite{yue3,yue_prl}. Two central ingredients are necessary. First, it is assumed that the Adam-Gibbs model for viscosity (equ. (\ref{adam_gibbs})), $\eta=\eta_\infty\exp(A/TS_c)$, holds in the temperature range under consideration, and that the corresponding barrier height $A$ is a slowly varying function with composition. This means that only the configurational entropy $S_c$ will contain the temperature and composition dependence. Secondly, it is assumed that the expression (\ref{entropy_naumis}) relating the configurational entropy to topological degrees of freedom (floppy mode density) is valid. A strong support to this approach is provided by a test of the MYEGA viscosity modelling curve (equ. (\ref{myega})) which uses these two basic assumptions and has been tested with success over more than a hundred of different glass-forming liquids \cite{mauro_pnas}.
\begin{figure}
\begin{center}
\hspace{0.5cm}
\end{center}
\caption{\label{mauro_gupta_fig} Left: Temperature behavior of the Mauro-Gupta function \cite{mauro_gupta} (solid curves, equ. (\ref{q0})) for bond-bending bridging oxygen (BO) in densified sodium silicates, compared to a direct calculation \cite{dens_prl} from MD simulations (see below). The data can be also fitted with a random bond function \cite{broken_constraints} (equ. (\ref{q1})). Calculations of topological constraint theory usually involve a step-like function (thick black curve) having an onset temperature, e.g. here at $T_\alpha\simeq$2900~K. Right: Prediction of the glass transition temperature $T_g$ of binary alkali phosphate glasses (1-x)R$_2$-(1-x)P$_2$O$_5$ (R=Li, Na, Cs) from temperature dependent constraints \cite{yue2}. Permission from AIP Publishing LLC 2015. 
}
\end{figure}
By furthermore stating that $T_g$ is a reference temperature at which $\eta(T_g(x),x)$=10$^{12}$~Pa.s for any composition, equs. (\ref{adam_gibbs}) and (\ref{entropy_naumis}) can be used to write:
\begin{eqnarray}
\label{sc_xr}
{\frac {T_g(x)}{T_g(x_R)}}=\frac {S_c(T_g(x_R),x_R)}{S_c(T_g(x),x)}}={\frac {f(T_g(x_R),x_R)}{f(T_g(x),x)}={\frac {3-n_c(T_g(x_R),x_R)}{3-n_c(T_g(x),x)}}
\end{eqnarray}
and $T_g(x)$ can be determined with composition $x$ from a reference compound having a composition $x_R$ and a glass transition $T_g(x_R)$, knowing the number of topological degrees of freedom (i.e. 3-$n_c$) for compositions $x$ and $x_c$ from equation (\ref{eq2}), and the behavior of the step functions $q(T)$.
\par
Using the expression for $S_c(T_g(x),x)$ in equ. (\ref{sc_xr}), and the definition of fragility (equ. (\ref{frag})), one can furthermore extract an expression for the fragility index ${\cal M}$ as a function of composition:
\begin{eqnarray}
\label{mauro_m}
{\cal M}(x)={\cal M}_0\biggl(1+{\frac {\partial \ln{S_c(T,x)}}{\partial \ln T}}\biggr|_{T=T_g(x)}\biggl)={\cal M}_0\biggl(1+{\frac {\partial \ln{f(T,x)}}{\partial \ln T}}\biggr|_{T=T_g(x)}\biggl)
\end{eqnarray}
Typical applications for the prediction of the glass transition temperature concern simple chalcogenides \cite{mauro_gupta}, borates \cite{mauro_gupta1}, borosilicates \cite{yue4} phosphates \cite{yue3}, or borophosphate glasses \cite{yue1} (Figure \ref{mauro_gupta_fig}, right). 
Equation (\ref{mauro_m}) usually leads to a good reproduction of fragility data with composition, but requires a certain number of onset temperatures $T_\alpha$  that can be estimated from basic assumptions, or which act as parameters for the theory. Such onset temperatures furthermore can be related to the corresponding activation energy \cite{mauro_gupta} (equ. (\ref{q0})) to break a constraint {\em via}:
\begin{eqnarray}
\Delta=-k_BT_\alpha\biggl[1-2^{-1/(\nu t_{obs})}\biggr]
\end{eqnarray} 
The agreement of such predictive laws for fragility is usually excellent (Fig. \ref{mauro_fragility}a), and calculations have been stressed to be robust against parameter sensitivity \cite{mauro_gupta1,yue1,yue2,yue3,yue4}. 
\par
\begin{figure}
\begin{center}
\end{center}
\caption{\label{mauro_fragility} Modelling the composition dependence of the liquid fragility index (m) for borosilicate glasses \cite{yue4}. a) Prediction of the fragility index (solid line) as a function of composition, compared to experimental data measured from DSC experiments. (b) Effect of the constraint contribution on the computed fragility with composition: Boron $\beta_B$ and silicon $\beta_{Si}$ BB constraints, BS constraints ($\alpha$) and non-bridging oxygen related constraints ($\mu$). $m_0$ is the reference fragility ${\cal M}_0$ appearing in equ. (\ref{mauro_m}). Reprinted from J. Phys. Chem B {\bf 115}, 12930 (2011).}
\end{figure}
What is learned from such fragility predictions ? First, since the constraint count that evaluates the fragility index is performed on models that reproduce the change in local structure under composition change, the temperature dependent constraint approach provides a top-down validation of such structural models. Furthermore, as noticed from Fig. \ref{mauro_fragility}b, one has the opportunity to probe what aspects of interactions contribute the most to the evolution of fragility with composition. In the represented example of borosilicates \cite{yue4} (Fig. \ref{mauro_fragility}b), ${\cal M}$ is found to be mostly driven by the bond angle interactions involved in silicon  ($\beta_{Si}$) and boron ($\beta_B$) that constrain the bond-bending motion of the glass forming liquid (Fig. \ref{mauro_fragility}b), whereas the BS interactions contribute only at large silicon to boron ratio. 
\par
With the same formalism, the heat capacity change $\Delta C_p$ at the glass transition can be calculated and compared to measurements accessed from DSC. Using the temperature dependence of constraints, it is assumed at a first stage that the major contribution to heat capacity at the glass transition arises from configurational contributions $C_{p,conf}$ that can be related to $S_c$, so that one has: $\Delta C_p$=$C_{pl}-C_{pg}\simeq C_{p,conf}$. The heat capacity change can then be written as a function of the configurational enthalpy and $S_c$:
\begin{eqnarray}
\Delta C_p=\biggl({\frac {\partial H_{conf}}{\partial T}}\biggr)_P=\biggl({\frac {\partial H_{conf}}{\partial \ln S_{conf}}}\biggr)_P\biggl({\frac {\partial \ln S_{conf}}{\partial T}}\biggr)_P
\end{eqnarray}
and using the Adam-Gibbs expression (\ref{adam_gibbs}) and the derived expression for fragility (\ref{mauro_m}), one can furthermore write the jump of the heat capacity at the glass transition with composition $x_i$:
\begin{eqnarray}
\label{cp_mauro}
\Delta C_p(x_i,T_g(x_i))={\frac {1}{T_g(x_i)}}\biggl({\frac {\partial H_{conf}}{\partial \ln S_{conf}}}\biggr)_{P,T=T_g}\biggl({\frac {{\cal M}(x_i)}{{\cal M}_0}}-1\biggr)
\end{eqnarray}
Equation (\ref{cp_mauro}) can be recast in a more compact form given that $S_c(T_g)$ is inversely proportional to $T_g$, and by assuming that $\partial H_{conf}$/$\partial S_{c}$ is by definition equal to the configurational temperature at constant pressure \cite{mauro_ceram} which is close to $T_g$. This, ultimately leads to a decomposition dependent prediction of the heat capacity jump at $T_g$:
\begin{eqnarray}
\label{dcp}
\Delta C_p(x_i,T_g(x_i))\simeq{\frac {[A(x_i)]_R}{T_g(x_i)}}\biggl({\frac {{\cal M}(x_i)}{{\cal M}_0}}-1\biggr)
\end{eqnarray}
 Once again, and similarly to relationships on $T_g$ (equ. (\ref{sc_xr})) or the fragility index (equ. (\ref{mauro_m})), the heat capacity jump is evaluated with respect to some reference composition because the parameter $[A(x_i)]_R$ appearing in equ. (\ref{dcp}) connects the configurational entropy $S_c(T_g)$ with $T_g$ for a reference composition $x_R$.
Applications have been performed on the same glassy systems \cite{yue4}, and successfully compared to experimental measurements (Fig. \ref{dcp_fig}).
\begin{figure}
\begin{center}
\end{center}
\caption{\label{dcp_fig} Composition dependence of the change in the isobaric heat capacity $\Delta C_p$ during the glass transition in a sodium borate glass \cite{yue4}. Comparison between experimental data (filled squares) and the model calculations of $\Delta C_p$ from equ. (\ref{dcp}). Reprinted from J. Phys. Chem B {\bf 115}, 12930 (2011).
}
\end{figure}
Again, such predictions have the merit to accurately reproduce experimental data, and to provide some insight into the validity of structural models that can be checked independently from a variety of other experimental (e.g. spectroscopic) probes.

\subsection{MD based dependent constraints}

A more general and alternative approach to topological constraint theory can be proposed in order to establish the number of constraints $n_c$($x$,$T$,P) for any thermodynamic condition, including under pressure. This is achieved by using Molecular Dynamics (MD) which also permits establishing correlations with thermodynamic and dynamic properties independently characterized from such atomic scale simulations. In all approaches -classical or First Principles (FPMD) using e.g. a Car-Parrinello scheme \cite{micol1}- Newton's equation of motion is solved for a system of N atoms or ions, representing a given material. Forces are either evaluated from a model interaction potential which has been fitted to recover some materials properties, or directly calculated from the electronic density in case of a quantum mechanical treatment using density functional theory (DFT). Recent applications have permitted to describe very accurately the structural and dynamic properties of most archetypal network-forming systems (Ge-Se \cite{micol1,micol2,micol3}, SiO$_2$ \cite{wilson_prl}, GeO$_2$ \cite{salmongeo2,pre06}, B$_2$O$_3$ \cite{axelle,axelle1}, As-Se \cite{bauchy1,bauchy_asse,micol5,micol6}, As-Ge-Se \cite{wangasgese}, Si-Se \cite{sise1,sise2}...), in the glassy or liquid state, and at ambient or densified conditions. A similar achievement has been realized on modified glasses such as alkali silicates \cite{du1,du2,du3,du4,du5}, soda-lime silicates \cite{lime1,lime2}, borosilicates \cite{boro1,boro2} or alumino-silicates \cite{alumino}.
\par
\begin{figure}
\begin{center}
\end{center}
\caption{Method of constraint counting from MD-generated configurations. Large (small)
radial (a) or angular (b) excursions around a mean value are characterized by large
(small) standard deviations on bonds or angles representing broken
(intact) constraints.}
\label{defini}
\end{figure}

The way topological constraints can be extracted from atomic scale trajectories relies essentially on the recorded radial and angular motion of atoms that connects directly to the enumeration of BS and BB constraining interactions which are the relevant ones for the identification of flexible to rigid transitions. Instead of treating mathematically the forces and querying about motion which is the standard procedure of MD simulations for obtaining trajectories, as in classical mechanics, an alternative scheme is followed. Here, the atomic motion associated with angles or bonds can be related to the absence of a restoring force (Fig. \ref{defini}), this strategy being somewhat different from the "{\em Culture of force}" analyzed by F. Wilcek \cite{wilcek} given that one does not necesseraly need to formulate the physical origin of the forces to extract constraints. 
\par
In the case of atomic scale systems, since one attempts to enumerate BS and BB constraints, one is actually not seeking for motion arising from large radial and angular excursions, but for the opposite behavior and also for atoms displaying a small motion (vibration) that maintains corresponding bonds and angles fixed around their mean value. These can ultimately be identified with a BS or BB interaction constraining the network structure at the molecular level. Having generated the atomic scale configurations at different thermodynamic conditions from MD, a structural analysis is applied in relation with 
the constraint counting of Rigidity Theory such as sketched in equ. (\ref{eq1}).

\begin{figure}
\begin{center}
\end{center}
\caption{\label{stretch}\footnotesize Decomposition of partial pair correlation functions g$_i$(r) into neighbor distributions in amorphous As$_2$Se$_3$ \cite{micol6}. The inset show the positions (first moments) of the neighbor distributions and their standard deviations (second moments, represented as error bars), indicating that As and Se have 1.5 and 1 BS constraints, respectively.}
\end{figure}

\subsection{Bond-stretching}
To obtain  the number of BS interactions, one focuses on neighbour distribution functions (NDFs) around a given atom $i$ (see Fig. \ref{stretch}). A set of NDFs can be defined by fixing the neighbor number $n$ (first, second etc) during bond lifetime, the sum of all NDFs yielding the usual $i$-centred pair correlation function g$_i$(r) whose integration up to the first minimum gives the coordination numbers r$_i$, and hence the corresponding number of bond-stretching constraints r$_i$/2 \cite{bauchy_asse,micol6,cons1,cons2,cons3}. Figure \ref{stretch} shows an application to amorphous As$_2$Se$_3$ \cite{bauchy_asse}. In As$_2$Se$_3$, three NDFs (colored curves) contribute to the first peak
of the As-centred pair correlation function g$_{As}$(r), very well separated from the second shell of neighbours, and indicative of the presence of three neighbours around an As atom. It is to be noticed that a fourth NDF is present at the minimum of g$_{As}$(r), indicating that a small fraction of 4-fold As atoms should be present in the glass \cite{micol6}, typically less than 10\%. The separation between first and second shell of neighbours can be also characterized by plotting
the NDF peak positions as a function of the neighbour number (inset of Fig. \ref{stretch}). For e.g. As$_2$Se$_3$, but also for tetrahedral glasses \cite{cons2} one remarks that there is a clear gap in distance between the distributions belonging to the first and the second neighbor shell. These NDFs belonging to the first shell display furthermore a much lower radial excursion (error bars, see inset of Fig. \ref{stretch}) as compared to the NDFs of the next (second shell) neighbor distributions. From this simple example (As$_2$Se$_3$), one determines $r_{As}=3$ and $r_{Se}=2$ leading to 1.5 and 1 BS constraints, a result that is expected from a constraint count based solely on the 8-${\cal N}$ rule \cite{mft1,r4}, ${\cal N}$ being the number of s and p electrons. However, for certain systems for which this rule does not apply (telluride network glasses, see below) or in densified systems with non-monotonic evolutions of coordination numbers under pressure \cite{dens_prl}, such MD based constraint counting algorithms provide, indeed, a neat estimate \cite{cons1} of BS constraints without relying on crude assumptions. 
\begin{figure}
\begin{center}
\end{center}
\caption{\label{bend1}\footnotesize From top to bottom: oxygen, selenium and germanium partial
bond angle distributions (PBAD) in GeO$_2$ and GeSe$_2$ for an arbitrary N=6 \cite{cons2} leading to 15 possible PBADs. The colored curves correspond to PBADs having the lowest standard deviation(s) $\sigma_\theta$. 
The sharp peaks at $\theta\simeq$ 40$^o$ correspond to the hard-core repulsion. The constrained angle around oxygen in germania (panel c) is found to be centred at 135$^o$, close to the value obtained from experiments \cite{r118}. All other angles display broad variations and correspond to angles defined by next-nearest neighbor shells. Permission from American Physical Society.}
\end{figure}

\subsection{Bond-bending}

\subsubsection{Average behavior}
The bond-bending (BB) constraint counting from MD simulations is based on partial bond angle distributions (PBADs) $P({\theta_{ij}})$ (or $P(\theta)$ in the following) and defined as follows \cite{cons1,cons2}: for each type of a central atom $0$, the N first neighbours $i$ are selected, leading to N(N-1)/2 possible angles $i0j$ ($i$=1..N-1, $j$=2..N), i.e. 102, 103, 203, etc. The standard deviation $\sigma_{\theta}$ of each distribution P($\theta_{ij}$) gives a quantitative estimate of the angular excursion around a mean angular value (Fig. \ref{defini}b), and provides a measure of the bond-bending strength. Small values for $\sigma_{\theta}$ correspond to an intact bond-bending constraint which maintains a rigid angle at a fixed value, whereas large $\sigma_{\theta}$ correspond to a bond-bending weakness giving rise to an ineffective constraint.
\par
Figure \ref{bend1} shows the PBADs for glassy GeSe$_2$ and GeO$_2$ \cite{cons2}. Broad angular distributions are found in most of the situations, but a certain number of sharper distributions (colored) can also be found, and these are identified with intact angular constraints because these arise from a small motion around an average bond angle.  For the special case of tetrahedral glasses, only six angles have nearly identical and sharp distributions, and these are the six angles defining the tetrahedra with a mean value that is centred close to the angle of $\bar \theta$=109$^o$. From such N(N-1)/2 different PBADs, a second moment (or standard deviation) can be computed for an arbitrary set of triplets (i0j) with (i,j=1..N). 
Figure \ref{bend2} shows corresponding results for the standard deviations $\sigma_\theta$ in stoichiometric oxide glasses \cite{cons2}. 
\begin{figure}
\begin{center}
\vspace{0.5cm}
\end{center}
\caption{\label{bend2}\footnotesize Oxygen and Si/Ge standard deviations computed from 15 PBADs in vitreous germania and silica \cite{cons2}. The labels on the x-axis refer to all possible triplets $i$0$j$ between a central atom 0 and two neighbors $i$ and $j$. For all systems, the PBADs relative to the Group IV (Si, Ge) atom have a low standard deviation $\sigma_{\theta}$, of the order of 10-20$^o$ when the first four neighbours are considered. One finds e.g. $\sigma_{Ge}\simeq 7^o$ for the PBAD 102 of GeO$_2$, which is substantially smaller as compared to those computed from other distributions (105, 106, etc.) which have $\sigma_\theta\simeq$40$^o$. Permission from American Physical Society.}
\end{figure}
Such glasses have their standard deviations nearly equal for the six relevant (Ge,Si) distributions, which are associated with bending motions around the tetrahedral angle of 109$^o$. A slightly different situation occurs in glasses subject to stress, i.e. densified silicates \cite{dens_prl} or stoichiometric chalcogenides \cite{bauchy_asse,micol1} which exhibit an increased angular bending motion of tetrahedra, as discussed below. 
\begin{figure}
\begin{center}
\end{center}
\caption{\label{a-inst}\footnotesize Time evolution (in MD steps) of two typical angles in a liquid sodium silicate \cite{cons3} defined by either the first two oxygen neighbors around a silicon atom (O$_1$-Si-$O_2$, 102) or by neighbor 1 and neighbor 5 (O$_1$-Si-O$_5$, 105). Large variations are obtained for angles with an ineffective constraint.}
\end{figure}

\subsubsection{Individual constraints}
An additional way of analyzing angular constraints is to follow a given angle individually during the course of the MD simulation (Figure \ref{a-inst}). For each individual atom $k$, the angular motion over the time trajectory then leads to a single bond angle distribution P$_k$($\theta$) characterized by a mean ${\bar \theta}_k$ (the first moment of the distribution), and a second moment (or standard deviation $\sigma_{\theta_k}$). The latter represents, once again, a measure of the strength of the underlying BB interaction. If $\sigma_{\theta_k}$ is large (one has usually $\sigma_{\theta_k}>$15-20$^o$ \cite{cons3}), it suggests that the BB restoring force which maintains the angle fixed around its mean value ${\bar \theta}_k$ is ineffective. As a result, the corresponding BB topological constraint will be broken, and will not contribute to network rigidity. Ensemble averages then lead to a distribution f($\sigma$) of standard deviations which can be analyzed and followed under different thermodynamic conditions.
\par
This alternative scheme following constraints individually permits to separate effects which may arise from disorder, from those which are originated by the radial or angular motion and which enter in the constraint counting analysis. In fact, when averaged simultaneously over time and space, $\sigma_\theta$ can simply be larger because of an increased angle and bond length variability induced by an increased bond disorder which will broaden corresponding bond angle/bond length distributions. By following angles and distances with time, this drawback can be avoided. 
Figure \ref{bimodale} shows the distribution f($\sigma$) of angular standard deviations for a bridging oxygen in a sodium silicate liquid with increasing temperatures \cite{cons3}, and the assignement of the peaks can be made from the inspection of the two limiting temperatures. At elevated temperatures (4000~K), all constraints are, indeed, broken by thermal activation so that f($\sigma$) displays a broad distribution centred at a large standard deviation (25$^o$). On the opposite, at low temperature (300~K) $\sigma_\theta$ values display a sharp distribution ($\sigma<$10$^o$), indicating that corresponding BB constraints are active. Interestingly, there is a temperature interval at T$\simeq$2000~K at which one can have a mixture of both types of constraints -effective and ineffective- and the corresponding fraction of intact BB constraints can be computed (inset of Fig. \ref{bimodale}). It exhibits a broad step-like behavior having all the features of the Mauro-Gupta q(T) function \cite{mauro_gupta} introduced previosuly (see also Fig. \ref{mauro_gupta_fig}). 

\begin{figure}
\vspace{1cm}
\begin{center}
\end{center}
\caption{\label{bimodale}\footnotesize Behaviour of the bridging oxygen (BO) centred standard deviation distributions f($\sigma$) in a sodium silicate liquid \cite{cons3}. Note the bimodal distribution occuring at T$\simeq$2000 K. The broken line defines a boundary between broken and intact constraint population, estimated to be about $\sigma_\theta$=15$^o$ at low temperature. Gaussian fits (red curves) are shown for selected temperatures.The inset shows the fraction q(T) of intact oxygen constraints as a function of temperature. The solid curve is a fit using the Mauro-Gupta function \cite{mauro_gupta} of equ. (\ref{q0}).}
\end{figure}
\par
Such methods based on angular standard deviations have also proven to be efficient in order to enumerate the fraction of tetrahedra in amorphous telluride networks \cite{tell1,tell2}. To calculate the population of GeTe$_{4/2}$ or SiTe$_{4/2}$ tetrahedra in binary Ge-Te, Si-Te or ternary Ge-Si-Te glasses, one detects atoms having six low standard deviations $\sigma_\theta$ around an Ge/Si atom. Once such atoms are identified, it is found that the corresponding average angle is equal to $\langle \bar \theta \rangle\simeq$ 109$^o$, and a corresponding bond angle distribution for the whole system peaks at 109$^o$.
\par
Having set the basis of Molecular Dynamics based topological constraint counting, we now review certain results obtained within this new framework, and how this connects to aspects of glassy relaxation. 

\section{Rigidity and dynamics with composition}

Applications have been performed on a variety of systems \cite{cons1,cons2,cons3}. We focus in the forthcoming on Ge-Se and silicate network glasses and liquids which are probably the most well documented alloys in the field of rigidity transitions. 

\subsection{Topological constraints}

The enumeration of constraints on realistic models of Ge-Se glasses \cite{micol1} and liquids \cite{micol3} shows that six Ge standard deviations have a low value ($\sigma_{Ge}\simeq$~ 10$^o$), i.e. four times smaller than all the other angles. One thus recovers the result found for the stoichiometric oxides (SiO$_2$, GeO$_2$, see Fig. \ref{bend2}). A more detailed inspection has revealed that there is a clear difference between compositions (10, 20, and 25\% Ge) having the six standard deviations $\sigma_{Ge}$ nearly equal, and compositions belonging to the stressed rigid phase (33\%, 40\%) which have an increased value of $\sigma_{Ge}$ for selected angles. For such systems as well as for the isochemical compounds (Ge$_x$S$_{100-x}$ \cite{chakra}, Si$_x$Se$_{100-x}$ \cite{selven}) the flexible phase has been found to be defined for 0$\leq x\leq$20~\% and the stressed rigid phase for $x\geq$25~\%, the limit of the glass forming region being somewhat larger than 33~\%.
\par
As the Ge content is increased, the intra-tetrahedral angular motion is growing for selected angles, as detected for e.g. GeSe$_2$ by the important growth of standard deviations involving the fourth neighbor of the Ge atom. When the six standard deviations $\sigma_{Ge}$ defining the tetrahedra are represented as a function of the Ge content (Fig. \ref{sigma_gese}, left), it is found, indeed, that the angular motion involving the fourth neighbour (PBADs 104,204,304) exhibits a substantial increase once the system is in the stressed rigid phase, while the others (102, 103, 203) are left with a similar angular excursion close to the one found for the oxides (SiO$_2$, GeO$_2$) where it was concluded \cite{cons2} that tetrahedra were rigid.
This underscores the fact the quantity $\sigma_{Ge}$ is an indicator of stressed rigidity \cite{micol3}. Moreover, the presence of stress will lead to asymetric intra-tetrahedral bending involving an increased motion for selected triplets of atoms, and this indicates that some BB constraints have softened to accomodate stress. A similar situation is encountered in densified silicates \cite{NatComm} or in hydrated calcium silicate networks \cite{bauchy_prl,bauchy_cement} for which angular motion associated with the tetrahedra SiO$_{4/2}$ undergoes a substantial change with pressure or composition.
\par
The origin of this softening can be sketched from a simple bar network when stretching motion is considered instead (Fig. \ref{sigma_gese}, left), and this connects to the well-known relationship between stressed rigidity and bond mismatch in highly connected covalent networks \cite{r119}. In such systems, atoms having a given coordination number can indeed not fulfill all their bonds at the same length because of a too important connectivity that prevents from a full relaxation towards identical lengths. In the simplified bar structures sketched in Fig. \ref{sigma_gese}, all bars can have the same length in flexible (0-20~\% at Ge) and isostatic networks (20-25~\% at. Ge) but once the structure becomes stressed rigid, at least one bar (e.g. the red bar in Fig. \ref{sigma_gese}) must have a different length. A similar argument holds for angles. In the stressed rigid Ge-Se, because of the high network connectivity, GeSe$_{4/2}$ tetrahedra must accomodate for the redundant cross-links which force softer interactions \cite{kamita} (i.e. angles) to adapt and to break a corresponding constraint. This leads to increased angular excursions for atomic Se-Ge-Se triplets (Fig. \ref{sigma_gese}) involving the farthest (fourth) neighbour of a central Ge atom. 
\par
\begin{figure}
\begin{center}
\vspace{0.5cm}
\end{center}
\caption{\label{sigma_gese}\footnotesize Left: Standard deviations $\sigma_{Ge}$ as a function of Ge composition in the Ge-Se system \cite{cons2}, split into a contribution involving the fourth neighbour (red line, average of 104, 204 and 304) and the other contributions (black line). The shaded area corresponds to the Boolchand (isostatic) intermediate phase \cite{gese} (see below). A simple bar structure represents the nature of the different elastic phases (see text for details). Right: Distribution of Ge angular standard deviations using individual constraints \cite{micol1}. The total distributions have been split, depending on the neighbor rank : angles involving the first three neighbors (102,103,203, black symbols), and the fourth neighbor (104,204,304, red symbols). The solid curves are Gaussian fits which serve to estimate the population of broken constraints at high $x$ content. Permission from American Physical Society.}
\end{figure}
When such systems are analyzed from individual constraints (Fig. \ref{sigma_gese} \cite{micol1}, right), it is also seen that some Ge centred angles have softened once the glass has become stressed rigid. For the flexible GeSe$_9$ (10~\%) and the isostatic GeSe$_4$ (20~\%, $n_c$=3 in a mean field count) compositions, a single distribution f($\sigma$) for all six angles (102,...304) is found, located at a low value ($\sigma\simeq$ 8-9$^o$) which indicates a weak angular intra-tetrahedral motion. However, at higher Ge content corresponding angles involving the fourth neighbor (104,204,304) partially soften and produce a bimodal distribution (red curve), indicative that some angular constraints ($\sigma\simeq$ 22$^o$) are now broken. An enumeration shows that the fraction $\xi$ of broken constraints is about 17.2\% and 21.4\% for GeSe$_2$ and Ge$_2$Se$_3$, respectively \cite{micol1}. This implies a reduction of the number of Ge BB constraints by a quantity $\Delta$n$_c$=3$x\xi$, i.e. 0.17 and 0.26 for the aforementioned compositions so that n$_c$ reduces from 3.67 (the mean-field estimate \cite{mft1}) to 3.5, and from 4.00 to 3.74 for GeSe$_2$ and Ge$_2$Se$_3$, respectively.
\par
\subsection{Behavior in the liquid phase}
How do constraints in chalcogenide melts behave at high temperature, and how do the evolution of such constraint link with relaxation ? Using MD simulations, the number of topological constraints has been investigated as a function of temperature in a certain number of glass-forming systems such as GeSe$_2$ \cite{mmi_book}, Ge-Te \cite{tell2}, Ge-Si-Te \cite{tell1} or SiO$_2$-2SiO$_2$ \cite{dens_prl,cons3,NatComm}. Figure \ref{sigma_T} shows the analysis for 300~K and 1050~K for GeSe$_2$, using either an average count (left) or indiviual constraints giving the distribution f($\sigma$) (right) \cite{mmi_book}.
\begin{figure}
\begin{center}
\end{center}
\caption{\label{sigma_T}\footnotesize Effect of temperature on standard deviations in liquid (1050~K, red) and amorphous (300~K, blue) GeSe$_2$. Left: Standard deviation $\sigma_{Ge}$ extracted from the partial bond angle distributions (PBAD). Right: Distribution of Ge angular standard deviations using individual constraints. The total distributions are split, depending on the neighbor rank : angles involving the first three neighbors (102,103,203), and the fourth neighbor (104,204,304, shifted).}
\end{figure}
Changes in constraints are weak and indicate that a counting at low temperature holds to some extent in a high temperature liquid, and close to the glass transition temperature. The corresponding calculated fraction of broken constraints \cite{mmi_book} has been found to be of the same order which highlight the fact that thermal effects on topological constraints are weak in these chalcogenides. This conclusion is consistent with recent results from neutron spin-echo spectroscopy showing that the rigidity concept (and the underlying constraint count) can be extended from the glass to the liquid \cite{bychkov_gese}. Parameters giving the temperature dependence of the relaxation patterns of binary chalcogenide melts have, indeed, shown to be linearly dependent with the mean coordination number $\bar r$, which represents a measure of a low temperature network for which the rigidity analysis assumes that all constraints are intact \cite{r6,mft1}. This seems also in line with results in Ge-Se using liquid-state NMR \cite{nmr0} (see also Fig. \ref{sen_nmr}), which emphasize that relaxational phenomena in the liquid are linked with the constraint count performed at low temperature.
\par
This situation actually contrasts with the findings obtained for oxides (Figure \ref{bimodale}) which exhibit a much more proncounced evolution of $n_c$ with temperature (Fig. \ref{mauro_gupta_fig}, left) or under a combined change in temperature and pressure (GeO$_2$, \cite{pacaud}). Furthermore, for such oxides, it has been found \cite{EPL13} that the distribution of constraints is not randomly distributed (Fig. \ref{hetero}, left), and corresponding liquids display a heterogeneous distribution with zones of thermally activated broken constraints that can be increased with temperature at constant pressure.
In addition, the spatial extent of these flexible regions shows a percolative behavior at a characteristic temperature $T_{onset}$ (Fig. \ref{hetero}, right) which is deeply connected to flexible to rigid transitions, and influences the fragility of the glass-forming liquid \cite{EPL13}. Here, the temperature at which constraints become homogeneously distributed across the liquid strcuture is found to depend both on pressure and temperature, with a minimum found at $T_{onset}$ for a certain pressure interval.

\begin{figure}
\begin{center}
\end{center}
\caption{\label{hetero} Left: Contour plot of the the bridging oxygen standard deviation (13$^o<$~$\sigma<$~20$^o$) in a (P=0, 2000~K) sodium silicate liquid \cite{EPL13}. Yellow zones indicate regions of broken constraints. Right: size of the broken constraint clusters (in simulation cell length unit) as a function of temperature for different isobars. The inset shows the evolution of the onset temperature T$_{onset}$ (in kK) as a function of pressure.
}
\end{figure}
\subsection{Isostatic relaxation}
Isostatic glasses are found to relax very differently from other glass-forming liquids, and the behavior of transport properties appears to be also different, as recently demonstrated for a densified silicate glass \cite{NatComm} using molecular simulations. It relies essentially on the computation of viscosity using the Green-Kubo (GK) formalism \cite{r107} which is based on the calculation of the stress tensor auto-correlation function, given by:
\begin{eqnarray}
\eta=\frac{1}{k_BTV}\int_0^{\infty}\langle P_{\alpha\beta}(t) P_{\alpha\beta}(0) \rangle
\label{equf}
\end{eqnarray}
using off-diagonal components $\alpha\beta$ ($\alpha,\beta$)=(x,y,z) of the molecular stress tensor $P_{\alpha\beta}(t)$ defined by:
\begin{eqnarray}
P_{\alpha\beta}=\sum_{i=1}^{N}m_i v_i^{\alpha} v_i^{\beta} +
\sum_{i=1}^N \sum_{j>i}^N F_{ij}^{\alpha} r_{ij}^{\beta}\quad
\alpha \neq \beta,
\label{equ2}
\end{eqnarray}
where the brackets in equ. (\ref{equf}) refer to an average over the whole system. In equ. (\ref{equ2}), $m_i$ is the mass of atom $i$, and F$_{ij}^{\alpha}$ is the component $\alpha$ of the force between the ions $i$ and $j$, $r_{ij}^{\beta}$ and $v_i^\beta$ being the $\beta$ component of the distance between two atoms $i$ and $j$, and the velocity of atom $i$, respectively. 
\par
\begin{figure}
\begin{center}
\end{center}
\caption{\label{activ} a) Calculated activation energies $E_A$ for diffusion (red) and viscosity (black) as a function of the number of constraints $n_c$ \cite{NatComm}, derived from an Arrhenius plot of oxygen diffusivity and Green-Kubo calculated viscosity in a densified sodium silicate. b) Calculated relaxation time $\tau$ as a function of $n_c$ determined from a separate evaluation of the viscosity and the instantaneous shear modulus \cite{bauchy20}.
}
\end{figure}
When such calculated viscosities are investigated \cite{du4} at fixed pressure/density as a function of inverse temperature in an Arrhenius plot, a linear behavior is obtained which allows to extract an activation energy $E_A$. A similar procedure can be realized for diffusivity \cite{bauchy2011}. Both activation energies and diffusivities are found to display a minimum with pressure \cite{bauchy1,bauchy2} (Fig. \ref{compar}). When the number of constraints is independently calculated \cite{dens_prl,NatComm}, it has been detected that the minimum in $E_A$ coincides with $n_c$=3 (Fig. \ref{activ}a) suggesting that isostatic networks will lead to a singular relaxation behavior with weaker energy barriers, as also detected experimentally in e.g. Ge-As-Se \cite{r18} (Fig. \ref{tatsumisago}) for which the condition $n_c=3$ coincides with the mean coordination number of $\bar r$=2.4 defining the condition of isostaticity. Given that one has ${\cal M}$=E$_A\ln_{10}$ 2/k$_B$T$_g$, the combination of a minimum in $E_A$ and a smooth or constant behavior of T$_g$ with thermodynamic conditions (pressure, composition) might lead to a minimum in fragility. Such observations have been made for certain chalcogenide melts \cite{r13,asse_punit,lucas_asse} for which minima in $E_A$ and ${\cal M}$ coincide, and, under certain assumptions regarding structure, the link with the isostatic nature of the network could be established \cite{gese,chakra,selven}. 
\par
These conclusions actually parallel those made from a simplified Kirkwood-Keating model of the glass transition \cite{JPCM2010} showing that isostatic glass-forming liquids have an activation energy for relaxation time which is minimum. In addition, a calculated relaxation time $\tau$ in the region of the glass transition for the same densified silicate (2000~K, \cite{NatComm}) has been found to evolve similarly. A deep minimum is found, indeed, in the relaxation time ($\tau\simeq$~2-3 ps) in the region where the system is nearly isostatic (3.0$<n_c<$3.2, Fig. \ref{activ}b).

\subsection{Reversibility windows}

\begin{figure}
\begin{center}
\end{center}
\caption{\label{hystero} a) Volume-temperature dependence during a cooling (blue) and heating (black) cycle for selected pressures in a liquid silicate \cite{NatComm} across the glass transition. The volumes have been rescaled with respect to their evolution in the glassy state V$_{glass}$. Curves at 0 GPa and 5 GPa have been shifted by multiples of 0.1. The cycle leads to a hysteresis which is due to structural relaxation but is also controled by rigidity. The inset shows the hysteresis area of the enthalpic ($A_H$) and volumetric ($A_V$, red curve, right axis extracted from the main panel) hysteresis as a function of the applied pressure P, defining a reversibility window (gray zone). b) Calculated total number of constraints per atom. The horizontal red line represents the isostatic line $n_c$=3, and separates flexible from stressed rigid networks. c) Calculated number of oxygen stretching- and bending constraints $n_c^{BS}$ (black) and $n_c^{BB}$ (red) as a function of pressure. The gray zone (reversibility window) in panel b and c refer to the one defined in the inset of panel a. 
}
\end{figure}

\subsubsection{MD signature} Isostatic glasses furthermore display reversibility windows, i.e. a tendency to display a minimum of thermal changes at the glass transition which is obviously linked with the particular relaxation behavior (Fig. \ref{activ}). When MD numerical cycles are performed across the glass transition from a high temperature liquid, one finds a hysteresis between the cooling and heating curve (Fig. \ref{hystero}a) in a similar fashion to the salient experimental phenomenology of the glass transition (Fig. \ref{thermal}). This behavior simply reflects the off-equilibrium nature of glasses that are able to slowly relax at $T<T_g$, and decrease volume or enthalpy as the glass is heated back to the liquid phase. However, it has been found \cite{NatComm} that for selected thermodynamic conditions (pressure) and fixed cooling/heating rate the hysteresis curves become minuscule, and the cooling/heating curves nearly overlap. 
When the area $A_H$ ($A_V$) of the enthalpy (volume) hysteresis is investigated as a function of pressure or density (inset Fig. \ref{hystero}a), a deep minimum is found which reveals a so-called {\em reversibility window} (RW) \cite{NatComm,PRB2015}.
\par
These thermal anomalies are actually linked with the isostatic nature of the glass-forming liquid, as detected from a MD based constraint count (Fig. \ref{hystero}b,c). A calculation of the total number of constraints shows a plateau-like behavior at a value $n_c$=3 between ~3 GPa and 12 GPa, which can be put in parallel with the evolution of the hysteresis areas. The detail shows that angular (BB) adaptation drives the mechanical evolution of the liquid under pressure because BS constraints increase due to the conversion \cite{natcomm1} of silica-like tetrahedral order which prevails at ambient conditions, into octahedral order which dominates at elevated pressure, and which is typical of the short range order of the crystalline stishovite polymorph \cite{natcomm2}. However, at a pressure of about 3 GPa, the system attains an obvious threshold, and further compression leads to a decrease of the number of BB constraints which indicates that some of the angular interactions have soften. Upon further compression, this evolution holds up to a pressure of about 12 GPa beyond which an important growth takes place. The results indicate an obvious correlation between the RW threshold pressures and those obtained from the constraint count, while identifying the isostatic nature of RW.

\begin{figure}
\begin{center}
\end{center}
\caption{\label{ip_exp} Measured non-reversing heat flow $\Delta H_{nr}$ as a function of modifier content in telluro-vanadate (TeO$_2$-V$_2$O$_5$, \cite{hmo}) and borate glasses (B$_2$O$_3$-M$_2$O, M=Li, Na \cite{borates}). Note the square well behavior of $\Delta H_{nr}$ with composition, and the nearly constant value of $\Delta H_{nr}$ over select intervals in composition.
} 
\end{figure}

\subsubsection{Experimental signature from calorimetry}
There is actually a strong experimental support for these findings connecting RW with the isostatic nature of the network structure, and a vast literature has been accumulated on this topic during the last fifteen years. One of the most direct signatures of reversibility windows having a nearly one to one correspondance with the result from MD (inset of Fig. \ref{hystero}a) comes from mDSC measurements (equ. (\ref{dotH}), Fig. \ref{prb00_mdsc}) \cite{tgGeSe} which exhibit a minimum (Fig. \ref{ip_exp}) or even a vanishing (in selected cases, see Fig. \ref{ip_exp}a) of the non-reversing enthalpy $\Delta H_{nr}$ \cite{ip_joam}. The sharp boundaries allow one to define a compositional window displaying these enthalpic anomalies (e.g. 23.5~\%~$<x<$~26.5~\% in TeO$_2$-V$_2$O$_5$, Fig. \ref{ip_exp}a, \cite{hmo}), and these can be also evidenced to a lesser extent from the total heat flow and the heat capacity jump at the glass transition (Fig. \ref{tatsumisago}, \cite{r18,varshneya1}).
\par
The use of such calorimetric methods to detect the nearly reversible character of the glass transition has not been without controversy, due, in part, because the intrinsic measurement of $\Delta H_{nr}$ depends on the imposed frequency, and relates to the imaginary part of the heat capacity $C_p"(\omega)$ \cite{kobac,descamps1}. Frequency corrections \cite{mauro} have to be taken into account in order to avoid the spurious effects arising from the frequency-dependence of the specific heat \cite{nagel1}.  
Even with this frequency correction on the non-reversing heat flow leading to a neat measurement of $\Delta H_{nr}$, results have been challenged by several authors who have argued that conclusions drawn from the observed anomalies (Fig. \ref{ip_exp}) might well be the result of a measurement artefact \cite{lucas_geasse,lucas_geass,con1,con2,con3}. However, $\Delta H_{nr}$ appears to be not only sensitive to impurities and inhomogeneities \cite{ijags} but also to the relaxation state of the glass \cite{ijags,mdsc_asse} so that the accurate detection, measurement and reproduction of $\Delta H_{nr}$ represent a true experimental challenge. Furthermore, it has been demonstrated that conclusions against the detection of a RW were based on samples of unproven homogeneity \cite{bhag,mdsc_asse}, as exemplified from the dependence of the fragility with the reaction time (Fig. \ref{frag_homo}).

\begin{figure}
\begin{center}
\end{center}
\caption{\label{rw} Location of experimental reversibility windows driven by composition for different chalcogenide and modified oxide glass systems \cite{NatComm}:  Si-Se \cite{selven}, Ge-Se \cite{gese}, Ge-S \cite{chakra}, As-Se \cite{asse_punit}, As-S \cite{ass_ip}, P-Se \cite{pse_ip}, P-S \cite{ps_ip}, Ge-Se-I \cite{gesei_ip}, Ge-S-I \cite{gesi_ip}, Ge-As-Se \cite{wang00}, Ge-As-S \cite{geass_ip}, Ge-P-Se \cite{gepse_ip}, Ge-P-S \cite{geps_ip}, Ge-Sb-Se \cite{gesbse_ip}, Si-Ge-Te \cite{tell2,sigete_ip}, Ge-Te-In-Ag  \cite{geteinag_ip}, SiO$_2$-M$_2$O (M=Na,K) \cite{amin}, GeO$_2$-M$_2$O (M=Li,K,Cs) \cite{germanate_ip}, GeO$_2$-Na$_2$O \cite{geo2}, AgPO$_3$-AgI \cite{novita1}, TeO$_2$-V$_2$O$_5$ \cite{hmo} and B$_2$O$_3$-M$_2$O (M=Li, Na) \cite{borates}. In the same families of modified oxides (e.g. borates, see Fig. \ref{ip_exp}b), there is an effect of the cation size. Using the 8-${\cal N}$ (octet) rule, the location of RWs can be represented in select systems (bottom panel) as a function of the number of constraints $n_c$ using the mean-field estimate of $n_c$ (equ. (\ref{nc0})). Permission from American Physical Society.
}
\end{figure}
\par
A large number of network glasses -chalcogenides, oxides- display RWs, and these are summarized in Fig. \ref{rw}. These represent systems which cover various bonding types, ranging from ionic (silicates \cite{amin}), iono-covalent, covalent (Ge-Se, \cite{gese}), or semi-metallic (Ge-Te-In-Ag \cite{geteinag_ip}). In certain of these systems, e.g. for the  simple binary network glasses such as Ge$_x$S$_{1-x}$ or Si$_x$Se$_{1-x}$, the experimental boundaries of the RW are found to be all very close \cite{gese,chakra,selven}, i.e. located between 20~\%~$<x<$~25~\%, and aspects of topology fully control the evolution of rigidity with composition, given that there is a weak effect in case of isovalent Ge/Si or S/Se substitution. This compositional interval defining the RW connects to the mean-field estimate of the isostatic criterion (equ. (\ref{nc_final})) satisfying $n_c=3$ because coordination numbers of Ge/Si and S/Se can be determined from the 8-${\cal N}$ (octet) rule to yield an estimate of the constraints $n_c$=2+5$x$ using equ. (\ref{nc0}). In fact, for these IV-VI glasses, the lower boundary of the RW ($x_c$=20~\%) coincides with the Phillips-Thorpe \cite{r6,mft1} mean-field rigidity transition $n_c$=3 and $\bar r$=2.4. 
\par  
For most of the systems however, uncertainties persist regarding the constraint count (equ. (\ref{nc0})) derived from the local structures and geometries. Coordination numbers and related active/inactive constraints must be derived from specific structural models, and this becomes already apparent when Group V selenides/sulphides are considered (Fig. \ref{rw}) because different RW locations are found for isovalent compounds, e.g. differences emerge between As- and P-bearing chalcogenides, and between sulphides and selenides (e.g. P$_x$S$_{1-x}$ and P$_x$Se$_{1-x}$, \cite{geps_ip}). Local structural features have been put forward to explain the trends due to chemistry \cite{asse_punit,ass_ip,pse_ip,ps_ip}, as well as the special effect of sulfur segregation in sulphide-rich glasses , and these have also served for the characterization of related ternaries \cite{wang00,geass_ip,gepse_ip,geps_ip}. The validity of these structural models is still debated in the literature, although rather well established in some cases from spectroscopic studies \cite{ps_ip,pse_ip}. The above statements seem to remain valid when the tellurides are considered. Because of the increased electronic delocalization of the Te atoms, Group IV and V atoms not necessarily follow the 8-${\cal N}$ rule and lead to mixed local geometries that are now composition dependent \cite{tell3,tell4}, e.g. sp$^3$ tetrahedral and defect-octahedral for Ge atoms, so that a proper constraint count must rely on accurate simulations, in conjunction \cite{cons2,cons3} with MD based constraint counting algorithms such as those derived above. RW have also been measured in modified oxides for which the connectivity change is realized by the addition of modifiers which depolymerize the network structure (Fig. \ref{zach}). As a result, the same phenomenology is found, and RWs have been detected between the two possible end limits of networks or elastic phases, i.e. strongly depolymerized and flexible (e.g. pyrosilicates SiO$_2$-2Na$_2$O) or highly connected and stressed rigid (e.g. silica-rich silicates).
 
\subsubsection{Alternative signatures of RWs}

The presence of a peculiar relaxation phenomena that induces RWs for select compositions leads to various other anomalous behaviours - maxima or a minima in physical properties- in the glassy state. These provide other alternative and complementary evidence of the RW signature from calorimetric (mDSC) measurements. Fig. \ref{review_ip} displays a survey of some of such properties for three families of modified glasses with widely different chemical bonding, although they display similar features in terms of rigidity: covalent Ge$_x$Se$_{1-x}$ \cite{gese,fei}, ionic (1-$x$)AgPO$_3$-$x$AgI \cite{novita1,novita2} and iono-covalent (1-$x$)GeO$_2$-$x$Na$_2$O glasses \cite{geo2}. When the atomic sizes are comparable (e.g. r$_{Ge}$=1.22~\AA, and r$_{Se}$=1.17~\AA\ for the covalent radius in Ge-Se), it has been suggested that glasses will display an increased tendency to space-filling because of the isostatic nature of the networks (i.e. absence of stress \cite{fei}), which manifests by a minimum in the molar volume (Fig. \ref{review_ip}a and c), a salient feature that has been reported for various systems \cite{r13,asse_punit,ip,gese,tell1,hmo,geo2,bourgel}. The stress-free nature of such RWs has been detected from pressure experiments \cite{fei} showing the vanishing of a threshold pressure (Fig. \ref{review_ip}a) prior to a pressure-induced Raman peak shift. This peak shift serves usually to quantify and to measure residual stresses in crystals. Ionic conductors (Fig. \ref{review_ip}b) display an onset of ionic conduction only in compositions belonging to the flexible phase, i.e. when the network can be more easily deformed at a local level because of the presence of floppy modes \cite{novita1} which promote mobility. This leads to an exponential increase of the conductivity. However, it is to be noticed that in RWs an intermediate conductive r\'egime sets in, which also shows an anomalous behavior for a typical jump distance associated with dynamics \cite{prl2010}. Other quite different probes have also revealed the signature of RWs such as dc permittivity (Fig. \ref{review_ip}b) \cite{novita1} or the frequency \cite{geo2} associated with the imaginary part the complex dielectric function (IR-TO, Fig. \ref{review_ip}c). 

\begin{figure}
\begin{center}
\end{center}
\caption{\label{review_ip} Different quantities showing an anomalous behavior in reversibility windows \cite{bauchy2}. a) Raman threshold pressure \cite{fei} and molar volume \cite{gese} (right axis) in Ge$_x$Se$_{100-x}$ as a function of Ge composition. b) Ionic conduction and zero-frequency permittivity (right axis) in (100-$x$)AgPO$_3$-$x$AgI as a function of AgI composition \cite{novita1}. c) IR-TO vibrational frequency and molar volume (right axis) in (100-$x$)GeO$_2$-$x$Na$_2$O as a function of Na$_2$O composition \cite{geo2}. The gray areas correspond to the reversibility windows determined from calorimetry (mDSC), such as in Fig. \ref{ip_exp}. Permission from J. Wiley and Sons 2015.
} 
\end{figure}

\subsubsection{Insight from models: evidence for an elastic intermediate phase}

A certain number of scenarios have been proposed to describe the observed behaviors depicted in Figs. \ref{ip_exp} and \ref{review_ip}, and some emphasize the role of fluctuations \cite{thorpi1,barre,PRB2003,mousseau_ip} in the emergence of a double threshold/transition that define an intermediate phase (IP) between the flexible and the stressed rigid phase. Alternatively, mean-field aspects of jamming have been considered, and here fluctuations in e.g. coordinations are thought to be limited, but atoms are coupled spatially via elasticity and can organize locally into distinct configurations that may promote an IP. 
\par
Given the link between isostaticity \cite{NatComm} and reversibility at the glass transition (Figs. \ref{hystero} and \ref{rw}), and following the path based on coordination fluctuations several authors have attempted to modify the modelling of the initial mean-field theory \cite{r6,mft1} that leads to a solitary phase transition when $n_c=3$ (or $\bar r$=2.4 if all BS and BB constraints are considered as intact). These contributions usually assume that amorphous networks will adapt during the cooling through the glass transition, similarly to the angular adaptation revealed from MD \cite{micol1,NatComm}, in order to avoid stress from additional cross-linking elements. 
\par
Using a graph-theoretical approach, Thorpe and co-workers \cite{thorpi1,thorpi2} have developed an algorithm (a Pebble Game, \cite{thorpi3}) that takes into account the non-local characteristics of rigidity, and allows to calculate the number of floppy modes, to locate over-constrained zones of an amorphous network, and ultimately identify stressed rigid rigid clusters for simple bar-joint networks. In the case of simulated self-organized or adaptive networks, the addition of bonds in a network with increasing average connectivity will be accepted only if this leads to isostatically rigid clusters, so that the emergence of stressed rigid clusters is delayed.  However, with a steady increase of the connectivity, the network will undergo percolation of rigidity (a rigidity transition at $\bar r_{c1}$) which leads to an unstressed (isostatic) structure (Fig. \ref{thorpi_fig}). The addition
of new bonds will contribute to the occurrence of stressed rigid clusters that finally percolate at a second transition ($\bar r_{c2}$), identified with a stress transition, and both transitions define, indeed, a window in connectivity $\Delta \bar r$=$\bar r_{c2}-\bar r_{c1}$, and and IP. 
\par
\begin{figure}
\begin{center}
\end{center}
\caption{\label{thorpi_fig} Evidence for a stress-free intermediate phase from the Pebble Game analysis (adapted from \cite{thorpi1,thorpi2}). Fraction of sites on isostatically rigid and stressed rigid percolating cluster in a self-organized network as a function of the network mean coordination number $\bar r$. The intermediate phase which is rigid but unstressed exists in these classes of models between 2.375$\leq \bar r\leq$ 2.392, and coalesces in random networks. This generic behavior is also observed from a spring network \cite{wyart0} (top) showing regions which are flexible (blue), isostatic (marginally constrained, green) and stressed rigid (red). Permission from American Physical Society.} 
\end{figure}
Other approaches have built on the same idea, using either a spin cavity method \cite{barre} or cluster expansions \cite{PRB2003,JNCS2007}. These theories lead to a solitary floppy to rigid transition in absence of self-organization, and to an intermediate phase corresponding to a window in composition/connectivity in which the network is able to adapt in order to lower the stress due to constraints. However, some of these models do not take into account the fact that rigid regions cost energy and, thus, corresponds to an infinite temperature. Also, the Pebble Game \cite{thorpi1} and the cavity method \cite{barre} apply on T=0 networks which have infinite energy barriers for bond change/removal. Thermal effects have been included \cite{mousseau_ip,mousseau_ip2} and an equilibrated self-organized IP has been recovered for two-dimensional lattice-based models. An important outcome from these models is that an increased sensitivity upon single bond addition or removal exists close to the IP, and this suggests that the system is maintained in a critical state on the rigid-floppy boundary throughout the IP. 
\par
Instead, using the phenomenology of the elasticity of soft spheres and jamming transitions, Wyart \cite{wyart0} and co-workers have shown that the RW could occur in a certain number of physical situations by considering a lattice spring model for rigidity transitions with weak noncovalent (Van der Waals) interactions \cite{wyart1,wyart2,wyart3}. It reveals that the temperature considerably affects the way an amorphous network becomes rigid under a coordination number increase, and the existence of an isostatic reversibility window not only depends on $T$ but also on the relative strength of the weak forces. In a strong force r\'egime, a RW can be found which is revealed by a finite width in the probability to have a rigid cluster spanning the system, driven by fluctuations in coordination, similarly to the results of the Pebble Game (\cite{thorpi3}, Fig. \ref{thorpi_fig}). However, when weak interactions are present, the RW disappears below a certain temperature suggesting that the transitions become mean-field at low temperature and coalesce. Furthermore, weak interactions lead to an energy cost for coordination number fluctuations, which decay at finite temperature. 
These results are partially supported by MD simulations \cite{NatComm} taking into account long-range interactions (Coulomb, Van der Waals) allowing to probe the weak-force r\'egime. Here coordination fluctuations are found to be small given the weak abundance of five-fold Si atoms (10-20~\%) \cite{du4}, and fluctuations are essentially found in angular constraints which show a non-random distribution (Fig. \ref{hetero}, \cite{EPL13}). However, the vibrational analysis \cite{wyart0} suggests that in the IP vibrational modes are similar to the anomalous modes observed in packings of particles near jamming, thus providing also an interesting connection with the jamming transition \cite{nagel00} that might also be embedded in the anomalous variation of the molar volume (Fig. \ref{review_ip}a,c). 
\par
This mean-field scenario of the IP is also the one followed \cite{mouk} in a rigidity percolation model on a Bethe lattice \cite{mouk1,mouk2,mouk3} that is based on a binary random bond network with a possibility of having two types of degrees of freedom. Under certain conditions, two discontinuous transitions are found, and the associated IP displays an enhanced isostaticity at the flexible boundary. As a result, the entire IP has a low density of redundant bonds and has, therefore, a low self-stress. The double transition solution is found to depend on the coordination and the degrees-of-freedom contrast, and might be directly comparable to experiments although important coordination contrasts do not necessarily correspond to situations encountered in e.g. chalcogenides \cite{gese}. 
\par
Although some other models \cite{con3,con4} with a weaker theoretical basis have argued that the existence of the IP remains elusive, albeit contradicted by the variety of experimental signatures, there is a strong theoretical and numerical indication that RW or IP glasses display a particular relaxation kinetics manifesting in $\Delta H_{nr}$ that leads to anomalous properties in different physical properties (Fig. \ref{review_ip}). These findings have actually a much more universal ground because links between the RW and protein folding \cite{protein}, high-temperature superconductors \cite{htsc} or computational phase transitions \cite{monasson} have been stressed. Such deep analogies simply underscore the fact that a complex network with external constraints/conditions has the ability to lower its energy by adapting internal thermodynamic variables. 

\section{Numerical methods}

As already mentioned in the previous sections, MD simulation is the method of choice to investigate aspects of glassy relaxation in relationship to structure, this relationship being central for the case of network-forming liquids. Rather than presenting the basis of computer simulations (see Refs. \cite{md1,md2,md3}), we discuss which tools have been developed for an increased understanding of glassy relaxation. 
\par
In an MD simulation, the trajectories (i.e. the positions ${\bf r}_i(t)$ of N particles with $i$=1..N) serve as starting point for further investigations regarding e.g. relaxation. Here ${\bf r}_i(t)$ are obtained by solving Newton's equations of motion for a given system using e.g. the well-known Verlet algorithm:
\begin{eqnarray}
{\bf r}_i(t+\Delta t)={\bf r}_i(t)+{\bf v}_i(t)\Delta t+{\frac {(\Delta t)^2}{2}}{\bf F}_i(t) 
\end{eqnarray}
where:
\begin{eqnarray}
{\bf v}_i(t+\Delta t)={\bf v}_i(t)+{\frac {\Delta t}{2}}\biggl[{\bf F}_i(t)+{\bf F}_i(t+\Delta t)\biggr] 
\end{eqnarray}
and ${\bf F}_i(t)$ is the force acting on atom $i$, and is derived from an interaction potential that has been fitted in order to reproduce some of the properties of the system of interest. $\Delta t$ is the time step for the integration of the equations of motion (typically 1~fs in classical MD \cite{md3}), and usually several orders of magnitude lower than the typical atomic vibrational frequency. There are intrinsic limitations with the MD method which concern timescales and size. For the latter aspect, with the available computer power, one is able to investigate systems of up to 10$^7$ atoms, whereas the timescale will be limited to the $\mu$s domain. This means that the glass transition can be only partially addressed using these methods, and is limited to the liquid-to-supercooled domain, i.e. to temperatures having, to the best, a relaxation time of $\tau\simeq 1~\mu$s. 
\par
From these trajectories \{${\bf r}_i(t)$\}, different properties of the supercooled liquid can be directly calculated, at least in principle, while also connecting with aspects of structure or constraints (Fig. \ref{defini}). This is an important reason, and this has motivated an important body of research in the very recent years. 
\subsection{Dynamic observables}
\subsubsection{Diffusion and viscosity} 

A useful means for the investigation of the dynamics of the glass-forming liquid is given by the investigation of the mean-square displacement of an atom of type $\alpha$: 
\begin{equation}
\langle r^2(t) \rangle =\frac{1}{N_{\alpha}} \sum_{i=1}^{N_{\alpha}}
\langle |{\bf r}_i(t)-{\bf r}_i(0)|^2\rangle\quad,
\label{msdd}
\end{equation}
where the bracketts indicate ensemble averages. The behavior of $\langle r^2(t) \rangle$ with time and temperature displays some generic behaviors. At high temperature and short times, the motion of the atoms is usually governed by a ballistic r\'egime for which $\langle r^2(t) \rangle$ scales as $t^2$. At long times the dependence of $\langle r^2(t) \rangle$ becomes linear (Fig. \ref{diffusion_fig}, left) and signals the onset of diffusion, with a diffusion constant that follows Einstein's relation $\lim_{t\to \infty} \langle r^2(t)\rangle/6t$=D. In a multicomponent liquid, one can, thus, have access to the diffusion constants D for different species, and these are represented in Fig. \ref{diffusion_fig} (right) for e.g. sodium silicates in an Arrhenius plot.   
\begin{figure}
\caption{\label{diffusion_fig} Left: Time dependence of the silicon mean squared displacement for different temperatures in liquid silica \cite{horbach0}. Permission from American Physical Society. Right: Computed diffusion constants D$_{Na}$, D$_{Si}$ and $D_{O}$ in a sodium silicate liquid as a function of inverse temperature (blue curves and symbols), compared to experimental data for D$_{Na}$ (see Ref. \cite{du4} for references) and to the simulated values of D$_{Na}$, D$_{Si}$ and D$_O$ using an alternative potential (red curves and symbols, Horbach et al. \cite{horbach}. Permission from Elsevier 2015.}
\end{figure}
The present figure is quite instructive because it also signals that a change in the force field \cite{bks,teter} used for the MD simulations can lead to behaviors that can be very different, and diffusion constants can differ by at least one order of magnitude, and may, therefore, disagree in some cases with experimental measurements.
\par
With decreasing temperature and the slowing down of the dynamics, the mean square displacement reveals some additional features (Fig. \ref{diffusion_fig} left) because $\langle r^2(t) \rangle$ still extends to the diffusive regime for the longest simulation times, but in addition shows a plateau-like behavior. This feature appearing at intermediate times is due a cage effect created by neighbouring atoms which trap the tagged atom during a certain time interval (e.g. 0.1~ps-100~ps for 2750~K silica, see Fig. \ref{diffusion_fig}, left), the typical distance associated with this phenomenon being of the order of a fraction of \AA\ ($\langle r^2(t) \rangle\simeq$0.1-1\AA), i.e. somewhat smaller than a typical bond distance. For sufficiently long times however, the atom is able to escape from that cage, and diffusion sets on.
\par
Once the diffusion constant is determined, it has been found that most of these simulated network-forming liquids display an Arrhenius dependence (Fig. \ref{diffusion_fig}, right) for the diffusivity \cite{bauchy_asse,pre06,du4,horbach0,horbach,joam_gese} leading to an estimate of an activation energy $E_A$ (e.g. Fig. \ref{compar}) that is found to be close to experimental findings (e.g. $E_A$=4.66~eV in silica \cite{horbach0}, compared to the experimental 4.70~eV \cite{brebec1,brebec2}). For selected systems, simulation data exhibit a significant curvature in diffusivity at higher temperatures \cite{pre06,horbach0} that has been interpreted as a reminiscence of the more fragile behavior of the liquids \cite{horbach0} once $T$ increases.
\par
At high temperature, the evolution of diffusivity parallels the one found for calculated viscosities $\eta$ using the Green-Kubo formalism and equs. (\ref{equf})-(\ref{equ2}) which is also found to be Arrhenius-like, and corresponding activation energies $E_A$ are similar (see Fig. \ref{activ}). This simply reveals that the Stokes-Einstein relationship holds :
\begin{eqnarray}
\label{stok1}
D\eta={\frac {T}{6\pi R}},
\end{eqnarray}
where $R$ is the particle radius ($\simeq$\AA) moving in a fluid. Alternatively, the phenomenological Eyring equation can be used 
\begin{eqnarray}
\label{eyring}
\eta={\frac {k_BT}{\lambda D}}.
\end{eqnarray}
Here $\lambda$ represents a typical jump distance in the liquid, of the order of a bond distance. It has been shown \cite{shimizu1,shimizu2} that the Eyring equation works well with viscous liquids such as silicates with a high silica contents provided that $\lambda$ is taken as the oxygen-oxygen mean distance, a result that was also checked numerically for two other silicate liquids \cite{du4}. 

\subsubsection{Van Hove correlation function}
An additional signature of the dynamics is given by the Van Hove correlation functions which probes in real space, rather than the average value (equ. (\ref{msdd})), the distribution of distances over which the particle has moved during a time $t$. This is conveniently quantified by the self-part of the Van Hove correlation function defined as:
\begin{eqnarray}
\label{vanhove_def}
{\cal G}_s^\alpha(r,t)={\frac {1}{N_\alpha}}\Biggl\langle\sum_{k=1}^{N_\alpha}\delta(r-\vert{\bf r}_k(0)-{\bf r}_k(t)\vert)\Biggr\rangle
\end{eqnarray}
where $\delta(r)$ is the Dirac function. This function ${\cal G}_s^\alpha(r,t)$ is the probability density of finding an atom $\alpha$ at time $t$ knowing that this atom was at the origin ($r=0$) at time $t=0$. By probing the probability that an atom has moved by this distance $r$, one is, therefore, able to gather additional information about dynamics. Figure \ref{vanhove_asse} shows such a function (4$\pi r^2{\cal G}_s^{Se}(r,t)$) for liquid As$_2$Se$_3$ \cite{bauchy_asse} at fixed temperature for different times.
\begin{figure}
\begin{center}
\end{center}
\caption{\label{vanhove_asse} Calculated self part of the Van Hove correlation function 4$\pi r^2{\cal G}_s^{Se}(r,t)$ at 800~K for three selected compositions in As-Se liquids \cite{bauchy_asse}: flexible As$_{20}$Se$_{80}$ (black), intermediate As$_{35}$Se$_{65}$ (red) and As$_{40}$Se$_{60}$ (blue). The function is represented for selected times: 12~fs, 0.12~ps, 1.2~ps, 12~ps. Permission from AIP Publishing LLC 2015.}
\end{figure}
Note that because of the isotropic nature of the system, the angular integration can be performed leading to the term 4$\pi r^2$. It is seen that for very short times (12~fs), 4$\pi r^2{\cal G}_s^{Se}(r,t)$ nearly reduces to the Dirac function as it should be \cite{hansen}, given the definition of ${\cal G}_s(r,t)$ (equ. (\ref{vanhove_def})). For increased times however, the Se atoms now experience larger distances for a given time, and for $t=12~ps$, atoms move over distances typical of second nearest neighbor distances (4-5~\AA). The second important characteristics that appears from Fig. \ref{vanhove_asse} is that the distribution is not Gaussian for long times as would be expected for an ordinary liquid for which relaxation phenomena are neglible \cite{hansen}. In this simple case, the mean square displacement is, by definition, the second moment of the Van Hove function which behaves as:
\begin{eqnarray}
\label{glotz_eq}
{\cal G}_s(r,t)={\frac {\exp\biggl[-{\frac {r^2}{4Dt}}\biggr]}{\biggl(4\pi Dt\biggr)^{3/2}}}
\end{eqnarray} 
Here, the function appears to be much wider with tails in the long time limit that have been revealed by a series of simulations of network forming liquids (silica \cite{kob_vh}, As$_2$Se$_3$ \cite{bauchy_asse}, densified silicates \cite{bauchy2011}).
For short times however, the Van Hove function is made of a single Gaussian distribution that is shifted to the right with increasing time, and the location of the maximum evolves as $t^2$ which arises from the ballistic behavior of $\langle r^2(t) \rangle$ (Fig. \ref{diffusion_fig}, left). This characteristics does not apply at intermediate times but it recovered at very long times, for which ${\cal G}_s(r,t)$ is again given by a Gaussian. A convenient way to characterize the departure from such distributions is given by the non-Gaussian parameter \cite{non_gaussian}:
\begin{eqnarray}
\label{non_gauss}
\alpha_2={\frac {3\langle r^4(t) \rangle}{5\langle r^2(t) \rangle^2}}-1
\end{eqnarray}
which becomes non-zero at intermediate times (Fig. \ref{glotz_fig}) when $\langle r^2(t) \rangle$ exhibits a plateau-like behavior (Fig. \ref{diffusion_fig}, left), and which is directly related to the cage effect when neighbouring atoms act as a trap for a moving particle. Current investigations have focused on glass-forming liquids such as water \cite{nongauss1}, silica \cite{nongauss2b,sciortino_ag3,nongauss2} or alumino-silicates \cite{vanhoang}, and have established the correlation between the onset of the $\beta$-relaxation plateau (Fig. \ref{stretchf}) and departure ($\alpha_2\neq$ 0) from a Gaussian distribution in the $r$-dependence of ${\cal G}_s(r,t)$. In addition, the large $r$ tail seen in the Van Hove correlation function \cite{nongauss4,nongauss5} can be also rather well described \cite{kob_vh1} by an exponential decay of the form ${\cal G}_s(r,t)\simeq$exp(-r/$\lambda(t)$) which signals that viscous liquids will differ quite markedly from a high temperature liquid exhibiting a standard Fickian diffusion and a Van Hove function of the form of equ. (\ref{glotz_eq}). One of the main conclusions of such studies is that the dynamics of the atoms at long times and low temperature, i.e. those which contribute to the tail of ${\cal G}_s(r,t)$ have a non-trivial dynamics that can be further characterized using dynamic heterogeneities (see below), and seem to contain some universal features that are common to network glasses, colloids, grains or simple sphere systems \cite{kob_vh1}.

\begin{figure}
\begin{center}
\end{center}
\caption{\label{glotz_fig} Non-Gaussian parameter $\alpha_2$ for oxygen and silicon atoms at various temperatures in liquid silica \cite{nongauss2b}. The insets show the temperature dependence of the maximum of $\alpha_2$ which follows an Arrhenius law. Permission from American Physical Society 2015.}
\end{figure}
\subsubsection{Intermediate scattering function}

As described above, scattering experiments using e.g. neutron diffraction (Fig. \ref{kargl_fig}) are performed in reciprocal space, and can, therefore be compared to the calculated analogue of the intermediate scattering function (equ. (\ref{kargl_eq})) which uses direcly the positions ${\bf r}_i(t)$ obtained from the MD trajectory. 
\par
Such functions actually display the same phenomenology as the experimental ones, i.e. they exhibit a single Debye-like decay at high temperature, and lead to a $\beta$-relaxation plateau at lower $T$ which extends beyond the available computer timescale at low temperature (Fig. \ref{isf}, left). For intermediate temperatures however, the structural ($\alpha$) relaxation can be investigated and its characteristics ($\tau$, $\beta$, see Fig. \ref{stretchf}) determined as a function of the wavevector, temperature, etc. and correlated with other calculated structural properties, e.g. $\tau$ being a decreasing function of the wavevector (Fig. \ref{isf}, right). 
\par
Fig. \ref{isf} shows such an example in liquid silica \cite{kobijo} and liquid GeSe$_2$ for different temperatures and wavevectors. It is seen that $F_s(k,t)$ behaves very similarly to the schematic figure represented (Fig. \ref{stretchf}). As the temperature effects are considered, it is seen that $F_s(k,t)$ rapidly decays to zero at high temperature, and also reproduces the anticipated Debye single exponential. At low temperature, the usual two-step relaxation process is found that permits to detect an $\alpha$-relaxation for the longest times. A rescaling of the x-axis using an $\alpha$-relaxation time $\tau(T)$ defined by $F_s(k,\tau(T))$ = e$^{-1}$ usually permits to detect the two temperature r\'egimes that appear as a function of temperature. At high temperatures, such curves $F_s(k,t)$ fall at long times rather well onto a master curve that is accurately reproduced by a simple exponential. At low temperature, all curves also nearly overlap, but are reproduced this time by a stretched exponential \cite{kobijo}.

\begin{figure}
\begin{center}
\end{center}
\caption{\label{isf} Left: Time dependence of the oxygen-related intermediate scattering function $F_s(k,t)$ at different temperatures investigated \cite{kobijo}. The wave-vector $q$ is 1.7~\AA$^{-1}$, the location of the first peak in the structure factor. Permission of the American Physical Society. Right: same function $F_s(k,t)$ at fixed temperature (1050~K) in liquid GeSe$_2$ for different wavevectors $k$. Note that the timescale of GeSe$_2$ is signifcantly reduced because of a different modelling scheme ({\em ab initio} MD)}
\end{figure}
\par
The exponential appearing in the definition of the intermediate scattering function (equ. (\ref{kargl_eq})) can actually be expanded in $\langle r^2(t) \rangle$ and connects back to the $\beta$-relaxation and effects due to non-Gaussian dynamics. In fact, equ. (\ref{kargl_eq}) rewrites \cite{expand}:
\begin{eqnarray}
\label{expansion}
F_s(k,t)=\exp\biggl(-{\frac {k^2\langle r^2(t) \rangle}{6}}\biggr)\times
\biggl[1+{\frac {1}{2}}\biggl(-{\frac {k^2\langle r^2(t) \rangle}{6}}\biggr)^2\alpha_2(t)+...\biggr]
\end{eqnarray}
where $\alpha_2(t)$ is the non-Gaussian parameter given in equ. (\ref{non_gauss}) that can be accessed from measurements/calculations of $F_s(k,t)$ at different wavevectors (Fig. \ref{isf}, right). A certain number of limiting cases are intresting and useful for further analysis. For instance, in the time interval where $F_s(k,t)\gg$ e$^{-1}$, the quantity $k^2\langle r^2(t) \rangle/6$ is small, and the intermediate scattering function reduces to a single exponential that is equal to $\exp[-k^2\langle r^2(t)\rangle/6]$ and can be directly obtained from the Fourier transform of the "Fickian" Van Hove function (equ. (\ref{glotz_eq})), a condition that is also met when $\ln[F_s(k,t)]/k^2$ is independent of $k$. 
\par
Given the timescale involved, investigations of the glassy relaxation using the intermediate scattering functions have been essentially made on model glasses (soft, hard) sphere glasses or model network glasses \cite{coslov1,coslov2}, and for selected cases on oxides: borosilicates \cite{boro1} silicates at ambient \cite{isf_kob1,isf_kob2} or under pressure \cite{NatComm}, borates \cite{isf1}, calcium alumino-silicates \cite{isf3} because classical MD simulations can be performed with confidence. In this case, the dynamics explored on timescales (ns-$\mu$s) which are of the order of the timescale probed in neutron diffraction experiments \cite{kargl}. This situation contrasts with the one encountered in chalcogenides for which {\em ab initio} simulations are necessary \cite{masso} to account for structural defects and for charge transfer defining the covalent bonds (see however Fig. \ref{isf}, right), which considerably reduces the timescale ($\simeq$100~ps). The dependence of $F_s(k,t)$ with wavevector shows that probing the relaxation on larger lengthscales (i.e. smaller $k$) leads to a reduced dynamics, i.e. $F_s(k,t)$ decays more slowly, and eventually does not fall to zero for the largest computation time.

\subsection{Dynamic heterogeneities}

An inspection of the local displacements during simulations (Fig. \ref{dh_intro}) highlight the fact that the relaxation at the atomic scale is not homogeneous, and evidence has been found numerically that the dynamics is made of vibrations around well-defined positions followed by jumps once atoms have been able to escape from cages. In this respect, the existence of non-Gaussian dynamics \cite{kob_vh1} that connects to the $\beta$-relaxation plateau of the function $F_s(k,t)$ represents  a strong indication that one has a distribution of relaxing events that vary with space and time, and emphasize the central role of dynamic fluctuations in the viscous slowing down.

\begin{figure}
\begin{center}
\end{center}
\caption{\label{dh_intro} Atomic snapshot of liquid (1050~K) GeSe$_2$ \cite{n480} showing the particle displacements over 88~ps. Different color codes, from dark blue (1~\AA) to yellow-red (12~\AA) indicate that the displacement of the particles is not homogeneously distributed.}
\end{figure}

\subsubsection{Space and time fluctuations}
Detailed features about this slowing down have emerged in the more recent years, and studies provide strong evidence for the existence of these dynamic fluctuations in time and space, now also known as "{\em dynamic heterogeneity}" \cite{ediger}. Simply speaking, one attempts both from experimental and theoretical measurements and signatures \cite{nongauss4,hetero0,hetero00,hetero000} to quantify the fact that regions in the glass-forming liquid can have different relaxation rates to equilibration, and that these rates will evolve in a non trivial way with time and temperature. This is thought to lead to a rater obvious origin for the non-exponentiality of the $\alpha$-relaxation given that the KWW stretched exponential can be developed in series of exponentials with different typical relaxation times, and might indicate that the relaxation is locally exponential but with a spatial distribution that is complex and non linear. There is, however, experimental and theoretical evidence \cite{ediger} showing that even the local dynamics can be non-exponential as well, which increases the complexity.
\begin{figure}
\begin{center}
\end{center}
\caption{\label{dh_1} Time resolved squared displacement of individual Ge atoms in liquid (1050~K, \cite{n480}) and glassy (300~K, inset) GeSe$_2$. It is seen that individual trajectories are made of long periods of vibrations and cage-like motions with a reduced spatial extent but jumps can be noticed.}
\end{figure}
\par
On this issue, an insightful picture is again provided by MD simulations which show that while the mean-square displacement of a given species displays a smooth behavior with time (Fig. \ref{diffusion_fig}, left), and will, ultimately, provide some information about diffusion, there is evidence for species-dependent individual jumps that result, on average, in the spatial distribution of the Van Hove function. These salient features depicted in Fig. \ref{dh_1} are found for a variety of simple supercooled liquids \cite{hetero1,hetero2,hetero3}, and not only reveal that such events are intermittent with waiting times between successive jumps statistically distributed, but also that they strongly differ from one particle to another. 

\subsubsection{Four-point correlation functions}

An inspection of the single events depicted in Fig. \ref{dh_1} that lead to distributions in jump distances encoded in the function ${\cal G}_s(r,t)$ (Fig. \ref{vanhove_asse}) indicates that such spatio-temporal fluctuations cannot be described from ensemble-averaged measurements or calculations given that correlations between space and time fluctuations need to be considered. This also tells us such an inhomogeneous dynamics driven by mobile particles needs a generalization of mobility correlation functions, and current development have led to the definition of four-point correlation functions \cite{k41,k42} that focus on the statistical analysis of space and time deviations from the average behavior (for technical details, see e.g. \cite{hetero1,hetero_rev1}). There are also alternative approaches \cite{hetero000,dauchot,hurley} some of them focusing rather on the quantity: 
\begin{equation}
\label{five}
Q(t)=\sum_{i=1}^{N}\sum_{j=1}^{N}w\vert({\bf r}_{i}(0)-{\bf r}_{j}(t)\vert)
\end{equation}
which is a measure of the degree to which a configuration at time $t$ still overlaps the initial arrangement, the degree of overlapping being established from a window function $w({\bf r})$ (where $w({\bf r}) = 1$ if  $\vert {\bf r} \vert \le a$ and zero, otherwise). 
\par
In the simplest approach, one can define a mobility field $f_i(t)$ of the form:
\begin{eqnarray}
f({\bf r},t)=\sum_if_i(t)\delta({\bf r}-{\bf r}_i)
\end{eqnarray}
and its fluctuating part writes as $\delta f({\bf r},t)=f({\bf r},t)-\langle f({\bf r},t)\rangle$. This allows the definition of correlations over the fluctuations in both real and reciprocal space:
\begin{eqnarray}
\label{g4}
g_4({\bf r},t)=\langle\delta f({\bf 0},t)\delta f({\bf r},t)\rangle,
\end{eqnarray}
\begin{eqnarray}
\label{s4}
S_4({\bf k},t)=\langle\delta f({\bf k},t)\delta f({\bf -k},t)\rangle
\end{eqnarray}
$g_4({\bf r},t)$ depends only on the time $t$ and the distance ${\bf r}$, and is termed "{\em four-point}" because it measures correlations of motion arising at two points, 0 and {\bf r}, between 0 and t, and also connects to the variance of the mobility field $f_i(t)$. A dynamic susceptibility can be introduced from equ. (\ref{g4}):
\begin{eqnarray}
\chi_4(t)=\rho\int d^3rg_4({\bf r},t)
\end{eqnarray}
The function $\chi_4(t)$ represents the volume on which structural relaxation processes are correlated, and has been computed for a certain number of soft-sphere glass-forming liquids \cite{heterob1,heterob2} and silica \cite{heterob3}. Note that an alternative definition of $\chi_4(t)$ can be used from the overlapping function used in equ. (\ref{five}) \cite{hetero000}:
\begin{eqnarray}
\chi_4(t)={\frac {\beta V}{N}}\biggl[\langle Q^2(t)\rangle-\langle Q(t)\rangle^2\biggr].
\end{eqnarray}
The behavior of $\chi_4(t)$ with time appears to have some generic behaviour because for each temperature $\chi_4(t)$ displays a maximum at a peak position that corresponds to the relaxation time \cite{heterob3} of the system. As the temperature is decreased, the intensity of $\chi_4(t)$ increases (Fig. \ref{dh_3}) which signals the growing typical lengthscale involved in dynamic heterogeneities (space and time fluctuations), whereas the shift of the peak position to longer times reveals the increase of the relaxation time \cite{heterob3}.  

\begin{figure}
\begin{center}
\end{center}
\caption{\label{dh_3} Time dependence and temperature of $\chi_4(t)$ in a densified liquid 2 SiO$_2$-Na$_2$O (2000 K, red symbols), and in a Lennard-Jones liquid  (adapted from Ref. \cite{hetero000}). Temperatures have been rescaled in order to correspond to LJ argon \cite{wahn}.  As the temperature decreases, the position $t_{max}$ of the peak in $\chi_4(t)$ monotonically increases, and shifts to longer times. The behavior reveals an Arrhenius-like behavior (inset).}
\end{figure}
Given that the growth of the intensity of $\chi_4(t_{max})$ indicates a dynamics becoming increasingly spatially heterogeneous, with decreasing temperature a corresponding dynamic lengthscale $\xi_4$ can be accessed from the low wavevector region of $S_4(q,t)$ (equ. \ref{s4}) which shows an increase in intensity in the limit $q\rightarrow 0$ \cite{hetero000,heterob1,xi1,xi2} but, in absence of large systems \cite{xi3,xi4}, this limit is hardy attainable so that $\xi_4$ might be accessed from the numerical/experimental data using a low-$q$ functional form inspired from the analysis of static and dynamic density fluctuations in the Orstein-Zernike theory of liquid-gas transition \cite{OZ}.
\begin{eqnarray}
S_4(k,t)\simeq {\frac {S_0}{1+k^2\xi_4^2}}
\end{eqnarray}. 
\par
In supercooled silica, an analysis using the framework of four point correlation functions in different Ensembles \cite{heterob1,heterob3} shows that $\xi_4$ is an increasing function of the relaxation time. Furthermore, there is a strong indication \cite{science} that these multi-point dynamic susceptibilities can be accessed experimentally since temperature and density variations of averaged correlations of a mobility field $f(t)$ contribute to $\chi_4(t)$:
\begin{eqnarray} 
\label{sciki}
\chi_4(t)&=&\chi_4^{NVE}(t)+{\frac {k_B}{c_v}}T^2\biggl[{\frac {\partial \langle f(t)\rangle}{\partial T}}\biggr]^2+\rho^3k_BT\kappa_T\biggl[{\frac {\partial \langle f(t)\rangle}{\partial \rho}}\biggr]^2\\
&=&\chi_4^{NVE}(t)+{\frac {k_B}{c_v}}T^2\chi_T^2(t)+\rho^3k_BT\kappa_T\chi_\rho^2
\end{eqnarray}
where the susceptibilities $\chi_T$ and $\chi_\rho$ arise from the fluctuations induced by energy and density, respectively. At fixed density and low temperature $\chi_4^{NVE}(t)$ is much smaller, and $\chi_4(t)$ is dominated by the contribution due to $\chi_T$ (Figure \ref{chi4_fig}) that can be measured from the temperature variation of system averaged correlations $\langle f(t)\rangle$.
\begin{figure}
\begin{center}
\end{center}
\caption{\label{chi4_fig} Left: Time dependence of $\chi_4(t)$ in BKS silica \cite{heterob3}. At low temperature (large relaxation time), one has $\chi_4^{NVE}\simeq T^2\chi^2_T/c_v$ (see equ. (\ref{sciki})). Permission from AIP Publishing LLC 2015.}
\end{figure}
Other MD simulations on liquid silica have revealed \cite{hetero2} that the structural relaxation dynamics is spatially heterogeneous but cannot be
understood as a statistical bond breaking process which is thought to be the dominant process to viscous flow \cite{ediger}. In addition, the high particle mobility seems to propagates continuously through the melt. It has been furthermore demonstrated that, on intermediate time scales, a small fraction of oxygen and silicon atoms deviate from a Gaussian behaviour and are more mobile than expected from (equation (\ref{glotz_eq})). These highly mobile particles form transient clusters larger than those resulting from random statistics, indicating also that the dynamics is spatially heterogeneous \cite{nongauss2b}. From a Monte Carlo study of silica \cite{pre_berthier} the emergence of a heterogeneous dynamics is also suggested, and thought to be connected to a decoupling between translational diffusion and structural relaxation (see below), and to a growing four-point dynamic susceptibility. However, dynamic heterogeneity appears to be less pronounced than in more fragile glass-forming models, although not of a qualitatively different nature \cite{nongauss2b}.

\subsubsection{Stokes-Einstein breakdown}

The presence of intermittent dynamics with atoms having a different motion depending on time and space (Fig. \ref{dh_1}), leads to the occurence of a non-Gaussian diffusion that can be detected when $F_s(k,t)$ and the first term of the expansion of (equ. \ref{expansion}), $\exp[\langle r^2(t)\rangle]/6Dt$, are directly compared. While both will nearly overlap at the high temperature, there is a progressive deviation setting on as the system approaches $T_g$, and this indicates that the relaxation time (or the viscosity) derived from $F_s(k,t)$ decouple from diffusivity (given by $\langle r^2(t)\rangle$), and have a different behaviour with temperature, a feature also known as Stokes-Einstein breakdown, i.e. the product $\eta.D$ ceases to be constant (equs. (\ref{stok1}),(\ref{eyring})) or ceases to fulfill the dispersion relation $\tau=1/q^2D$. This decoupling of transport coefficients is usually tracked from the Stokes-Einstein ratio $R_{SE}\equiv D\eta/T$ or the Debye-Stokes-Einstein ratio $R_{DSE}\equiv \eta/\tau T$, and a separate calculation or measurement of both $\tau$ and $D$ has acknowledged the decoupling at low temperature in several systems, from organic \cite{se1,se1b,se1c}, hydrogen- bonded \cite{se2} to metallic liquids \cite{se2b,se2t,se2q}. 
\par
\begin{figure}
\begin{center}
\end{center}
\caption{\label{dh_2} Decoupling between Green-Kubo (GK+scaling) and Stokes-Einstein (SER) calculated viscosities using the calculated diffusivities in liquid GeTe \cite{se3}. Permission from J. Wiley and Sons 2015.}
\end{figure}

In structural glass-forming liquids (e.g. GeTe \cite{se3}), it has been found that a very high atomic mobility ($D\simeq$10$^{-6}$~cm$^2$.s$^{-1}$) remains important down to T$\simeq$ T$_g$, indicating the breakdown of the Stokes-Einstein relationship (Fig. \ref{dh_2}) that connects, indeed, with dynamic heterogeneities, as also suggested from crystallization measurements on a similar system (Ge-Sb-Te \cite{se3b}). For this particular GeTe system, the high atomic mobility results from zones of fast and slow moving atoms, the former containing a large fraction of homopolar (GeGe) defects \cite{se4}. 
\par
In order to quantify the decoupling and temperature evolution of both $R_{SE}$ and $R_{DSE}$, a fractional Stokes-Einstein relationship has been introduced \cite{se1,se1b,se1c}, i.e. $D$ now scales as $\tau^{-\zeta}$ where $\zeta$ relates to the characteristic spatio-temporal lengthscales involved in the hetergeneous dynamics, and a typical value has been found to be of about $0.82-0.85$ for different glass formers \cite{se1,se4b,se5}. Here, $\zeta$ has been proposed to derive from temperature dependent scaling exponents of both diffusivity and relaxation time, respectively \cite{se5b}.
However, the validity of this fractional Stokes-Einstein relationship has been questioned from a separate investigation of a series of silicate liquids \cite{se6} which emphasize the two fundamentally different mechanisms governing viscous flow and conductivity/diffusivity. Separate fits of resistivity and viscosity curves lead, indeed, to different temperature dependences that can be appropriately modelled by the AM (equ. (\ref{avramov_equ}), \cite{avramov}) and MYEGA (equ. (\ref{myega}), \cite{mauro_pnas}) functional forms, respectively, and which lead to a decoupling of diffusivity and viscosity at low temperature without invoking the need of a fractional Stokes-Einstein relation. Alson there is no general agreement on the temperature region over which decoupling of transport coefficients is supposed to onset. While building on the notion fractional SE relationship, a systematic study of glass-forming liquids including B$_2$O$_3$, SiO$_2$, GeO$_2$ and soda-lime silicate glasses \cite{se4b} indicates, indeed, that the breakdown of the SE relationship should occur at much higher temperatures, i.e. at viscosities of about 10$^{2}$ Pa.s, a value that is 8-10 decades lower that the one found for $T_g$.

\subsection{Energy landscapes}

As already mentioned in different examples above \cite{sciortino_ag1,sciortino_ag2,sciortino_ag3}, numerical simulations also allow to study in detail the link between thermodynamics and glassy relaxation by using the framework of energy landscapes. This school of thought traces back to the seminal contribution of Goldstein \cite{pel1} who identified what aspects of glassy dynamics connect to the relevant features of the topography of landscapes: saddles, minima, peaks, basins with an important emphasis on the description of potential energy barriers which contribute to the slowing down of the dynamics, and contribute, on the overall to a statistical definition of activated dynamics that is encoded in the previously introduced activation energy $E_A$ for viscosity or diffusion. This has led to the definition of the energy landscape picture in which a high temperature liquid is able to sample the entire phase space and, correspondingly, the energy landscape. Because the thermal energies will be of the same order as the heights of the potential energy barriers. As the temperature is lowered, the potential energy landscape will affect the dynamics and thermal energy fluctuations will still allow the liquid to make transitions over energy barriers from one local minimum to another, i.e. activated dynamics. There is a clear separation of the time scales for vibration within one minimum and for transitions from one minimum to another. Once quenched to a glass, the system will be stuck in some local minimum, given that the barrier heights are now much larger than the thermal fluctuations the rearrangement of atomic positions takes essentially place in small regions of the landscape. 

\begin{figure}
\begin{center}
\end{center}
\caption{\label{pel1} Potential energy landscape of a glass-forming liquid. Local energy minima (inherent structures, IS \cite{pel2}) contribute to the global shape of the landscape which also contains a deep minimum corresponding to the crystalline polymorph, and a local minimum with the lowest energy corresponding to an "ideal" glass. In the glass transition region, the landscape dominates the dynamics. The red dot signals a local minimum corresponding to an IS.}
\end{figure}

\subsubsection{Inherent structures} MD simulations appear to be very helpful in order to conveniently characterize the multidimensional potential energy hypersurface created by a large number of interacting atoms or molecules. The notion of "{\em inherent structure}" (IS) has been introduced \cite{heuer3,pel2,pel2b,pel2c}, and this notion permits to uniquely separate the complex landscape topography into individual "{\em basins}", each containing a local potential energy minimum or IS (Fig. \ref{pel1}). The search of ISs is usually performed by starting at an original point in configuration space (atomic positions), and performing a steepest descent minimization of the potential energy function by changing atomic coordinates locally until a local minimum (the IS) is found. Such a procedure can be repeated and used to characterize the entire configuration space by separating landscape into regions called basins, all points in a basin having the same IS. There are various ways to detect numerically such IS and to classify their characteristics, depth or curvature, the latter being related to the eigenmode frequencies of the system \cite{heuer2}. Also shared basin boundaries are defined by saddles or transition states which allow to move from one basin to another one and a variety of techniques have been developed in the recent years, such as activation -relaxation \cite{pel3,pel4}, steepest descents \cite{heuer1}, basin hopping global optimization \cite{pel5} or other graph-connected approaches \cite{pel6,pel7,pel8}, some of them applying only to clusters.  

\subsubsection{Light formalism}

For the case of glassy relaxation, the starting point is a system of $N$ particles interacting {\em via} a potential $V({\bf r})$. Following the Stillinger-Weber formalism (see e.g. \cite{heuer2,wales}), one usually decomposes the position-related contribution $Q(T,V)$ of the partition function :
\begin{eqnarray}
\label{zzero}
Z(T,V)={\frac {1}{N!\lambda^{3N}}}Q(V,T)={\frac {1}{N!\lambda^{3N}}}\int_V e^{-\beta V({\bf r}^N)}d{\bf r}^N
\end{eqnarray} 
into contributions arising from the local minima having an inherent structure energy $e_{IS}$:
\begin{eqnarray}
Q(T,V)=\sum_ie^{-\beta e_{IS_i}}\int_{basin\ i}e^{-\beta\Delta V({\bf r}^N)}d{\bf r}^N
\end{eqnarray}
where $\Delta V({\bf r}^N)\equiv V({\bf r}^N)-e_{IS}$ is the value of the energy in the local minimum (IS), and the sum exluces basins having some crystalline order. By averaging over all distinct basins having the same energy $e_{IS}$, and counting the number of basins of energy $e_{IS}$:
\begin{eqnarray}
\Omega(e_{IS})=\sum_i\delta_{e_{{IS}_i}e_{IS}}
\end{eqnarray}
one can write a partition function that is averaged over all distinct basins with the same energy $e_{IS}$:
\begin{eqnarray}
Q(e_{IS},T,V)={\frac {\sum_i\delta_{e_{{IS}_i}e_{IS}} e^{-\beta e_{IS_i}}\int_{basin\ i}e^{-\beta\Delta V({\bf r}^N)}d{\bf r}^N }{ \sum_i\delta_{e_{{IS}_i}e_{IS}} }}
\end{eqnarray}
which leads to a basin related partition function and free energy:
\begin{eqnarray}
Z(T,V)=\sum_{e_{IS}}\Omega(e_{IS})e^{-\beta f_{basin}(e_{IS},T,V)}=\sum_{e_{IS}}e^{-\beta [-TS_{conf}(e_{IS})+f_{basin}(e_{IS},T,V)]}
\end{eqnarray}
\begin{eqnarray}
-\beta f_{basin}(e_{IS},T,V)\equiv \ln\biggl[ {\frac {Q(e_{IS},T,V)}{\lambda^{3N}}}\biggl]
\end{eqnarray}
which leads to the definition of the configurational entropy $S_{conf}(e_{IS})$:
\begin{eqnarray}
S_{conf}(e_{IS})\equiv k_B\ln \Omega(e_{IS})
\end{eqnarray}
so that the free energy can be reduced to the free energy of the typical basins and the number of such basins with a given IS energy explored during at a temperature T.
\par
In practice and as mentioned before, IS configurations at temperature T are explored and enumerated by e.g. a steepest descent minimization \cite{sciortino_ag4,steep}. In addition, the full calculation of the configurational entropy and free energy needs an explicit account of the vibrational (curvature) contributions to each minimum, and a contribution from anharmonic effects that are determined by difference \cite{wales} from the calculated total free energy F(T,V) of the system (derived from equ. (\ref{zzero}) which is evaluated by thermodynamic integration \cite{heuer2}. 

\subsubsection{Main results}

A certain number of studies have focused on the link between the energy landscape characteristics, the inherent structures and the glassy relaxation of a supercooled liquid. Ultimately, connections with the dynamics can indeed be made, and reveal e.g. that the diffusivity obeys \cite{sciortino_ag1} an Adam-Gibbs like relationship (Fig. \ref{adam_gibbs_verif}, left).
\par
In supercooled silica, potential energy landscapes have been investigated \cite{sciortino_ag3,pelsi1,pelsi2,pelsi3}, and have revealed that the distribution of IS energies significantly
deviates from a Gaussian distribution \cite{sciortino_ag3}, a result that seem to be connected with the progressive formation of a defect free tetrahedral network which as a ground state for the system \cite{pelsi3}. As a result the configurational entropy $S_{conf}$ does not appear to extrapolate to zero at finite temperature \cite{sciortino_ag3,pelsi4}, and this suggests the absence of a finite Kauzmann temperature (Fig. \ref{scio}) at select conditions. 
\begin{figure}
\begin{center}
\end{center}
\caption{\label{scio}Temperature dependence of $S_{conf}$ of supercooled silica at different densities (adapted from \cite{sciortino_ag3}). Dashed lines represent possible extrapolations, and may indicate the possibility of an entropy crisis only for selected densities.}
\end{figure}
Another key result is that for small systems the typical timescale involved in the pseudo-periodic motion between two adjacent inherent structures can be very long \cite{pelsi5}, and for certain systems of about 8 times the average relaxation time \cite{pelsi6}. However, this type of local dynamics does not contribute to the structural relaxation of the supercooled liquid, but at low temperature there are techniques \cite{pel3} which activate the dynamics in order to escape from such large basins connecting two IS with a low energy barrier. Using such an IS analysis, the viscosity can be decomposed \cite{lacks1} into a structural contribution that is associated with energy minima, and a vibrational contribution, the former leading to strain activated relaxation, while the latter is purely Newtonian, and this has also implications for the fragility behavior \cite{lacks2}. 
\par
Given that chalcogenides are usually studied from {\em ab initio} simulations that lead to system sizes that are considerably smaller (see however \cite{n480}), such potential energy landscapes approaches have not been considered and applied to these materials so far. Furthermore, such relatively small systems would be relevant for understanding supercooled liquids only at high temperature given that correlated motions of particles grow as the temperature is lowered in the landscape-influenced r\'egime \cite{pelb1}. Studies on transitions in small systems have shown, indeed, that a system with a small number of atoms can be trapped in metabasins with a wide variety of energies and lifetimes at temperatures in the landscape-influenced region \cite{pelb2,pelb3}, and because hops between such metabasins are correlated, only a limited number of particles will introduce a bias in the dynamics. 

\section{Ageing}\label{ageing}

Although relaxation times of glasses exceed common observation timescales, physical properties still continue to evolve with time at temperatures below $T_g$. An increase of the observation time will, therefore, permit to detect the equilibration of the system at lower temperatures. Much below $T_g$ however, such equilibrium relaxation times become so huge that they are clearly out of reach, e.g., one has an increase to relaxation times of the order of 10$^{14}$s, given that $\tau(T/T_g)/\tau(1)=\exp[E_A/T]$, and assuming an Arrhenius activation energy of about 1~eV, and a glass transition temperature $T_g\simeq500~K$ both values being typical of network glasses. In ageing experiments, one therefore focuses on temperature intervals which are close to $T_g$, i.e. $T/T_g\simeq0.8$. For such temperatures, physical ageing can be followed over months and years, as detailed below. 
\par
The experimental protocol for an ageing experiment is well established. As depicted in Fig. \ref{thermal}, an equilibrated supercooled liquid is abruptly quenched to a temperature $T_w\leq T_g$ at $t_w$=0 which corresponds to the beginning of the experiments. Physical properties are then recorded as a function of time but, because of the off-equilibrium nature of the system, such properties will also depend in a non trivial fashion on the waiting (ageing) time $t_w$ before the measurements is performed. In fact, while left unpertubated, the glass will continue to relax and will attempt to reach thermal equilibrium, and the way it relaxes depends on the temperature $T_w$ at which the ageing experiment is performed. As a result, the measurement will not only depend on the time $t_w$ but also on $T_w$.   
\subsection{Time-correlators}


Mean-field glass models \cite{cug1,cug2} originally designed for spin glasses \cite{cug3} have been introduced in the context of ageing, and have emphasized the central role played by broken ergodicity. In such approaches, thermal equilibrium cannot be reached, and ageing results from a downhill motion of an energy landscape that becomes increasingly flat. One major result of such approaches is that the time-translational invariance typical of ergodic systems is broken so that time correlators for any observable must be defined in the ageing r\'egime, and these now depend explicitely on both times $t$ and $t_w$, as do the response functions of the system. Mathematically, such complex time evolutions can be cast into two-time dependent fucntions, namely i) a two-time correlation function defined by:
\begin{eqnarray}
C(t,t_w)=\langle A(t)A(t_w)\rangle - \langle A(t)\rangle\langle A(t_w)\rangle,
\end{eqnarray}
with $t\leq t_w$, and where $A(t)$ is a typical observable (e.g. the intermediate scattering function $F_s(k,t)$), and ii) the response function $G(t,t_w)$ given by:
\begin{eqnarray}
G(t,t_w)=\biggl({\frac {\delta \langle A(t)\rangle}{\delta h(t_w)}}\biggr)_{h=0}
\end{eqnarray}
where $h(t_w$) is a conjugate field to the observable $A$, and the bracketts indicate averages over the thermal history. For instance, if $A(t)$ is the average 1D position of the particles $A(t)=1/N\sum_i\langle x_i(t)\rangle$, then the response function is computed from the perturbated Hamiltonian $H=H_0-h\sum_ix_i$. In systems in equilibrium, both two-time functions are related {\em via} the fluctuation-dissipation theorem \cite{cug4} which quantifies the relation between the fluctuations in a equilibrated liquid and the response to applied perturbations. This leads to:
\begin{eqnarray}
\label{fdr}
TG(t,t_w)=\ -\ {\frac {\partial C(t,t_w)}{\partial t}}
\end{eqnarray}
or, using the integrated response: 
\begin{eqnarray}
\label{fdr1}
\Delta C(t,t_w)=C(t,t)-C(t,t_w)=T\int_{t_w}^tG(t,t')dt'
\end{eqnarray}
One important property of equ. (\ref{fdr}) and of the two functions $G(t,t_w)$ and $C(t,t_w)$ is their time translation invariance at equilibrium, i.e. all depend only on $t-t_w$. 
Once the system becomes out of equilibrium at $T<T_g$, a major consequence is that fluctuation-dissipation theorem does not hold, this behaviour being fulfilled in glasses displaying ageing behavior for which the response functions and the FDT become waiting time-dependent (i.e. on $t_w$), and stop being invariant under $t-t_w$. A convenient way is to rewrite the FDT of equ. (\ref{fdr}) using an effective temperature $T_{eff}$ that also leads to the introduction of a fluctuation-dissipation ratio (FDR) $X(t,t_w)$:
\begin{eqnarray}
T=T_{eff}X(t,t_w)
\end{eqnarray}
and, by definition, on has at equilibrium $X(t,t_w)$=1 and $T_{eff}$=$T$. From a fundamental viewpoint, the introduction of an effective temperature has a profound implication for the meaning and the measurement of temperature in a system under aging. In fact, since the temperature is related to the timescale (equilibration) and to particle velocities (equipartition), its measurement in a glassy state for which both relaxation times and velocities are time and spatiallly dependent, (heterogeneous dynamics) poses a true challenge \cite{cug4b}, the difficulty being to define an appropriate thermometer. For certain simple models, additional degrees of freedom representing the thermometer can be coupled in a simple fashion to an observable of interest, and this allows to relate the measured temperature to the correlation function and the integrated response \cite{cug4c}. While this definition of $T_{eff}$ has been found to be rather appropriate in a certain number of systems or models \cite{cug5,cug6} or investigated within the energy landscape formalism \cite{cug7}, its robustness has been also questioned \cite{cug10,cug11}. We refer the reader to reviews that specifically focus on this topic \cite{review_ageing1,review_ageing2}.
\begin{figure}
\begin{center}
\end{center}
\caption{\label{ageing1} Response versus correlation calculated in the aging r\'egime of a silica glass \cite{prl_berthier}. At short times, the plots converge to a straight line of slope 1, whereas at large times (small values of $C(t,t')$) a slope $X\simeq$0.51 is found, yielding an effective temperature of $\simeq$4900 K. Permission from American Physical Society.}
\end{figure}
\subsection{Insight from trap models}
A simple way to understand the physics of ageing is directly derived from the Goldstein picture of energy landscapes, and uses trap models \cite{bouchaud1,bouchaud2,bouchaud3,bouchaud4} in which particles can move from one local minimum of the complex landscapes to another. Here, these  local minima are seen as metastable configurations with high energy barriers so that such minima can act as traps and hold of the glassy system during a certain finite time.
\par
The central question of the approach is to ask what could be the distribution of such trapping times, and a simple answer can be brought assuming that there exists a "{\em percolation}" energy level $f_0$ below which the minima are disconnected. For $f>f_0$, it is possible to hop between any two states given that the energy barrier is equal to $\Delta E=f_0-f$, and the system can relax to lower energies. 
\par
An interesting outcome is the existence of a cross-over between two ageing r\'egimes illustrated by e.g. a magnetization function $m(t)$ of a spin system \cite{bouchaud2} that is exponential:
\begin{eqnarray}
m(t)=m_0\ e^{-{\frac {\gamma}{1-x}}(t/t_w)^{1-x}},
\end{eqnarray}
and is a power law $m(t)=(t/t_w)^{-\gamma}$ for $t\ll t_w$ and $t\gg t_w$, respectively. Here, $x$ characterizes the distribution of free-energies in the glassy phase \cite{bouchaud1}, and $\gamma$ is related to the probability to relax when leaving a trap. In the short time domain, these models reproduce the stretched exponential decay typical of $\alpha$-relaxation, and are also found to depend on $t/t_w$ only. On a more general ground, such off-equilibrium statistical mechanics models capture some of the salient features of the dynamics of aging. Here, the phase space is seen as a large collection of metastable states which induce a broad distribution of lifetimes. When the average lifetime of these metastable states diverges, all the physical observables are dominated by the properties of the deepest state. 
\par
A certain number of models with different distributions of trap depths (Boltzmann \cite{bouchaud5}, Gaussian \cite{bouchaud1}) lead to the usual features of glassy relaxation, i.e. a power-law for simple dynamic variables \cite{bouchaud6}, a super-Arrhenius behavior for the relaxation with a diverging temperature $T_0$ \cite{bouchaud4} or a stretched exponential decay.

\subsection{Molecular Dynamics Simulations}

Computer simulations can directly probe the enunciated breakdown of the fluctuation-dissipation ratio (equ. \ref{fdr}), and a separate computation of the integrated response and the correlation function leads to a straight line with slope -1/$T$ (equ. \ref{fdr1}, Fig. \ref{ageing1}) in an equilibrated liquid \cite{cug5,prl_berthier,cugbis1}, and -1/$T_{eff}$ in a system subject to ageing, the latter situation occuring for small values of the correlation functions. A break in slope permits the detection of ageing r\'egimes, and  provides an approximate limit of equilibration. Such simulations also allow to verify the behavior predicted from trap models \cite{vollmayr0000} or to connect to dynamic heteregeneities, and for a selected number of strong and fragile glass-forming systems (including silica), the calculated 4-point density susceptibility $\chi_4(t,t_w)$ and the dynamic correlation length
$\xi_4(t,t_w)$ have qualitatively the same behaviors as a function of $t_w$ and $t-t_w$ \cite{vollmayr000}. Also, the tails of the displacement distributions, show a qualitatively different evolution with $t_w$ in the case of more fragile liquids, and this evolution appears to be associated with the particles which have diffused the
most. 
\par
The local aging dynamics can also be characterized, and this can eventually connect back to dynamic heterogeneities for model structural glasses \cite{vollmayr00}. In silica, aging seems to result from single particle trajectories and jump events corresponding to the escape of an atom from the cage formed by its neighbors \cite{vollmayr}. It has been found that the only $t_w$-dependent microscopic quantity is the number of jumping particles per unit time \cite{vollmayr0}, and this quantity has been found to decreases with age. The second key finding is that for such strong glass-forming systems but also in more fragile ones \cite{vollmayr1,vollmayr2}, neither the distribution of jump lengths that relates to the self-part of the Van Hove correlation functions (equ. (\ref{vanhove_def})) nor the distribution of times spent in the cage are $t_w$ dependent. Except silica, we are not aware of any other MD study on ageing in network-forming glasses.

\subsection{Applications}
\subsubsection{Oxides}
Ageing effects in oxides and, particularly silicates, have been first reported from the seminal work of Joule \cite{joule1,joule2} on silicate glass thermometers, and then studied in detail by Nemilov \cite{aging_golo5,nemilov} and Nemilov and Johari \cite{gold_nemilov} who investigated changes of various physico-chemical properties for ageing times ranging from hours to decades. From these studies, it has been stated that the completeness of aging of glass at any age is determined by its aging rate after about one year \cite{nemilov}, and nearly forty glasses have been ranked with respect to this criterium. For the special case of silicates, one should keep in mind that glass transition temperatures in such systems are quite high so that aging experiments performed at ambient conditions only show small variations given the important difference with $T_g$. However, this seems to be contradicted by the observed changes in e.g. density with time \cite{gold_nemilov} which have been found to have characteristic relaxation times much shorter than those of an $\alpha$-relaxation process. On this basis, it has been proposed that the structural changes occur on timescales typical of a $\beta$-relaxation that is typical of a cage-like dynamics (Fig. \ref{stretchf}). Densification is then seen as the result of angular changes between SiO$_{4/2}$ tetrahedra which induce local strained regions in the glass, and a subsequent dissipation of this strain energy, the former process being much slower and determining the kinetics of ageing. The link with structural features driving the effect of aging have been also detected on phosphate \cite{siliag1} or silica-based glasses \cite{siliag2} from spectroscopic studies.
\par
The use of x-ray photon correlation spectroscopy appears to be an interesting probe for the investigation of structural relaxation processes, and by using different thermal histories (cooling rates), one can observe a complex hierarchy of dynamic processes that are characterized by distinct aging regimes. These features are seen in metallic glasses \cite{xray20,xray21} but also in silicates \cite{rruffle}), and find strong analogies with the aging dynamics of more softer glassy materials \cite{weeks}, while also pointing stress relaxation as a universal mechanism driving the relaxation dynamics of out-of-equilibrium systems. This has been also acknowledged for a borosilicate glass showing stress relaxation under aging at $T/T_g\simeq$0.3 \cite{stress5bc} but contradicts the qualitative mechanism sketched from density changes with aging \cite{gold_nemilov}.

\subsubsection{Chalcogenides}
Because of their relatively low glass transition temperature allowing for ageing experiments at ambient temperature, there is a quite important body of literature on the effect of ageing on various thermal, mechanical, and dielectric properties in polymers \cite{aging_golo1,aging_golo3,aging_golo4,aging_golo6,aging_golo7} that has inspired work on chalcogenide network glass-forming liquids such as As-Se or Ge-Se \cite{aging_golo2,golo_book}, mostly accessed experimentally, and from techniques such as DSC or mDSC. Readers interested in this topic with promising applications should refer to an excellent review article \cite{golo_book} which also contains more specific aspects of ageing such as thermally- or irradiation-induced effects.
\par 
There is a crucial dependence of $T_w$ on the maximum enthalpy $\Delta H_{\infty}$ that can be released upon heating an infinite aged sample, and Saiter \cite{saiteri} has proposed that it scales as:
\begin{eqnarray}
\label{saiter_eq}
\Delta H_{\infty}=\Delta C_p(T_g)\biggl(T_g(q^+)-T_w\biggl)
\end{eqnarray}
where $T_g$ is the glass transition determined by the scan with a fixed heating rate $q^+$, and $\Delta C_p$ is the jump in specific heat at $T_g$. From equ. (\ref{saiter_eq}), it becomes clear that the measurement of ageing will strongly depend on $T_w$, and eventually cancel in case of rejunevation when $T_w\simeq T_g$. Given this relationship (equ. (\ref{saiter_eq})), long-term physical ageing has been investigated in these chalcogenide networks \cite{long_term1,long_term2,long_term3}. Also, since the enthalpic loss is directly related to $T_g-T_w$, ageing is enhanced for compositions which have a lower $T_g$ (i.e. usually chalcogen-rich \cite{gese}) so that the enthalpic relaxation can be measured within a short period, and seems to follow a sigmoidal time dependence \cite{angell3}.
\par
There are aspects of structure that have been characterized in connection with this topic, in particular from Raman spectroscopy \cite{long_term1,stru_ageing1}, NMR \cite{stru_ageing2}, and all indicate the weak changes in short range order under ageing, also characterized from X-ray absorbtion \cite{stru_ageing2}, a conclusion that is consistent with a large dynamic correlation length \cite{hetero000}. When two types of glasses are compared (20 years aged and rejuvenated), Raman spectroscopy of As-Se \cite{long_term1} seem to indicate small changes changes in bond statistics (as for NMR), and for Se-rich glasses, it was concluded that Se ring-like structure collapse and lead to a reorganization of chain fragments between AsSe$_{3/2}$ units. On the overall, these different probes signal that As-Se-Se-As motifs convert into As-Se-As and As-Se-Se-Se-As fragments during ageing. This chain switching mechanism is actually supported from {\em ab initio} simulations of elemental Se \cite{stru_ageing3} which show fast ($\simeq$100~fs) changes in chain structure involving defect coordinations ($r_{Se}$=1,3) that give support to proposed valence alternation pairs for light induced structural changes \cite{stru_ageing4}.  

\subsubsection{Ageing in isostatic glasses}
Isostatic network glasses are found to display a significantly reduced tendency to ageing and this has been detected for a certain number of systems. Experiments on Ge-P-Se across the isostatic phase \cite{jpcm2005} shows that the deep and wide reversibility window in these chalcogenides sharpens and gets deeper as glass compositions outside the window age at 300 K over different periods (figure \ref{ageing_ip}a). 

\begin{figure}
\begin{center}
\end{center}
\caption{\label{ageing_ip} Ageing effect on the non-reversing heat enthalpy $\Delta H_{nr}$ in chalcogenide networks as a function of ageing. a) Ge-P-Se glasses (adapted from \cite{jpcm2005}). b) As-Se glasses (adapted from \cite{joam2002}) For window compositions (As$_{30}$Se$_{70}$), the evolution with waiting time is substantiall reduced.
}
\end{figure}
It is to be noted that the experimental protocol does not follow the one usually designed for ageing studies in e.g. spin glasses for which the system is maintained at a fixed $T/T_g$ \cite{bouchaud2}. Here, given that $T_g$ is a function of glass composition, the effective ageing temperature $T_w/T_g$ itself will also vary with composition. Floppy glasses which are below the IP window age significantly over a 3 month waiting period (Fig. \ref{ageing_ip}b), while stressed-rigid glasses (above the window) age somewhat slower, over $\simeq$ 5 months, an observation that directly results from the slower ageing kinetics connected with higher glass transition temperatures. In such IP glasses, there is weak evidence of ageing, even after a 5 month waiting period. Similar results have been found for As-Se glasses \cite{joam2002} with a weak evolution of the non-reversing enthalpy for reversible glasses, and contradicting results on this topic \cite{golo_asse} have been further analyzed \cite{mdsc_asse} and could be interpreted as a result of nanoscale phase separation resulting from light exposure. It has been furthermore pointed out that a proper ageing procedure (i.e. at fixed $T_w/T_g$ for all compositions) may not lead to the anomalous behavior observed in Figs. \ref{ageing_ip}b), and measurements on Ge-Se glasses using DSC could not reproduce the generic behavior proposed for isostatic glasses \cite{sen_ageing}.
\par
Finally, it must be emphasized that sophisticated experimental procedures, multiple cycles of cooling, heating and waiting times, modulation of the applied external fields, can lead to spectacular effects of aging in glassy materials, such as rejuvenation and memory \cite{golo_book}.

\section{Conclusions and perspectives}

At this stage, rather than summarizing the different topics covered in this article, we would like to emphasize that specific features typical of relaxation in supercooled liquids could benefit from the low temperature description of the corresponding glassy materials. This is particularly relevant in the case of network glasses. There has been recent efforts to bridge the gap between theoretical approaches and experimental methods used or derived from the liquid side and the glassy side of the glass transition \cite{fin1}. Progress have been slow in coming but more and more methods are being applied for this purpose, and this might be particularly crucial as one considers network glass-forming liquids.  
\par
Here we have reviewed the ongoing effort and studies that have been reported in the literature in order to characterize and understand the physics of the glass transition and related aging phenomena once a system is maintained at $T<T_g$. We have focused on the special case of network-forming glasses, i.e. on materials which are dominated by their low temperature underlying structure. These glasses are often thought to  systematically have low fragilities and to belong to the category of strong glass formers, albeit this is contradicted by various experimental measurements \cite{r13,asse_punit}. Most of the recent simulation work on inorganic supercooled liquids, and especially the one focusing on dynamic heterogeneities \cite{hetero2,heterob3} is restricted to liquid silica which has a fragility index of ${\cal M}$=20. Investigation of other typical network formers (B$_2$O$_3$, GeSe$_2$,...) are wellcome. While similar features with fragile liquids have been emphasized \cite{nongauss2b}, there is probably much to learn from an investigation of network glasses because an appropriate alloying allows to tune physical properties (e.g. structure) in a continuous fashion, that can in turn be connected or correlated to dynamic properties such as fragility. Studies of the compositional dependences and the detection of anomalies in dynamic or relaxation properties are, therefore, believed to represent an interesting and additional means to learn more on the glass transition phenomenon.
\par
In this respect, recent efforts \cite{mauro_gupta} attempting to derive approaches that use Rigidity Theory in a substantially revised version, represent attractive pathways for an improved quantitative description of glassy dynamics. Here, it is assumed that relaxation is controlled by aspects of structure, topology and/or rigidity, or, more generally, by features of the low temperature glass. This fact has been acknowledged by various authors (e.g. \cite{scopigno}), although methods and models are often based on equilibrium statistical mechanics that bear obvious limitations once they are applied to the glassy state. A promising way to combine both approaches, from the glass to the liquid state is provided by molecular dynamics based constraint counting \cite{cons2} which permits to calculate various properties from ensemble averages, and to connect to Rigidity Theory {\em via} an explicit account of the topological constraints. From a more applied viewpoint, given the general use of such simulations in the description of glassy materials, this recent extension now offers the possibility to rationalize the design of new families of other materials using as input the rigidity state of the underlying atomic network, as recently demonstrated \cite{bauchy_prl}, and the corresponding glassy dynamics can be investigated.
\par 
This link between network rigidity and the thermodynamics and relaxation of supercooled liquids seems to have an even more general ground as recently emphasized \cite{wyart1}, and these ideas can actually also be extended to other glass forming liquids including fragile ones \cite{scopigno}. Indeed, glass elasticity and the presence of soft elastic modes have been found to drive many aspects of glassy relaxation as mentioned throughout this review, and also to relate to thermodynamic changes across the glass transition. In fact, an abundance of such soft modes permits exploration of the phase space without large changes in energy \cite{naumis3}, and this ultimately leads to small changes in the specific heat. This, of course, connects back to the notion of floppy modes \cite{mft1} that are present in weakly connected (flexible) network glasses.
\par
Finally, it would be interesting to consider in a combined fashion aspects from the statistical mechanics of off-equilibrium systems \cite{review_ageing2}, molecular simulations and Rigidity Theory to investigate the ageing of chalcogenide glasses, and this would allow to describe these phenomena beyond the qualitative level (Fig. \ref{ageing_ip}). Given the number of important applications of chalcogenides in optoelectronics \cite{mitkova,golo_book}, the understanding and, eventually, the control of aging phenomena could improve the stability of devices using chalcogides as base material, as recently stressed in a study on optical phase change recording \cite{wuttig_fin}. Still, a certain number of challenges remain that are inherent to some of the methods employed : small system sizes due to {\em ab initio} simulations in order to treat correctly the covalent or semi-metallic bonding, and the short timescale of the MD simulations. Despite these limitations, such methods exhibit a certain number of promising results on this topic for the archetypal SiO$_2$ \cite{prl_berthier}, and might well be applied in a similar fashion to its parent chalcogenide system with a wealth of possible applications, directly derived from the very basic features of non-equilibrium statistical mechanics.

\section*{Acknowledgments}
Support from Agence Nationale de la Recherche (ANR) (Grant No. 09-BLAN-0109-01 and Grant No. 11-BS08-0012) is gratefully acknowledged. MM acknowledges support from the French-American Fulbright Commission, and from International Materials Institute (H. Jain). GENCI (Grand Equipement National de Calcul Intensif) is acknowledged for supercomputing access. The authors thank M. Bauchy, C. Massobrio, M. Wyart, S. Le Roux, M. Boero, C. Bichara, P. Boolchand, N. Sator, A. Pradel, B. Guillot, J.C. Mauro, M. Smedskjaer, S. Boshle, K. Gunasekera, J. Du, M. Malki, G.G. Naumis, P.S. Salmon, A. Pasquarello, Y. Yue, J.-Y. Raty, M. Salanne, G. Ferlat, J.C. Phillips, P. Simon, M. Salanne, D. De Sousa-Meneses, R. Vuilleumier, S. Ravindren, N. Mousseau, S. Chakraborty for stimulating discussions. 

\section*{References}

\end{document}